\documentclass[manuscript]{acmart}

\usepackage{colortbl}
\usepackage{multirow}
\usepackage{float}
\usepackage{tikz}
\usepackage{xcolor}
\usepackage{fancyvrb}
\usepackage{lscape}
\usepackage{geometry}
\usepackage{subcaption}

\usepackage{adjustbox}

\newlength{\RoundedBoxWidth}
\newsavebox{\GrayRoundedBox}
\newenvironment{GrayBox}[1][\dimexpr\linewidth-4.5ex]%
   {\setlength{\RoundedBoxWidth}{\dimexpr#1}
    \begin{lrbox}{\GrayRoundedBox}
       \begin{minipage}{\RoundedBoxWidth}}%
   {   \end{minipage}
    \end{lrbox}
    \begin{center}
    \begin{tikzpicture}%
       \draw node[draw=black,fill=black!10,rounded corners,%
             inner sep=2ex,text width=\RoundedBoxWidth]%
             {\usebox{\GrayRoundedBox}};
    \end{tikzpicture}
    \end{center}}

\newsavebox{\WhiteRoundedBox}
   {\setlength{\RoundedBoxWidth}{#1}%
    \begin{lrbox}{\WhiteRoundedBox}%
       \begin{minipage}{\RoundedBoxWidth}}%
   {   \end{minipage}%
    \end{lrbox}%
    \begin{center}%
    \begin{tikzpicture}%
       \draw node[draw=black,fill=white,rounded corners,%
             inner sep=2ex,text width=\RoundedBoxWidth]%
             {\usebox{\WhiteRoundedBox}};%
    \end{tikzpicture}%
    \end{center}}
    
\usepackage{listings}
\usepackage{subcaption}
\AtBeginDocument{\DeclareCaptionSubType{lstlisting}}
\definecolor{codegreen}{HTML}{567A0D}
\definecolor{codegray}{HTML}{999999}
\definecolor{codeblue}{HTML}{015493}
\definecolor{codeorange}{HTML}{B75301}
\definecolor{backcolour}{rgb}{1,1,1}

\lstdefinestyle{mystyle}{
    backgroundcolor=\color{backcolour},   
    commentstyle=\color{gray!60!black},     % #999999 for comments
    keywordstyle=\color{blue!80!black},     % #0000FF for keywords
    numberstyle=\tiny\color{codegray},
    stringstyle=\color{green!50!black},     % #008000 for strings
    identifierstyle=\color{black},          % #000000 for regular identifiers
    basicstyle=\ttfamily\smaller,
    breakatwhitespace=false,         
    breaklines=true,
    postbreak=\mbox{\textcolor{red}{$\hookrightarrow$}\space},
    captionpos=b,                    
    keepspaces=true,                 
    numbers=left,                    
    numbersep=5pt,                  
    showspaces=false,                
    showstringspaces=false,
    showtabs=false,                  
    tabsize=4,                       
    keywordsprefix={@},             
    moredelim=[is][\color{purple!80!black}]{[*}{*]},  % #800080 for special delimiters
    xleftmargin=9pt,
    morekeywords=[2]{print,len,range,str,int,float,list,dict,set,tuple},
    keywordstyle=[2]{\color{brown!70!black}},  % #AA4926 for built-in functions
    morekeywords=[1]{
        self,None,True,False,class,def,return,yield,
        if,else,elif,while,for,break,continue,
        import,from,as,global,nonlocal,lambda,
        try,except,finally,raise,assert,with,
        async,await,pass,del
    }
}
\newcommand\majorrev[1]{\textcolor{black}{#1}}
\newcommand\minorrev[1]{\textcolor{black}{#1}}
\AtBeginDocument{%
  }

\setcopyright{acmlicensed}
\copyrightyear{2018}
\acmYear{2018}
\acmDOI{XXXXXXX.XXXXXXX}

\acmConference[Conference acronym 'XX]{Make sure to enter the correct
  conference title from your rights confirmation emai}{June 03--05,
  2018}{Woodstock, NY}
\acmISBN{978-1-4503-XXXX-X/18/06}

\begin{document}

\title{Prompting Techniques for Secure Code Generation: A Systematic Investigation}

\author{Catherine Tony}
\affiliation{%
  \institution{Hamburg University of Technology}
  %\streetaddress{1 Th{\o}rv{\"a}ld Circle}
  \city{Hamburg}
  \country{Germany}}
\email{catherine.tony@tuhh.de}

\author{Nicol\'{a}s E. D\'{i}az Ferreyra}
\affiliation{%
  \institution{Hamburg University of Technology}
  \city{Hamburg}
  \country{Germany}
}
\email{nicolas.diaz-ferreyra@tuhh.de}

\author{Markus Mutas}
\affiliation{%
 \institution{Hamburg University of Technology}
 %\streetaddress{Rono-Hills}
 \city{Hamburg}
 %\state{Arunachal Pradesh}
 \country{Germany}}
\email{markus.mutas@tuhh.de}

\author{Salem Dhif}
\affiliation{%
  \institution{Hamburg University of Technology}
  %\streetaddress{30 Shuangqing Rd}
  \city{Hamburg}
  %\state{Beijing Shi}
  \country{Germany}}
  \email{salem.dhif@tuhh.de}

\author{Riccardo Scandariato}
\affiliation{%
  \institution{Hamburg University of Technology}
  %\streetaddress{8600 Datapoint Drive}
  \city{Hamburg}
  %\state{Texas}
  \country{Germany}}
  %\postcode{78229}}
\email{riccardo.scadariato@tuhh.de}

\renewcommand{\shortauthors}{Tony et al.}

\begin{abstract}
  Large Language Models (LLMs) are gaining momentum in software development with prompt-driven programming enabling developers to create code from natural language (NL) instructions. However, studies have questioned their ability to produce secure code and, thereby, the quality of prompt-generated software. Alongside, various prompting techniques that carefully tailor prompts have emerged to elicit optimal responses from LLMs. Still, the interplay between such prompting strategies and secure code generation remains under-explored and calls for further investigations. \textbf{Objective:} In this study, we investigate the impact of different prompting techniques on the security of code generated from NL instructions by LLMs. \textbf{Method}: First we perform a systematic literature review to identify the existing prompting techniques that can be used for code generation tasks. A subset of these techniques are evaluated on GPT-3, GPT-3.5, and GPT-4 models for secure code generation. For this, we used an existing dataset consisting of 150 NL security-relevant code-generation prompts. \textbf{Results:} Our work (i) classifies potential prompting techniques for code generation (ii) adapts and evaluates a subset of the identified techniques for secure code generation tasks and (iii) observes a reduction in security weaknesses across the tested LLMs, especially after using an existing technique called Recursive Criticism and Improvement (RCI), contributing valuable insights to the ongoing discourse on LLM-generated code security.
\end{abstract}



\received{4 May 2024}
\received[revised]{TBA}
\received[accepted]{TBA}

\maketitle

\section{Introduction}
\label{sec:intro}

Large Language Models (LLMs) have received major attention recently due to their high performance in solving Natural Language (NL) processing tasks. Alongside, their application to program synthesis has advanced significantly, allowing software developers to generate code from NL descriptions or prompts. Overall, this is achieved through vast training sets of code and documentation text extracted from open-source repositories. While this approach helps LLMs produce functional implementations, it offers no guarantees of correctness or quality, as it treats code simply as text, ignoring essential semantic information \cite{jain2022jigsaw}. Moreover, open-source projects are known for containing security flaws \cite{HazhirpasandGKB19,TonyFS22,WickertBBM21,WickertREDM19}, making LLM-generated code prone to security vulnerabilities \cite{Pearce2022, PearceA0DK22}. 

Recent investigations \cite{Vaithilingam0G22} show that developers are gradually showing a preference for AI-driven code assistants to initiate their coding process. These tools offer a valuable starting point, aiding in the development process and alleviating the need to search for information online. However, when utilizing such AI assistants powered by LLMs, developers often display an over-reliance behavior that involves optimistic assumptions regarding the correctness and security of the generated code without thorough questioning \cite{PerryS0B23}\cite{SarkarN0RP022}.  
 %Prior work has zoomed into the security shortcomings of LLMs, with a special focus on code completion \cite{PearceA0DK22, He2023} and vulnerability repair tasks \cite{Pearce2022}. 
Findings from a user study conducted by \citet{PerryS0B23} revealed that participants who had access to an AI assistant tended to produce insecure solutions more frequently compared to those who did not have access to such assistance. This emphasizes the importance of exploring avenues to strengthen the security incorporated by the LLMs in the code generated by them.

\textbf{Motivation:} 
Prompt engineering, the process of refining prompts to optimize the quality of responses generated by LLMs, has garnered significant attention following the emergence of LLMs like ChatGPT, BARD, and others. A variety of sophisticated prompting techniques have been developed for tasks such as text generation, classification, and problem-solving. Many of these techniques can be used by the end users to directly prompt or interact with LLM-powered tools and chatbots. Despite the abundance of research in this field, the correlation between such prompting strategies and secure code generation has not been thoroughly examined or documented in the existing literature. Specifically, the extent to which such techniques can guide LLMs towards producing secure implementations remains an open question. While models like GPT-3 continually advance, with each version improving upon its predecessor, the implications of these enhancements for security are unclear. This underscores the importance of investigating NL prompting techniques that have the potential to enhance the security of the code generated by LLMs.

In this work, we perform a literature review to identify potential prompting techniques that can be used for code generation followed by an in-depth analysis of the impact of these techniques on improving the security in LLM-generated code. 
For this, we elaborate on the following research questions (RQs):
%In this work, we explore (i) the extent to which mainstream LLMs such as GPT-3, and ChatGPT can produce secure implementations from NL coding task descriptions and (ii) whether carefully engineered prompts can improve the security of the generated code. F

\textbf{\textit{RQ1}: What are the existing prompting techniques that can be used for code generation?}
To answer this, we performed a systematic literature review of papers that introduced different prompting techniques that can be potentially used for code generation.

%\textbf{\textit{RQ2}}: Which of the existing prompt engineering techniques are best suited for secure code generation?
%We performed a preliminary study, where we experimented with the prompting techniques identified as suitable for code generation on 5 security-relevant code generation tasks to verify the feasibility and practicality of the prompt engineering techniques. We chose Python as the programming language for the experiments since it is a popular choice for developers. 

\textbf{\textit{RQ2}: What is the impact of different prompting techniques on the security of LLM-generated code?}
For this, we conducted an in-depth analysis using a subset of prompting techniques identified in the literature review. A dataset called \textit{LLMSecEval} \cite{Tony2023}, containing 150 NL prompts specifying coding tasks that could potentially lead to insecure code implementations, was used for our experiments. 
Experiments were conducted utilizing GPT-3, GPT-3.5, and GPT-4 models, due to their widespread usage and advanced natural language processing and coding capabilities, which are crucial for exploring various prompting techniques.
We mainly evaluated Python programs generated by the LLMs since it is one of the most popular choice of languages for developers\footnote{https://statisticstimes.com/tech/top-computer-languages.php}. \majorrev{Additionally, to further explore the generalizability of the findings obtained from Python to other programming languages, we also examined C code generated by GPT-4.} The code generated by the LLMs for the selected techniques was evaluated for security weaknesses \majorrev{using Bandit \cite{bandit} as the primary tool and CodeQL \cite{codeql} serving as a supplementary analysis.}
%using two static analysis tools called Bandit and CodeQL.

 Our findings reaffirm the fact that LLM-generated code contains a large number of security weaknesses mainly related to CWE-78, CWE-259, CWE-94, and CWE-330. We observed that integrating different prompting techniques has a positive impact on the security of code generated by LLMs, particularly noticeable in advanced models like GPT-4. Notably, a technique known as Recursive Criticism and Improvement (RCI) \cite{KimBM23} has exhibited significant potential in mitigating security weaknesses in the generated code. Furthermore, we have observed distinct changes in the coding behavior of the models when security specifications are introduced to the prompts, offering insights that can be utilized to refine prompting techniques for secure code generation.

\textbf{Contributions} 
This work makes the following contributions to the field of secure code generation using LLMs:
\begin{itemize}
    %\item To the extent of our knowledge, this is the first work that identifies and classifies existing prompting techniques and demonstrates how they can be used for code generation tasks.
    %\item We adapted a selection of the identified techniques for secure code generation tasks.
    %\item We benchmark the selected techniques for secure code generation and identify promising prompting techniques to prevent security vulnerabilities in LLM-generated code.
\item To the best of our knowledge, we present the first {\em systematic inventory of prompting techniques} that are suitable for code generation. Often, papers in this field make an arbitrary selection of a few techniques, e.g., based on convenience or because other referenced papers do the same. This paper highlights that a rich selection of techniques exists and incentivizes the community to explore the alternatives in their work.
\item To simplify this exploration, we have translated a selection of these generic prompting techniques into \majorrev{7} actionable templates \majorrev{(see Table \ref{tab:prompt-templates})} that can be reused by the community as is, or with some adaptations for (secure) code generation. This effort is expected to stimulate the use of the different prompting techniques, beyond the usual suspects.
\item We provide insights (and rankings) concerning the prompting techniques that are more promising for secure code generation with \majorrev{a focus on Python, but also exploring C}. Interestingly, to the extent of our knowledge, the most promising technique \majorrev{for both languages} has not been used in the related work for secure code generation (cf. the first point).
\end{itemize}

The rest of the paper is organized as follows: Section \ref{sec:rw&background} presents the existing work on using LLMs for (secure) code generation. Section \ref{sec:slr-methodology}  and \ref{sec:slr-results} present the approach used for the systematic literature review and the findings obtained from it. Following this, Sections \ref{sec:exp-methodology} and \ref{sec:exp-results} delve into the specifics of the security evaluation of code generated by LLMs using various prompting techniques and the results. Insights obtained from the results are elaborated in Section \ref{sec:discussion} followed by a discussion on the impact of data leakage on the results in Section \ref{sec:data-leakage}. Section \ref{sec:limitations} addresses the limitations, while Section \ref{sec:conclusion} brings the work to a close.

\section{Related Work}
\label{sec:rw&background}

This section presents prior research that delves into the use of LLMs for code generation and explores studies that assess the security aspects of code generated by LLMs.

\subsection{Code Generation Using LLMs}

There are several works (both published and unpublished) that evaluate the code generation capabilities of LLMs. The following are a few notable ones that are peer-reviewed.

A paper by Hendrycks et al. \cite{HendrycksBKMAGB21} evaluated the code generated by GPT-2 \cite{gpt-2}, GPT-3 \cite{GPT-3} and GPT-Neo using a benchmark dataset called APPS (Automated Program Progress Standard) \cite{HendrycksBKMAGB21} that consists of 10,000 NL coding problems along with corresponding test cases and ground truth solutions created by humans. For the evaluation, they employed the few-shot prompting technique where the model is provided with a set of <input-output> examples to demonstrate how to solve the problem. At the time of this study, they observed that the overall performance exhibited by the models was low based on the percentage of test cases passed. %However, they found that the syntax errors in the code reduced as the models were improved using techniques such as fine-tuning. 
In another study conducted by Austin et al. \cite{austin2022}, the authors explored the limitations of program synthesis carried out by language models trained at various scales, ranging from 244M to 137B parameters. To accomplish this, they created two datasets: the Mostly Basic Programming Problems (MBPP) dataset and the MathQA-Python dataset.
The MBPP dataset comprises problem statements, simple Python functions designed to solve these problems, and three corresponding test cases. On the other hand, the MathQA-Python dataset presents mathematical problems, multiple-choice answers for these problems, and Python implementations that produce the correct answers. Both datasets are created to verify the semantic correctness of the generated Python programs.
They also employed a few-shot prompting technique and their observations revealed a correlation between the increase in model size and improved performance.

Xu et al. \cite{Xu0NH22} conducted a comprehensive assessment of various LLMs, including Codex \cite{codex}, GPT-J, GPT-Neo, GPT-NeoX-20B \cite{gpt-neo}, CodeParrot \cite{codeparrot}, and PolyCoder (a model developed by the authors of this paper) for their code generation capabilities. Their evaluation focused on these models' performance using the HumanEval \cite{codex} dataset, which contains 164 distinct coding tasks presented as prompts with corresponding test cases. These prompts consist of incomplete code snippets paired with NL comments rather than a complete NL instruction describing the task.
In this study, they employed a zero-shot prompting technique. Zero-shot prompting entails not providing explicit <input-output> pairs to the LLMs to demonstrate how to approach the given task. 
Based on this study, Codex emerged as the top-performing model, outperforming all the other models in the evaluation.

A study by Zeng et al. \cite{ZengTZLZZ22} tried to understand how pre-trained models perform for program understanding and generation tasks by experimenting with 8 LLMs that include CodeBERT \cite{codebert}, GraphCodeBERT \cite{graphcodebert}, ContraCode \cite{contracode}, CodeGPT, PLBART \cite{plbart}, CodeTrans \cite{codetrans}, CoText \cite{cotext} and CodeT5 \cite{codet5} mainly using the CodeXGLUE \cite{CodexGlue} benchmark. This benchmark is a collection of datasets spread across 10 different code-related tasks. The dataset used for code generation tasks within this benchmark is known as Concode. The prompts in Concode encompass NL problem descriptions, structured in the form of Java Doc comments and class environments. The researchers employed zero-shot prompting to evaluate the models. %The quality of the code generated by these models was measured using  metrics such as BLEU-4, Accuracy, and CodeBLEU. 
The results of their experiments indicated that CodeT5 and CodeTrans consistently delivered the highest performance in code generation tasks.
In another work, an extensive literature review was conducted by Hou et al. \cite{LLM4SE} where they examine papers that present works done using LLMs for software engineering tasks. Their analysis reveals a growing emphasis on models from the GPT series, with GPT-4 \cite{GPT-4} gaining significant attention in studies related to code generation using LLMs.

Besides the aforementioned studies, there exist papers introducing code synthesis benchmarks like EvalPlus \cite{LiuXW023} and Multipl-E \cite{CassanoGNNPPYZAFGGJ23}, which assess the code generated by various LLMs. Furthermore, the papers that introduce different LLMs capable of performing code generation \cite{codex, codebert, codet5, CodeRL, incoder, palmcoder, Xu0NH22} task also perform evaluation of the code generated by their respective models. The prompting techniques employed in such studies are predominantly limited to either zero-shot or few-shot prompting.

% \vspace{2ex}
\begin{GrayBox}\small
\textbf{Motivation 1: }%\vspace{1ex}
Despite the extensive research in the domain of code generation by LLMs, there is a lack of papers that explore various prompting techniques other than \textit{zero-shot} and \textit{few-shot} prompting to enhance the code generation capabilities of LLMs.
\end{GrayBox}

\subsection{Security in LLM-Generated Code}
As mentioned earlier, prior work has elaborated on the security of code generated by LLMs. \citet{PearceA0DK22}, for instance, used 54 high-risk security scenarios containing incomplete code snippets (C and Python) to assess code completions produced by GitHub Copilot and observed that 40\% of them contained security vulnerabilities. However, a study by Asare et al. \cite{AsareNA23}, compared C/C++ code generated by human developers against the ones generated by Copilot and observed that Copilot is not as bad as humans in introducing vulnerabilities in code. 
The experiments in these studies were done using zero-shot prompts. In another work by Pearce et al. \cite{Pearce2022}, they tested the code repair capabilities of LLMs using various program repair scenarios. Overall, they concluded that Codex and Jurassic-1 \cite{Lieber2021} are capable of finding fixes for simple scenarios again under zero-shot settings. 
Jesse et al. \cite{JesseADM23} did a recent study where they examined if Codex and other LLMs generate simple, stupid bugs (SStuBs) and found that these models produce twice as many SStuBs as correct code. On the other hand, \cite{HeV23} proposed a learning approach for controlled code generation called SVEN. Such an approach, in which a boolean parameter is passed to enforce secure/insecure code generation, increased the number of secure code produced by an LLM called CodeGen by 25\%. Another study by \citet{yetistiren} assessed the quality (i.e., validity, correctness, reliability, security, and maintainability) of code generated by Copilot, Amazon CodeWhisperer, and ChatGPT using the \textit{HumanEval} dataset. Notably, no significant security vulnerabilities were found in the generated code. However, the authors acknowledge the limitations of their security evaluation, since the HumanEval dataset is designed to verify functional correctness rather than code security.

Delving further into the realm of secure code generation using LLMs, \citet{SandovalPNKGD23} investigated the impact of LLM on code security through a user study. The study involved 58 computer science students who were tasked with performing simple operations on a linked list using C programming language with a focus on memory-based vulnerabilities. They observed that the participants who used an AI assistant powered by Codex introduced security-critical bugs at a rate no higher than 10\% when compared to the control group indicating that the use of LLMs does not introduce new vulnerabilities. Nevertheless, it is essential to acknowledge that these findings may not be universally applicable to more complex programming tasks.
Contrary to the previous study \citet{PerryS0B23} observed different results in a study that explored developers' interactions with AI code assistants concerning security. Forty-seven participants were engaged with an AI assistant powered by Codex to fulfill five security-related programming tasks across Python, JavaScript, and C. The findings revealed that participants utilizing AI assistants were prone to generating insecure solutions more often than those without AI assistance in four out of five tasks. Typical issues encompassed the selection of unsafe libraries, incorrect library utilization, insufficient comprehension of edge cases involving external entities like file systems or databases, and inadequate sanitization of user input. 

Additionally, apart from empirical and user studies on LLMs, a systematic literature review conducted by \citet{Yao2023}, delves into the use of LLMs for security and privacy purposes. Their findings indicate a plethora of works employing LLMs in security-related tasks, such as coding, test case generation, bug detection, vulnerability detection, and fixing. These endeavors have positively influenced research within the security community. However, none of these studies thoroughly investigate different prompting techniques to enhance the secure code generation process.

%\vspace{2ex}
\begin{GrayBox}\small
\textbf{Motivation 2: }%\vspace{1ex}
Studies we have seen so far do not thoroughly explore the impact of prompting techniques to improve the security of the code generated by the LLMs. This underscores the need for further research to identify such techniques that can improve the secure code generation capabilities of LLMs.
\end{GrayBox}

\section{Methodology for Systematic Literature Review}
\label{sec:slr-methodology}

\begin{figure}[hbt!]
    \centering
    \includegraphics[width = 0.35\linewidth]{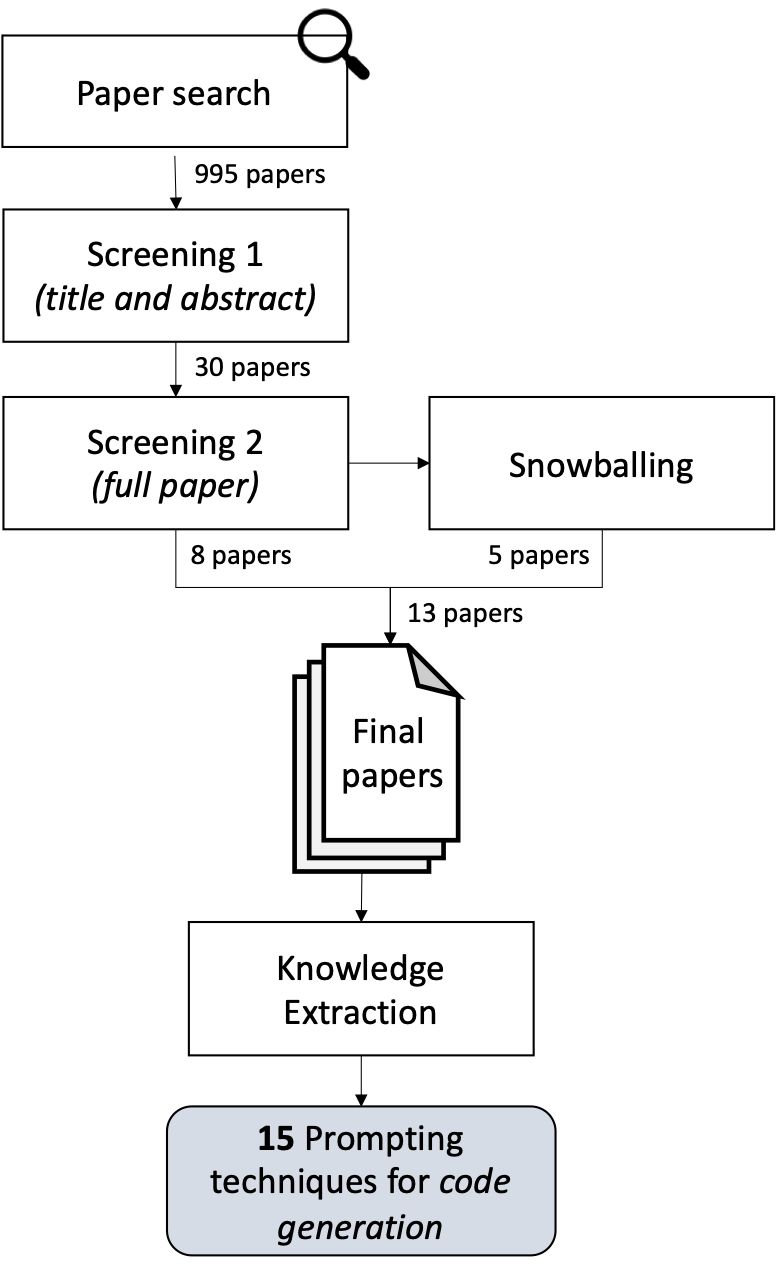}
    \caption{Steps followed for the SLR on prompting techniques that are suitable for code generation} 
    \label{fig:slr-fig}
\end{figure}

The goal of this review is to find prompting techniques that can be used for code-generation tasks using LLMs. However, there are only a limited number of prompting techniques explicitly designed for code generation. Therefore, we opted to review all prompting techniques introduced for generating textual content, presuming their potential transferability to code-generation tasks, given that code generation falls within the domain of textual content generation. The steps followed to perform the literature review are depicted in Figure \ref{fig:slr-fig}.
We used the \textit{Publish or Perish} tool \cite{harzing_2016} to retrieve papers from Google Scholar. Following the PICOC strategy \cite{PICOC}, the search query given below was employed to retrieve the relevant papers that introduce prompting techniques for textual content generation.

\begin{quote}
    \texttt{prompt*  AND (engineer* OR pattern* OR technique*) AND (language model* OR pre-trained model* OR llm* OR ptm*)}
\end{quote}

The search was conducted in October 2023. The results of this search were examined in their ranked order following the steps described below. 
%\textit{Publish or Perish} has a limit of 1000 results per search query, Hence, we extracted the first 1000 papers for review. 

\subsubsection*{\textbf{Paper Screening}}
The review process was done in two screening steps. In the \textit{first screening}, we looked at the title and abstract of the paper to decide if it was relevant to our study. If it is then it was shortlisted for the \textit{second screening}. In the second screening, we looked into the full paper to decide if it fits our criteria. The first and second screening was done based on the following inclusion and exclusion criteria. 

\noindent \textit{Inclusion Criteria:}
\begin{enumerate}
    \item [\textbf{IC1:}] Paper deals with prompting LLMs using one or more techniques 
    \item [\textbf{IC2:}] Paper is published since 2018: 
    \item [\textbf{IC3:}] Paper is written in English
    %\item [\textbf{IC4:}] Available in full text
\end{enumerate}

\noindent \textit{Exclusion Criteria:}
\begin{enumerate}
    \item [\textbf{EC1:}] Paper does not introduce new prompting techniques to query LLM 
    \item [\textbf{EC2:}] Paper deals with the generation of anything other than text and code (e.g: image, speech, and video data)
    \item [\textbf{EC3:}] Paper that presents prompting techniques that can not be used for generation tasks (e.g. techniques specific to classification tasks)
    \item [\textbf{EC4:}] Paper that presents automated prompt optimization techniques and frameworks (e.g. prompt tuning and black-box tuning)
    \item [\textbf{EC5:}] Paper that presents prompting technique for attacking the model (e.g. jailbreak prompts)
    \item [\textbf{EC6:}] Out of scope (e.g. techniques for medical science)
\end{enumerate}

\textbf{IC1} is the main criteria that we use to include papers in the review since our goal is to find papers that explore different ways to prompt LLMs to optimize the response. Significant developments in the field of LLMs started happening since the year 2018 (GPT-1\footnote{https://openai.com/research/language-unsupervised}, BERT \cite{DevlinCLT19}). Hence we defined \textbf{IC2} to look at relevant works on prompting techniques that emerged after this. \textbf{IC3} is a basic criterion that only includes papers written in English.

The exclusion criteria are designed to identify prompting techniques that can be used for code generation even though they are not specifically created for this task.
If any of the criteria outlined in EC are met, regardless of the IC, then a publication is disqualified from the review process.
There are several works that use LLMs for different tasks through prompting. However, many of these works adhere to basic prompting methods without introducing any novel techniques. As our objective is to identify and list novel prompting approaches, we employed \textbf{EC1} as the primary criterion for filtering out papers that rely on existing techniques. \textbf{EC2} excludes papers focusing on generating anything other than textual content, and by extension code. This decision is based on the differing training methodologies between LLMs handling non-textual data such as videos or images and those dealing with textual data. Consequently, we proceeded with the assumption that prompting techniques for non-textual data may not be suitable for code generation. \textbf{EC3} eliminates techniques that target problems with restrictive answers, such as yes/no questions, cloze-style questions, or multiple-choice questions. These techniques are excluded because they do not facilitate generation tasks like code generation. Automated prompt engineering techniques such as prompt tuning \cite{Wang0S22} and black-box tuning \cite{Han2023, SunSQHQ22} as well as automated frameworks that optimize prompts and LLM outputs \cite{YaoYZS00N23}\cite{ZhouMHPPCB23} are excluded from our list as they follow a very different methodology involving data training, learning, external tools or complex automated algorithms to improve prompts. Evaluating such techniques requires a different setup compared to non-automated prompting methods. Consequently, papers presenting these techniques are removed using \textbf{EC4}. Additionally, papers presenting various prompts and techniques aimed at attacking a model are excluded using \textbf{EC5}, as they are not suitable for code generation. Papers discussing topics outside of prompt engineering or belonging to irrelevant fields, such as medicine or construction, are eliminated using \textbf{EC6}.
 
We reached saturation at the mark of 358 search results, as we observed no new papers that passed the first screening process within over 100 results before that point.
Consequently, we concluded this stage upon reaching the 358th paper in the ranked results obtained from our search.
Following the first screening of titles and abstracts, 30 of them were chosen to undergo further evaluation. Out of the 358 papers, the majority were excluded based on \textbf{EC1}, which involves eliminating works that do not introduce a new prompting technique. Upon full review in the second screening step, 22 papers were excluded, leaving a selection of 8 relevant papers introducing novel prompting techniques.

\subsubsection*{\textbf{Snowballing}}
To ensure that we did not miss any other relevant papers, we also performed 3 rounds of backward snowballing \cite{Wohlin14}. Here we went through the references of the selected papers iteratively following the same two-step screening process as above until no new papers were obtained. From this, we obtained 5 additional relevant ones making the total number of relevant papers 13. 
%\subsubsection*{\textbf{Paper Finalization}}
Three papers under consideration were released on preprint servers like arXiv and have not undergone formal peer review. However, these preprint papers have been frequently cited with the least number of citations being 48. Hence we decided to retain those papers.

%This is not surprising given that prompt engineering represents an emergent field of study. Nevertheless, it is noteworthy that many of the preprint papers are frequently cited and describe widely recognized prompting techniques. Therefore, we chose to retain those preprint publications that have obtained a minimum of 30 citations. Following this criterion, the curated selection of relevant papers was narrowed down to 13. 

\subsubsection*{\textbf{Knowledge Extraction}}

Each final paper that introduced a prompting technique suitable for code generation was examined in detail. The primary objective was to extract the techniques themselves and pinpoint their key features. For this, we performed a lightweight thematic analysis with open coding as it offers a qualitative method for analyzing textual or qualitative data to interpret patterns or themes within the data \cite{thematic-analysis-1}\cite{thematic-analysis-2}. During this process, the first author extracted codes related to prompting techniques following an inductive approach. The themes that emerged from this coding were then discussed with two other authors to categorize and label the techniques.

In addition to this, attention was also directed towards details such as the LLMs on which the technique was tested, the specific tasks used for evaluation, and the datasets employed for this purpose. Furthermore, data regarding the year of publication, venue, and citation count at the time of the study were also extracted. This was aimed at creating a consolidated source of information beneficial to researchers and developers delving into prompting techniques for code generation.

\section{Prompting Techniques for Code Generation (\textbf{\textit{RQ1}})}
\label{sec:slr-results}

In this section, we present an overview of the selected prompting techniques from the SLR that are deemed suitable for code-generation tasks. Throughout our review, we encountered numerous prompting techniques. However, not all of them were selected to be in our final list as determined by our exclusion criteria. All results of this literature review, along with the techniques that were excluded from our consideration and the reason for their exclusion are documented in our replication package specified in Section \ref{sec:replication}. 

\subsection{Overview of the Selected Papers}
The information extracted from the 13 papers is presented in Table \ref{tab:slr-table}. The chosen papers are those that introduce novel prompting techniques. Among these, we identified 15 distinct techniques designed for textual content generation with potential applicability to code-generation tasks. Ten of these papers have undergone peer review, while the remaining have received at least 48 citations. Except for two papers \cite{ReynoldsM21}\cite{White2023a}, all have conducted experimental validation of their introduced prompting techniques. Only two of them \cite{MadaanTGHGW0DPY23} \cite{jiang2023selfplanning} have evaluated their techniques specifically for code generation tasks. The other techniques primarily target various reasoning tasks such as symbolic, logical, commonsense, and arithmetic. Among the papers that conducted experimental validation, ten out of eleven utilize OpenAI models indicating the prevalence of these models in research in this field.

Based on commonalities derived from the thematic analysis, we have labeled the techniques using 3 distinct properties related to their execution as shown in Table \ref{tab:slr-table}. They are \textit{\textbf{S}ingle/\textbf{M}ulti-step, \textbf{D}emonstrative/\textbf{N}on-demonstrative} and \textit{\textbf{L}inear/\textbf{P}arallel}. A technique that prompts the model in a single step, obtaining the final output with just one prompt, is referred to as a \textit{single-step} technique. Conversely, a technique requiring multiple prompts to generate the final output is termed a \textit{multi-step} technique. Single-step techniques are cost-effective compared to multi-step techniques as they necessitate only one prompt. Among the 15 techniques identified, 6 are single-step techniques, while the rest are multi-step techniques.
If a technique is executed by providing demonstrative examples of inputs and expected outputs for prompting the model, it is categorized as a \textit{demonstrative} technique. Conversely, a technique not requiring input-output examples is labeled as a \textit{non-demonstrative} technique. Six out of 15 techniques are non-demonstrative. 
\begin{landscape}
\begin{table}[]
\centering
\caption{Final list of prompting techniques obtained from the SLR that can potentially be used for code generation}
\label{tab:slr-table} \large
\resizebox{\columnwidth}{!}{%
\begin{tabular}{@{}l|lccc|lll|ccc@{}}
\toprule
\multirow{2}{*}{\textbf{Work}} & \multicolumn{4}{c|}{\textbf{Prompting Techniques}} & \multicolumn{3}{c|}{\textbf{Scope}} & \multicolumn{3}{c}{\textbf{Publication Details}} \\ \cmidrule(l){2-11} 
 & \textbf{Name} & \textbf{\begin{tabular}[c]{@{}c@{}}Single/\\ Multi-step\end{tabular}} & \textbf{\begin{tabular}[c]{@{}c@{}}Demonstrative/\\ Non-demonstrative\end{tabular}} & \textbf{\begin{tabular}[c]{@{}c@{}}Linear/\\ Parallel\end{tabular}} & \textbf{Evaluation Task(s)} & \textbf{LLM(s)} & \textbf{Dataset(s)} & \textbf{Year} & \textbf{Venue} & \textbf{Citations} \\ \midrule
\begin{tabular}[c]{@{}l@{}}Brown \\ et al. \cite{GPT-3}\end{tabular} & Zero-shot & S & N & L & \multirow{3}{*}{\begin{tabular}[c]{@{}l@{}}language modeling, question \\ answering, translation, com-\\ monsense, reading compreh-\\ ension, reasoning, inference \\ \& arithmetic\end{tabular}} &  & \begin{tabular}[c]{@{}l@{}}PTB \cite{PTB}, LAMBADA \cite{lambada}, StoryCloze \cite{storycloze}, \\ HellaSwag \cite{hellaswag}, Natural Questions \cite{naturalquestions}, \\ WebQuestions \cite{webquestions}, TriviaQA \cite{triviaqa},\end{tabular} &  &  &  \\
 & One-shot & S & D & L &  & GPT-3 & \begin{tabular}[c]{@{}l@{}}WMT \cite{wmt}, WinoGrande \cite{winogrande}, PIQA \cite{piqa}, ARC \cite{arc-da}, \\ UnifiedQA \cite{unifiedqa}, OpenBookQA \cite{openbookqa}, CoQA \cite{coqa}, \\ QuAC \cite{quac}, DROP \cite{drop}, SQuAD \cite{squad}, RACE \cite{race},\end{tabular} & 2020 & NeurIPS & 24786 \\
 & Few-shot & S & D & L &  &  & \begin{tabular}[c]{@{}l@{}}SUPERGLUE \cite{superglue}, RTE \cite{GPT-3}, ANLI \cite{anli}, \\ SAT analogy \cite{satanalogy}\end{tabular} &  &  &  \\
 &  & \multicolumn{1}{l}{} & \multicolumn{1}{l}{} & \multicolumn{1}{l|}{} &  &  &  &  &  &  \\
\begin{tabular}[c]{@{}l@{}}Reynolds\\ et al. \cite{ReynoldsM21}\end{tabular} & Memetic Proxy & S & N & L & N/A & N/A & N/A & 2021 & CHI & 544 \\
 &  & \multicolumn{1}{l}{} & \multicolumn{1}{l}{} & \multicolumn{1}{l|}{} &  &  &  &  &  &  \\
\begin{tabular}[c]{@{}l@{}}Kojima\\ et al. \cite{KojimaGRMI22}\end{tabular} & Zero-shot CoT & M & N & L & \begin{tabular}[c]{@{}l@{}}arithmetic, symbolic \& lo-\\ gical reasoning\end{tabular} & \begin{tabular}[c]{@{}l@{}}InstructGPT, \\ PaLM\end{tabular} & \begin{tabular}[c]{@{}l@{}}SingleEq \cite{singleeq}, AddSub \cite{addsub}, MultiArith \cite{multiarith}, \\ AQUARAT \cite{aquarat}, GSM8K \cite{gsm8k}, SVAMP \cite{svamp}, \\ Last Letter Concatenation \cite{Wei2022}, Coin Flip \cite{Wei2022}, \\ CommonsenseQA \cite{commonsenseqa}, StrategyQA \cite{strategyqa}, \\ BIG-bench effort \cite{bigbench-effort}\end{tabular} & 2022 & NeurIPS & 1901 \\
 &  & \multicolumn{1}{l}{} & \multicolumn{1}{l}{} & \multicolumn{1}{l|}{} &  &  &  &  &  &  \\
\begin{tabular}[c]{@{}l@{}}Lampinen\\ et al. \cite{LampinenDCMTCMW22}\end{tabular} & \begin{tabular}[c]{@{}l@{}}Few-shot \\ Explanation\end{tabular} & S & D & L & reasoning, inference & \begin{tabular}[c]{@{}l@{}}Decoder-only \\ Transformers \\ (1B to 280B \\ parameters)\end{tabular} & BigBench Effort \cite{bigbench-effort} & 2022 & \begin{tabular}[c]{@{}c@{}}EMNLP \\ Findings\end{tabular} & 197 \\
 &  & \multicolumn{1}{l}{} & \multicolumn{1}{l}{} & \multicolumn{1}{l|}{} &  &  &  &  &  &  \\
\begin{tabular}[c]{@{}l@{}}Wang \\ et al. \cite{WangWSLCNCZ23}\end{tabular} & Self-consistency & M & D & P & \begin{tabular}[c]{@{}l@{}}arithmetic \& commonsense\\ reasoning\end{tabular} & \begin{tabular}[c]{@{}l@{}}UL2, GPT-3, \\ LaMDA, PaLM\end{tabular} & \begin{tabular}[c]{@{}l@{}}GSM8K \cite{gsm8k}, SVAMP \cite{svamp}, AQuA \cite{aquarat}, \\ StrategyQA \cite{strategyqa} and ARC-challenge\end{tabular} & 2022 & ICLR & 626 \\
 &  & \multicolumn{1}{l}{} & \multicolumn{1}{l}{} & \multicolumn{1}{l|}{} &  &  &  &  &  &  \\
\begin{tabular}[c]{@{}l@{}}Wei \\ et al. \cite{Wei2022}\end{tabular} & Chain-of-Thought & M & D & L & \begin{tabular}[c]{@{}l@{}}symbolic \& commonsense\\ reasoning\end{tabular} & \begin{tabular}[c]{@{}l@{}}GPT-3, LaMDA, \\ PaLM\end{tabular} & \begin{tabular}[c]{@{}l@{}}CommonSenseQA \cite{commonsenseqa}, StrategyQA \cite{strategyqa}, \\ BigBench effort \cite{bigbench-effort}, SayCan \cite{saycan},  Last letter \\ concatenation \cite{Wei2022}, Coin flip \cite{Wei2022}\end{tabular} & 2022 & NeurIPS & 4584 \\
 &  & \multicolumn{1}{l}{} & \multicolumn{1}{l}{} & \multicolumn{1}{l|}{} &  &  &  &  &  &  \\
\begin{tabular}[c]{@{}l@{}}Zhou\\ et al. \cite{ZhouSHWS0SCBLC23}\end{tabular} & Least-to-Most & M & D & L & \begin{tabular}[c]{@{}l@{}}symbolic manipulation, \\ compositional generaliza-\\ tion, math reasoning\end{tabular} & GPT-3 & \begin{tabular}[c]{@{}l@{}}Last Letter Concatenation \cite{Wei2022}, SCAN \cite{scan}, \\ GSM8K \cite{gsm8k}, DROP \cite{drop}\end{tabular} & 2022 & ICLR & 672 \\
 &  & \multicolumn{1}{l}{} & \multicolumn{1}{l}{} & \multicolumn{1}{l|}{} &  &  &  &  &  &  \\
 \begin{tabular}[c]{@{}l@{}}Fu\\ et al. \cite{FuPSCK23}\end{tabular} & Complexity-based & M & D & P &  \begin{tabular}[c]{@{}l@{}}arithmetic, commonsense, \\ temporal \& referential\\ reasoning\end{tabular} & \begin{tabular}[c]{@{}l@{}}LaMDA, PaLM \\ Minerva, GPT-3 \\ Codex, DiVeRSe\end{tabular} & \begin{tabular}[c]{@{}l@{}}GSM8K \cite{gsm8k}, StrategyQA \cite{strategyqa}, \\ MathQA \cite{austin2022}, MultiArith \cite{multiarith},  Penguin \cite{suzgunSSGTCCLCZ23} \\ Date Understanding \cite{suzgunSSGTCCLCZ23}\end{tabular} & 2023 & ICLR & 194 \\
  &  & \multicolumn{1}{l}{} & \multicolumn{1}{l}{} & \multicolumn{1}{l|}{} &  &  &  &  &  &  \\
\begin{tabular}[c]{@{}l@{}}Jiang\\ et al. \cite{jiang2023selfplanning}\end{tabular} & Self-planning & M & D & L & code generation \& completion & Codex & \begin{tabular}[c]{@{}l@{}}MBPP-sanitized \cite{austin2022}, MBPP-ET \cite{mbpp-et}, HumanEval \cite{codex}, \\ HumanEval-X \cite{humaneval-x} and HumanEval-ET \cite{mbpp-et}\end{tabular} & 2023 & arXiv & 48 \\
 &  & \multicolumn{1}{l}{} & \multicolumn{1}{l}{} & \multicolumn{1}{l|}{} &  &  &  &  &  &  \\
\begin{tabular}[c]{@{}l@{}}Kim\\ et al. \cite{KimBM23}\end{tabular} & \begin{tabular}[c]{@{}l@{}}Recursive \\ Ciriticism and \\ Improvement\end{tabular} & M & N & L & \begin{tabular}[c]{@{}l@{}}arithmetic \& commonsense\\ reasoning\end{tabular} & \begin{tabular}[c]{@{}l@{}}InstructGPT3 +\\ RLHF\end{tabular} & \begin{tabular}[c]{@{}l@{}}SingleEq \cite{singleeq}, AddSub \cite{addsub}, MultiArith \cite{multiarith}, \\ AQuA \cite{aquarat}, GSM8K \cite{gsm8k}, SVAMP \cite{svamp}, \\ CommonSenseQA \cite{commonsenseqa} and StrategyQA \cite{strategyqa}\end{tabular} & 2023 & NeurIPS & 135 \\
 &  & \multicolumn{1}{l}{} & \multicolumn{1}{l}{} & \multicolumn{1}{l|}{} &  &  &  &  &  &  \\
\begin{tabular}[c]{@{}l@{}}Madaan\\ et al. \cite{MadaanTGHGW0DPY23}\end{tabular} & Self-Refine & M & D & L & \begin{tabular}[c]{@{}l@{}}sentiment reversal, dialog \\ response, code optimization, \\ code readability, math reasoning, \\ acronym generation, constrained \\ generation\end{tabular} & \begin{tabular}[c]{@{}l@{}}GPT-3.5, GPT-4, \\ Codex\end{tabular} & \begin{tabular}[c]{@{}l@{}}Yelp reviews \cite{yelp-reviews}, FED \cite{fed}, PIE \cite{pie}, CodeNet \cite{codenet}, \\ GSM8K \cite{gsm8k},  Acronyms,  CommonGen \cite{commongen}\end{tabular} & 2023 & NeurIPS & 430 \\
 &  & \multicolumn{1}{l}{} & \multicolumn{1}{l}{} & \multicolumn{1}{l|}{} &  &  &  &  &  &  \\
\begin{tabular}[c]{@{}l@{}}White\\ et al. \cite{White2023a}\end{tabular} & Persona & S & N & L & N/A & N/A & N/A & 2023 & arXiv & 580 \\
 &  & \multicolumn{1}{l}{} & \multicolumn{1}{l}{} & \multicolumn{1}{l|}{} &  &  &  &  &  &  \\
\begin{tabular}[c]{@{}l@{}}Zheng\\ et al. \cite{Zheng2023}\end{tabular} & Progressive-hint & M & N & L & arithmetic, reasoning & \begin{tabular}[c]{@{}l@{}}GPT-3, GPT-3.5-\\ Turbo, GPT-4\end{tabular} & \begin{tabular}[c]{@{}l@{}}AddSub \cite{addsub}, MultiArith \cite{multiarith}, SingleEQ \cite{singleeq}, \\ SVAMP \cite{svamp}, GSM8K \cite{gsm8k},  AQuA \cite{aquarat}, MATH \cite{math}\end{tabular} & 2023 & arXiv & 84 \\ \bottomrule
\end{tabular}%
}
\end{table}
\end{landscape}
Although demonstrative techniques may potentially yield desired outputs more effectively than non-demonstrative techniques, this depends on the availability of high-quality demonstrative examples. In real-world scenarios, especially in complex code generation tasks, obtaining such examples can be challenging.
Most techniques in our inventory involve conducting a single sequential interaction with the LLM. Here, the model is prompted, and its response is either used as the final output or serves as a basis for proceeding to the next step of prompting. These techniques are labeled as \textit{linear}.

Conversely, techniques that engage in multiple parallel chains of conversation with the model, and utilize the parallel responses generated by the model to either finalize the output or advance to the next parallel step of prompting are labeled as \textit{parallel}.
Among the 15 techniques examined, only two are classified as parallel. However, there are techniques outside of our list that employ parallel response generation or interactions, such as \textit{Ask Me Anything} \cite{AskMeAnything} and \textit{Tree-of-Thoughts} \cite{YaoYZS00N23} which were excluded due to their unsuitability for our use case. Therefore, we opted to maintain this label within our list of techniques.

A more detailed description of the individual techniques included in Table \ref{tab:slr-table} and how they can be used for code generation are presented in the following subsection.

\subsection{Classification of Prompting Techniques}
\begin{figure*}[hbt!]
    \centering
    \includegraphics[width = 0.9\linewidth]{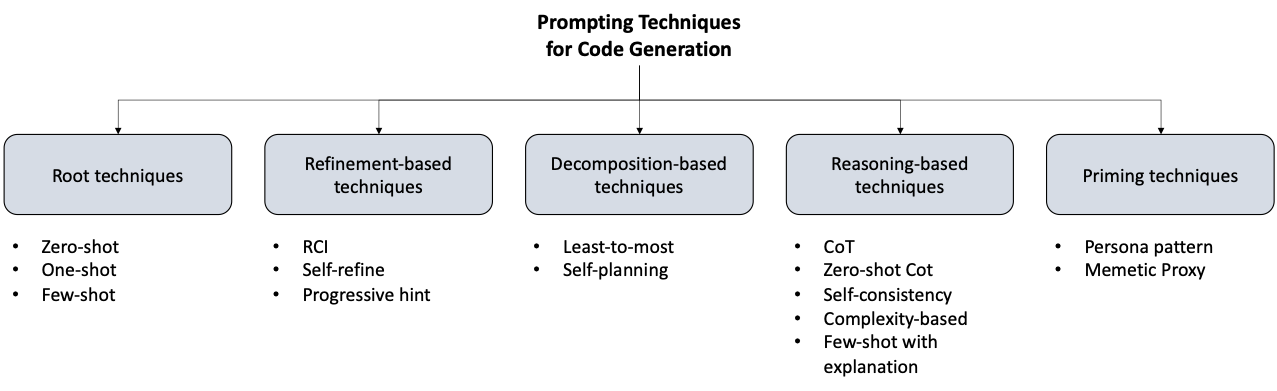}
    \caption{Classification of prompting techniques for code generation} 
    \label{fig:taxonomy}
\end{figure*}

Aside from the labels provided in Table \ref{tab:slr-table} (\textit{single/multi-step, demonstrative/non-demonstrative} and \textit{linear/parallel}), we also identified some other common characteristics based on the strategic design of different prompting techniques which we used to classify them into 5 different categories as shown in Figure \ref{fig:taxonomy}. Below we describe the categories and the techniques that belong to them, accompanied by demonstrations of how these techniques can be utilized for code generation tasks. The responses of the LLM depicted in these demonstrations were generated by ChatGPT (GPT-3.5), which is a conversational chatbot, in response to different prompting techniques.

\subsubsection{\textbf{Root Techniques}}

These are the foundational and most popular techniques based on which more advanced techniques are built. \textit{Zero-shot, one-shot,} and \textit{few-shot} prompting come under this category. 

\begin{figure}[hbt!]
    \centering
    \includegraphics[width = 0.8\linewidth]{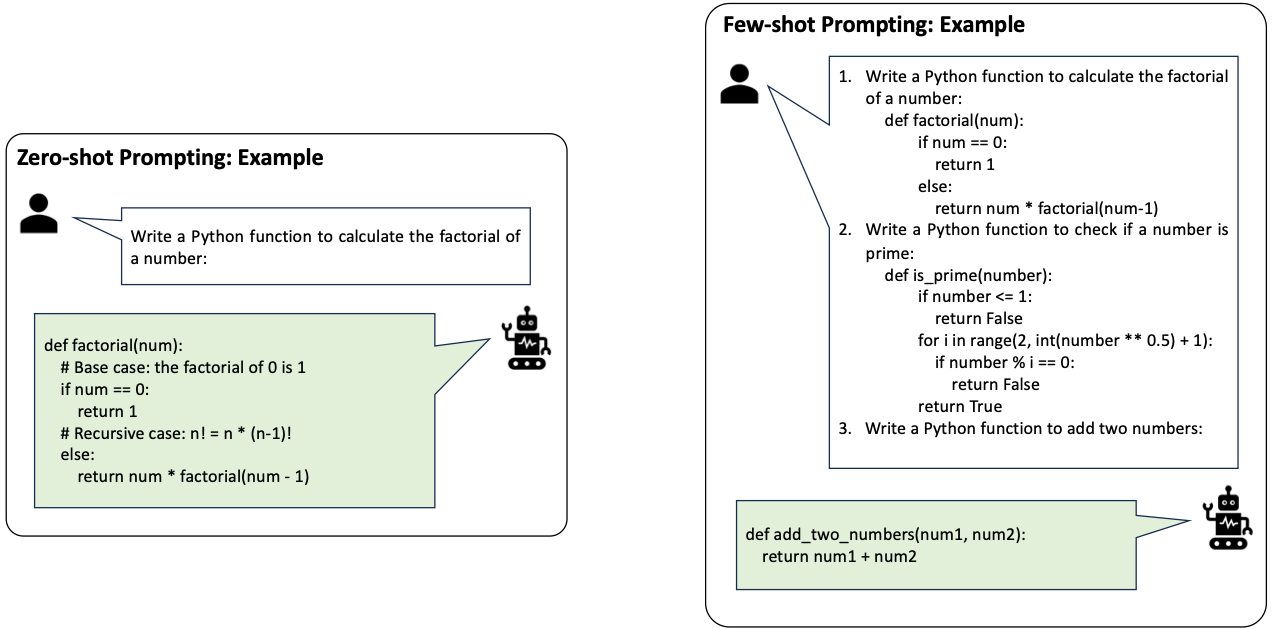}
    \caption{Zero-shot \textit{(left)} and few-shot \textit{(right)} prompting with ChatGPT for code generation.} 
    \label{fig:zero-shot-eg}
\end{figure}

\textbf{\textit{Zero-shot:}} In this technique a model is asked to perform a task without task-specific training or examples at the time of inference \cite{GPT-3}. In such cases, the model relies completely on the data it has seen during its pre-training to generate an appropriate response. In conversational LLMs such as ChatGPT, zero-shot prompting is possibly the most commonly used way of interaction by an average user. It has the advantage of not having to prepare a task-specific dataset of input-output demonstrations to generate desirable output. However, if the model has not seen data related to the task at hand in its training, then the performance of the model can be suboptimal with zero-shot prompting. This technique can be directly used for code generation tasks. Figure \ref{fig:zero-shot-eg} includes a demonstration of zero-shot prompting for a simple coding task and ChatGPT's response to it.

% \begin{figure*}[hbt!]
%     \centering
%     \includegraphics[width=0.4\linewidth]{figures/zero-shot-narrow.png}\hspace{2ex}
%     \includegraphics[width=0.4\linewidth]{figures/few-shot-narrow.png}
%     \caption{Zero-shot \textbf{\textit{(left)}} and few-shot \textbf{\textit{(right)}} prompting with ChatGPT for code generation tasks.}
%     \label{fig:heatmap}
% \end{figure*}

\textbf{\textit{One-shot/Few-shot:}} One-shot and few-shot prompting techniques \cite{GPT-3} are very similar to each other. In one-shot prompting, the model is given a single input-output example whereas in the few-shot prompting the model is given examples of the task at inference time as conditioning before providing the final input for which it is expected to produce the output. By supplying the model with both input and corresponding output samples, it gains the benefit of producing a response that closely aligns with the desired format. However, it can be a disadvantage when one does not have sufficient or relevant task-specific data in advance. An illustration demonstrating the application of few-shot prompting on ChatGPT for code generation can be seen in Figure \ref{fig:zero-shot-eg}. This example utilizes two few-shot examples. We have omitted a separate example of one-shot prompting since it closely resembles few-shot prompting but with only one demonstrative example.

% \begin{figure}[hbt!]
%     \centering
%     \includegraphics[width = 0.7\linewidth]{figures/few-shot-journal.png}
%     \caption{Few-shot prompting with ChatGPT using two demonstrative examples for a code generation task. } 
%     \label{fig:few-shot-eg}
% \end{figure}

\subsubsection{\textbf{Refinement-based Techniques}}
Techniques belonging to this category focus on improving, refining, or iterating the model outputs. They might involve feedback loops, user interactions, or model self-assessment to enhance the quality of the generated responses. The prompting techniques that come under this category include Recursive Criticism and Improvement (RCI), Self-refine, and Progressive Hint prompting.

\textbf{\textit{RCI: }}This prompting technique \cite{KimBM23} is built on the understanding that LLMs possess a strong capability to evaluate and recognize flaws in their own output. This technique involves a two-step process in addition to providing the initial input task. 
\begin{figure}[hbt!]
    \centering
    \includegraphics[width = 0.8\linewidth]{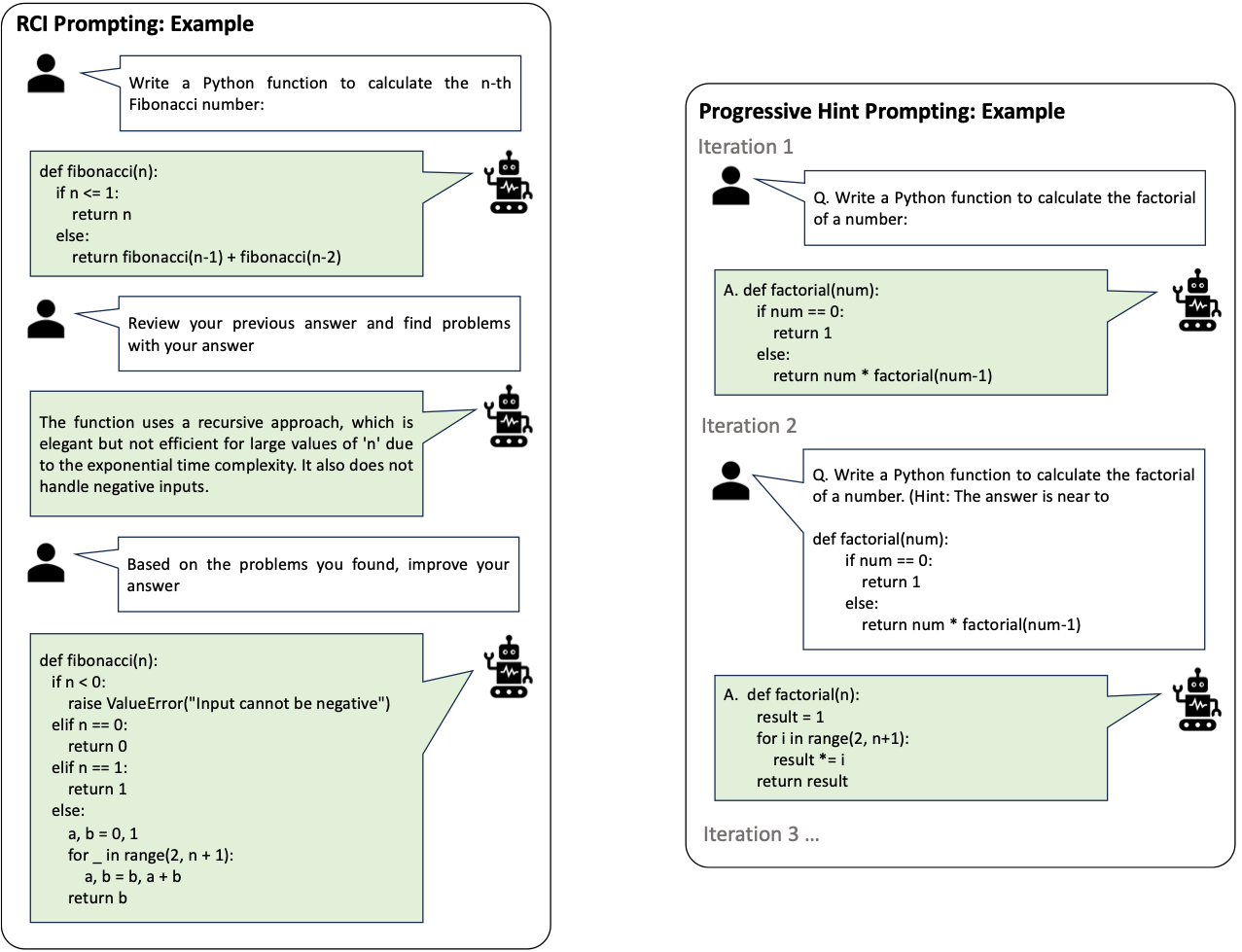}
    \caption{RCI \textit{(left)} and Progressive Hint \textit{(right)} prompting with ChatGPT for code generation.} 
    \label{fig:rci-eg}
\end{figure}
% \begin{figure*}[hbt!]
%     \centering
%     \includegraphics[width=0.4\linewidth]{figures/rci-narrow.png}\hspace{2ex}
%     \includegraphics[width=0.4\linewidth]{figures/progressive-hint-narrow.png}
%     \caption{RCI \textbf{\textit{(left)}} and Progressive-hint \textbf{\textit{(right)}} prompting with ChatGPT for code generation}
%     \label{fig:heatmap}
% \end{figure*}
Firstly, the LLM is prompted to analyze and critique its current response (for instance: \textit{"Review your previous answer and find problems with your answer"}). Subsequently, drawing from the critiques it has outlined, the LLM is then instructed to rectify the identified issues and revise its output accordingly (for example: \textit{"Based on the problems you found, improve your answer"}). This two-step process is repeated until a satisfactory output is obtained or until a predefined number of iterations is done. RCI has the advantage that it needs no task-specific expert data to generate desirable responses. However, this approach can be expensive due to the iterative nature of the process. An added disadvantage is that the success of this approach relies on the ability of the model to identify its own mistakes. A demonstration of one iteration of this technique used for a code generation task is shown in Figure \ref{fig:rci-eg}.

\begin{figure}[hbt!]
    \centering
    \includegraphics[width = 0.8\linewidth]{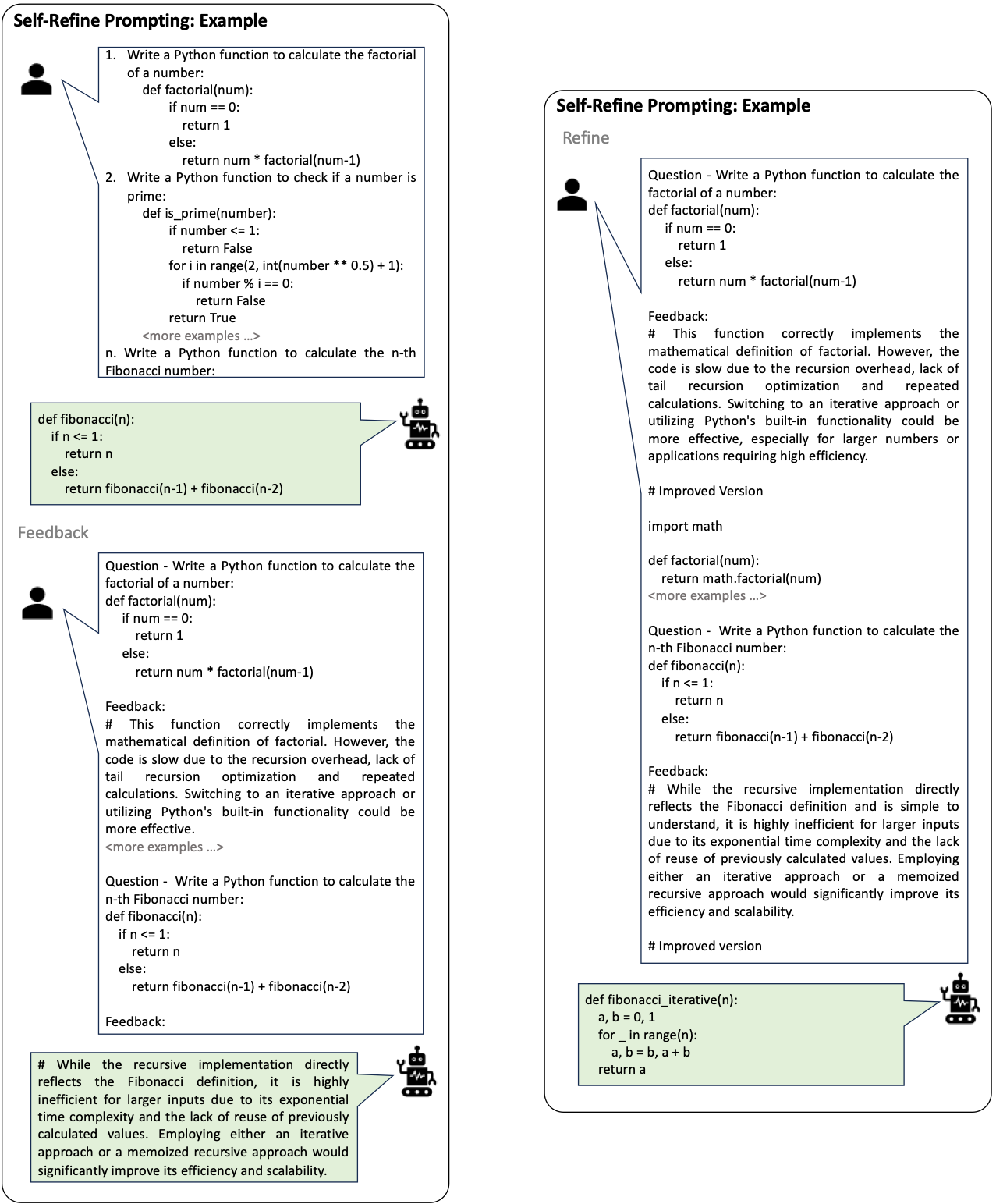}
    \caption{Feedback generation \textit{(left)} and refining  \textit{(right)} steps of Self-refining prompting with ChatGPT for code generation.}
    \label{fig:self-refine}
\end{figure}

\textbf{\textit{Self-refine:}}
This technique \cite{MadaanTGHGW0DPY23} is very similar to RCI. It uses 2 steps called \textit{feedback} and \textit{refine} in addition to an initial output generation step to generate high-quality output. 
The initial output from model M is generated using a task-specific prompt \textit{p\textsubscript{gen}} with few-shot <input, output> example pairs. Next, they use a prompt \textit{p\textsubscript{fb}} to generate feedback for the previously generated output by M. Few-shot examples are provided in this step in the form of <input, output, feedback> triplets. The next step is to refine the output based on the generated feedback. This is done using prompt \textit{p\textsubscript{refine}} that contains few-shot examples of refining outputs in the form of <input, output, feedback, refined> quadruples. An adaptation of this technique for code generation is shown in Figure \ref{fig:self-refine}.

%Since this approach is very similar to the RCI (Figure \ref{fig:rci-eg}) with the exception of providing few-shot examples with every step, a separate demonstration is not provided here.

\textbf{\textit{Progressive Hint:}} Progressive Hint prompting (PHP) \cite{Zheng2023} is another technique that iteratively refines the output from the LLM by providing increasingly informative hints in each iteration. The pipeline of this approach is divided into two stages. The first stage is called \textit{base answer and base prompt}.  In this stage, the model is provided with an input task with a basic prompt to which a base answer is generated. 

% \begin{figure}[hbt!]
%     \centering
%     \includegraphics[width = 0.7\linewidth]{figures/progressive-hint-journal.png}
%     \caption{Two iterations of progressive hint prompting using ChatGPT for a code generation task. } 
%     \label{fig:progressive-hint-eg}
% \end{figure}

The second stage is called \textit{subsequent answer and PHP} where the base prompt is combined with hints that are extracted from the previous answers (or base answer in this case). This is repeated until the answers from the model do not change. Figure \ref{fig:rci-eg} shows the interaction with ChatGPT for a simple coding task using this technique.
PHP can be combined with standard zero-shot prompting or sophisticated techniques such as CoT. This approach requires at least 2 iterations. The approach is not considered successful until the last 2 outputs from the model are the same.  This can become computationally expensive based on the task and the model. Additionally, the model can be misled if the hints provided stray too far away from the correct answer \cite{Zheng2023}. This approach can be theoretically used for code generation tasks as shown in Figure \ref{fig:rci-eg}. 

 \subsubsection{\textbf{Decomposition-based Techniques}}
 Techniques in this category break down complex tasks or prompts into simpler, more manageable pieces. Here, the language models perform multiple small tasks to incrementally build towards the final, complex solution, facilitating more accurate responses. The techniques under this category include \textit{least-to-most} and \textit{self-planning} prompting.

 \textbf{\textit{Least-to-most:}} This prompting technique \cite{ZhouSHWS0SCBLC23} is executed in two stages. In the \textit{decomposition} stage, the model is prompted to decompose the complex task into smaller sub-tasks. This prompt is delivered using a few-shot approach, where a few examples are presented to illustrate how larger tasks can be dissected into sub-tasks, followed by the actual complex task that needs to be addressed. The second stage is the \textit{sub-problem solving} stage where the model is asked to sequentially solve all the sub-problems or sub-tasks identified in the decomposition stage. Here also, few-shot examples demonstrating how sub-problems are solved are provided. Responses derived from solving each sub-task are integrated back into the original task description before presenting the subsequent sub-task to the model. This iterative process continues until all sub-tasks have been resolved, resulting in the final solution.

%  \begin{figure}[hbt!]
%     \centering
%     \includegraphics[width = 0.7\linewidth]{figures/least-to-most-journal.png}
%      %\includegraphics[width = 0.4\linewidth]{figures/least-to-most-2.png}
%     \caption{Decomposition \textit{(left)} and sub-problem solving \textit{(right)} stage of least-to-most prompting using ChatGPT for a code generation task. } 
%     \label{fig:least-to-most-eg1}
    
% \end{figure}

\begin{figure}[hbt!]
    \centering
    \includegraphics[width = 0.8\linewidth]{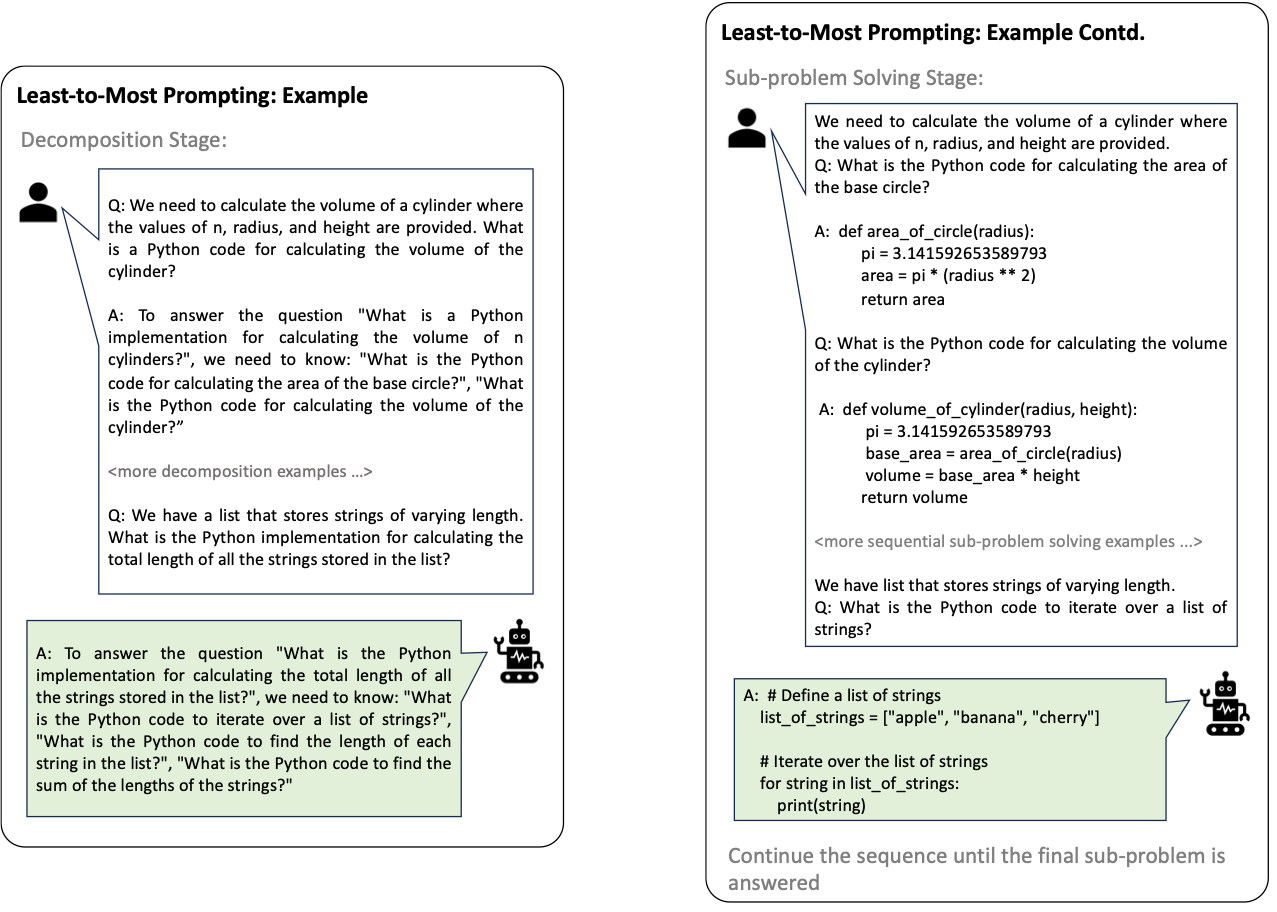}
    \caption{Decomposition \textit{(left)} and Sub-problem solving \textit{(right)} stage of Least-to-most prompting with ChatGPT for code generation} 
    \label{fig:least-to-most-eg}
\end{figure}

% \begin{figure*}[hbt!]
%     \centering
%     \includegraphics[width=0.4\linewidth]{figures/least-to-most-narrow-1.png}\hspace{2ex}
%     \includegraphics[width=0.4\linewidth]{figures/least-to-most-narrow-2.png}
%     \caption{Decomposition \textbf{\textit{(left)}} and Sub-problem solving \textbf{\textit{(right)}} phase of Least-to-most prompting with ChatGPT for code generation}
%     \label{fig:heatmap}
% \end{figure*}

 Least-to-most prompting technique can also be used in combination with CoT or self-consistency prompting techniques. Similar to other advanced prompting methods, this technique's drawback is the necessity to supply few-shot examples for both the decomposition of a complex task and the resolution of its sub-tasks. The resource demands can escalate with the increasing number of sub-tasks involved in the process. This approach can also be potentially used for code generation provided you have a sufficient dataset containing information on coding problem decompositions and solutions. Figure \ref{fig:least-to-most-eg} demonstrates how this technique can be used for code generation.

% \begin{figure}
%     \centering
%     \includegraphics[width = 0.4\linewidth]{figures/least-to-most-2.png}
%     \caption{Example demonstrating the sub-problem solving stage of least-to-most prompting for a code generation task. } 
%     \label{fig:least-to-most-eg1}
% \end{figure}

\begin{figure}[hbt!]
    \centering
    \includegraphics[width = 0.8\linewidth]{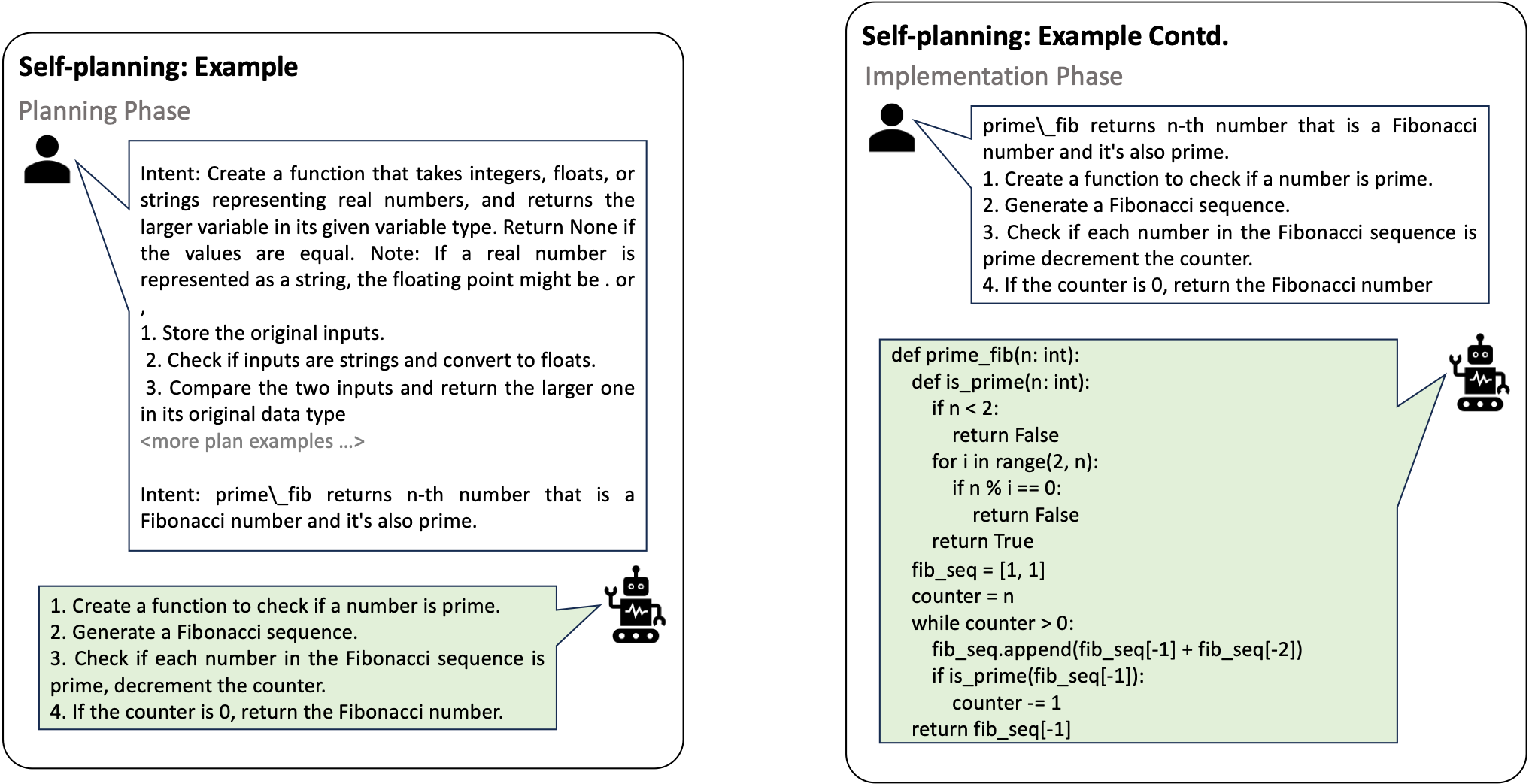}
    \caption{Planning \textit{(left)} and Implementation \textit{(right)} phase of Self-planning prompting for code generation (borrowed from \cite{jiang2023selfplanning}).} 
    \label{fig:self-planning-eg}
\end{figure}

\textbf{\textit{Self-planning:}} This prompting approach \cite{jiang2023selfplanning} is specifically designed for code generation problems. Hence no additional adaptation is required to tailor the technique for code generation tasks. Self-planning is carried out in two phases. The first one is the planning phase where the code generation task is decomposed into a plan of actions. This decomposition is done by the LLM itself. The LLM is provided with demonstrative examples of how to come up with plans to solve coding tasks before asking it to generate a plan for the task at hand. The action plan is structured as an ordered list of steps. The plan should always conclude with a return statement. The second phase is called the implementation phase wherein the LLM's formulated plan is integrated with the original task prompt. This integration prompts the LLM to adhere to its own outlined strategy when producing the final code snippet. An example demonstration of this prompting technique is shown in Figure \ref{fig:self-planning-eg}. This example is directly taken from the original paper itself.

% \begin{figure*}[hbt!]
%     \centering
%     \includegraphics[width=0.4\linewidth]{figures/self-planning-narrow-1.png}\hspace{2ex}
%     \includegraphics[width=0.4\linewidth]{figures/self-planning-narrow-2.png}
%     \caption{Planning \textbf{\textit{(left)}} and Implementation \textbf{\textit{(right)}} phase of Self-planning prompting with ChatGPT for code generation (borrowed from \cite{jiang2023selfplanning}).}
%     \label{fig:heatmap}
% \end{figure*}

\subsubsection{\textbf{Reasoning-based Techniques}}
Techniques that guide the model to employ and demonstrate logical reasoning for generating responses are categorized as reasoning-based techniques. Reasoning encompasses the act of drawing logical conclusions, evaluating arguments, and making inferences using the information at hand \cite{Huang2023a}. These methods emphasize the model's ability to engage in cognitive and logical processes. Rather than simplifying a task as in the case of decomposition-based techniques, these techniques encourage the model to follow a logical reasoning path and articulate its thought process. The techniques that come under this category are \textit{Chain-of-Thought, Zero shot Chain-of-Thought, Self-consistency} and \textit{Few-shot with Explanation}.

\begin{figure}[hbt!]
    \centering
    \includegraphics[width = 0.8\linewidth]{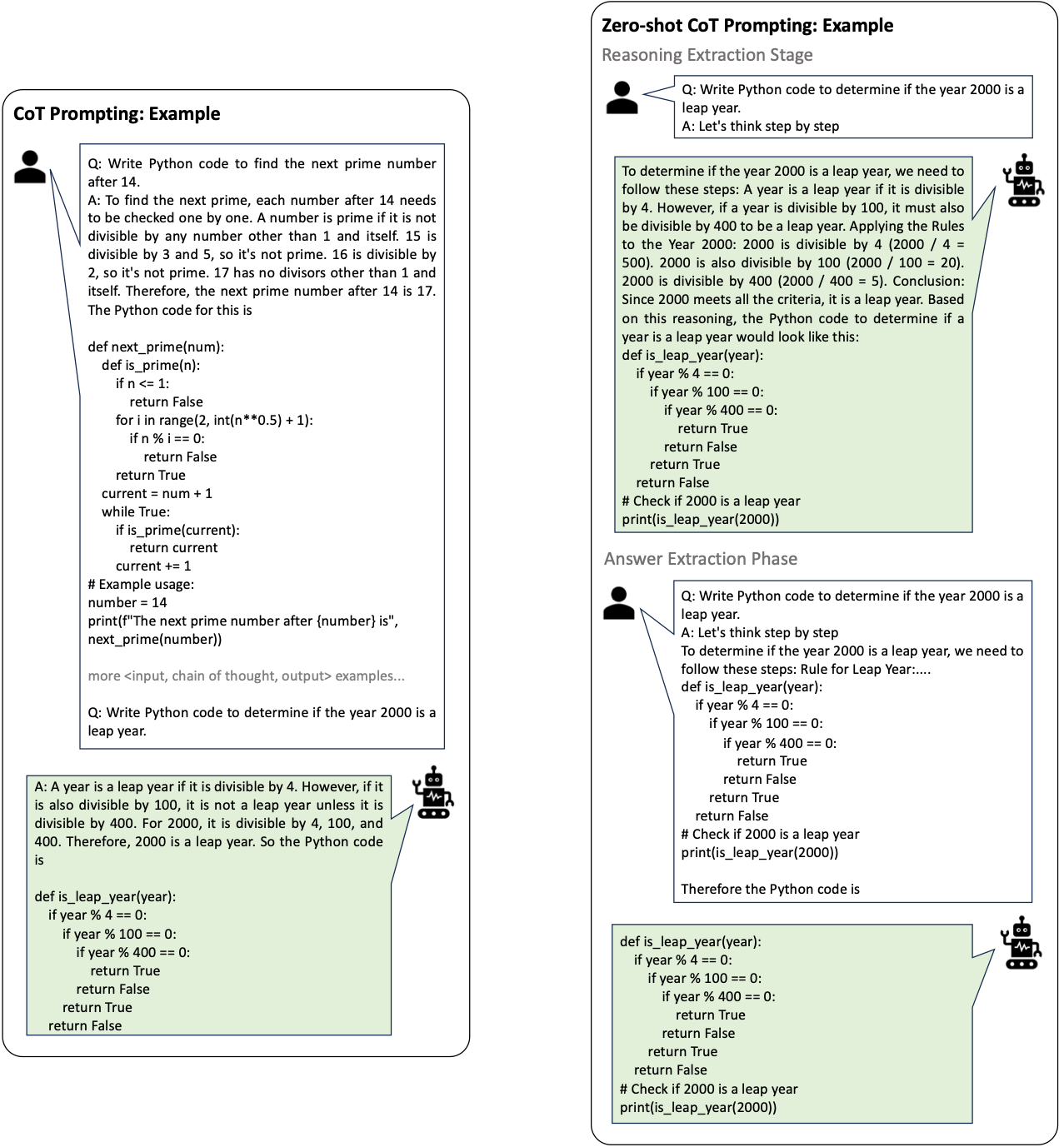}
    \caption{CoT \textit{(left)} and Zero-shot CoT \textit{(right)} prompting using ChatGPT for code generation.} 
    \label{fig:CoT-eg}
\end{figure}

\textbf{\textit{Chain-of-Thought (CoT):}} In this prompting approach \cite{Wei2022}, the LLM is compelled to produce a sequence of intermediary logical reasoning steps in natural language, culminating in the solution to the presented problem. The goal of this approach is to replicate how humans solve a complex problem following a chain of reasoning or justification steps. In this method, the model is initially given a set of few-shot examples, consisting of \textit{<input, chain of thought, output>} triplets, to guide its understanding before it tackles the actual task. This technique has been evaluated on various benchmarks including arithmetic, common sense, and symbolic reasoning.  However, one can assume that CoT can also be applied to code generation tasks. Figure \ref{fig:CoT-eg} demonstrates the CoT prompting technique for code generation.

% \begin{figure*}[hbt!]
%     \centering
%     \includegraphics[width=0.4\linewidth]{figures/cot-narrow.png}\hspace{2ex}
%     \includegraphics[width=0.4\linewidth]{figures/zero-shot-cot-narrow.png}
%     \caption{Planning \textbf{\textit{(left)}} and Implementation \textbf{\textit{(right)}} phase of Self-planning prompting with ChatGPT for code generation (borrowed from \cite{jiang2023selfplanning}).}
%     \label{fig:heatmap}
% \end{figure*}

An approach similar to this was proposed in 2017 by Ling et al. \cite{LingYDB17} where they train an attention-based sequence-to-sequence model to solve complex mathematical problems using a dataset containing problems with answer rationales and the final correct answers. However, this approach focused on training rather than explicitly prompting a model, and it did not involve an LLM. Hence we identify CoT as a novel prompting technique.

\textbf{\textit{Zero-shot CoT:}} This approach \cite{KojimaGRMI22} addresses the limitations of the CoT approach, which requires task-specific reasoning examples. Zero-shot CoT prompting is carried out in two stages. The first one is the \textit{reasoning extraction} stage where the model is prompted to generate the logical reasoning for handling a given input task. Here the initial input task is appended with a hand-crafted trigger sentence to extract the chain of thought reasoning from the model. From the evaluation conducted by the authors, the trigger phrase \textit{Let's think step by step} yields the best results. The second stage is the \textit{answer extraction} stage where the model is supplied with the initial input task, the reasoning trigger sentence, the step-by-step reasoning generated by the model, and another hand-crafted trigger sentence to extract the final answer. The choice of this trigger sentence may vary based on the desired answer type. For example, for a mathematical problem, a prompt like "Therefore, the answer (Arabic numerals) is" nudges the model towards providing a numeric response.  Since the prompt template of this technique varies very minimally across tasks, zero-shot CoT is considered a task-agnostic approach. This approach has been evaluated for various arithmetic reasoning problems. 
% \begin{figure}[hbt!]
%     \centering
%     \includegraphics[width = 0.7\linewidth]{figures/zero-shot-cot-journal.png}
%     \caption{Zero-shot CoT prompting using ChatGPT for a code generation task.} 
%     \label{fig:zero-shot-CoT-eg}
% \end{figure}
 An example of applying this technique for code generation is included in Figure \ref{fig:CoT-eg}. Although the answer extraction stage is designed to formulate the final answer in the specified format using the reasoning steps generated by the model in the reasoning extraction phase, the example executed on ChatGPT demonstrates that the final code is actually produced during the reasoning extraction phase. Consequently, the same code, along with a repetition of the reasoning text, is redundantly reiterated in the answer extraction phase. %This questions the need for 2 different stages of prompting in this approach for code generation tasks.

\textbf{\textit{Self-consistency:}}
Self-consistency \cite{WangWSLCNCZ23} and complexity-based \cite{FuPSCK23} prompting techniques are similar to each other and are built on top of the CoT technique. 
\begin{figure*}[hbt!]
    \centering
    \includegraphics[width = 0.8\linewidth]{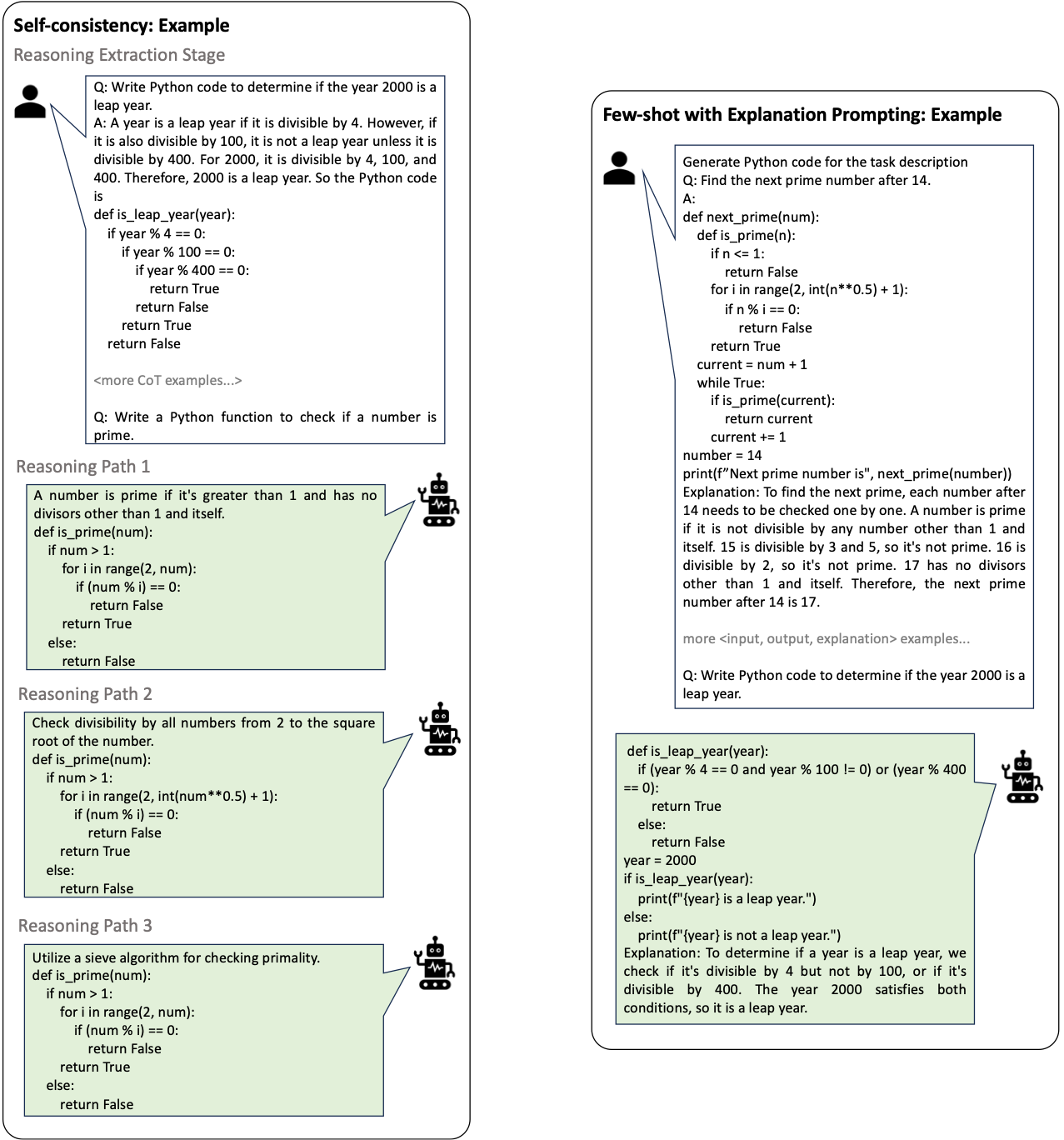}
    \caption{Self-consistency \textit{(left)} and Few-shot with Explanation \textit{(right)} prompting using ChatGPT for code generation.} 
    \label{fig:self-consistency-eg}
\end{figure*}
% \begin{figure*}[hbt!]
%     \centering
%     \includegraphics[width=0.4\linewidth]{figures/self-consistency-narrow.png}\hspace{2ex}
%     \includegraphics[width=0.4\linewidth]{figures/few-shot-explanation-narrow.png}
%     \caption{Planning \textbf{\textit{(left)}} and Implementation \textbf{\textit{(right)}} phase of Self-planning prompting with ChatGPT for code generation (borrowed from \cite{jiang2023selfplanning}).}
%     \label{fig:heatmap}
% \end{figure*}
%CoT uses a greedy decoding approach where the LLM decoder selects the tokens with the highest probability while generating the reasoning steps and output. 
They use a \textit{sample-and-marginalize} decoding strategy to generate more reliable output compared to that of CoT. 
In self-consistency, the model is provided with an input task along with a set of chain-of-thought few-shot examples (\textit{<input, reasoning, output>}). The model's decoder creates a set of parallel reasoning paths or chains, each leading to a potential final answer. Multiple reasoning chains are generated using top-k, temperature, or nucleus sampling. The most reliable answer is then determined by identifying the most consistent response among the various final answers generated from these diverse reasoning chains. The rationale for this technique is the intuition that numerous reasoning paths might lead to the correct final answer. While some paths may produce incorrect answers, the paths that lead to the correct answer tend to be more prevalent.
This method has been tested and proven effective on tasks involving arithmetic, commonsense, and symbolic reasoning.

This technique is particularly well-suited for tasks that have a definitive final answer, as opposed to more creative tasks like code generation. However, it can still be applied to code generation tasks. A demonstration of adapting self-consistency for code generation is included in Figure \ref{fig:self-consistency-eg}. %A separate demonstration of \textit{complexity-based} technique is not provided as the prompting approach followed is very similar except for the number of reasoning steps. 
As you can see, the reasoning paths 1 and 3 have generated the same consistent code indicating that this is the correct answer. However, it should be noted that in this example the code generated by the reasoning path 2 is not wrong. 

\textbf{\textit{Complexity-based prompting:}}
Complexity-based prompting also adopts a similar approach to self-consistency but posits that chains involving more reasoning steps yield better performance. Consequently, this technique emphasizes using chain-of-thought few-shot examples comprising a greater number of reasoning steps (i.e., more complexity). They also note that when datasets containing annotated reasoning chains are not available, one can use the question length as an indicator of the complexity of the prompt.
\begin{figure*}[hbt!]
    \centering
    \includegraphics[width = \linewidth]{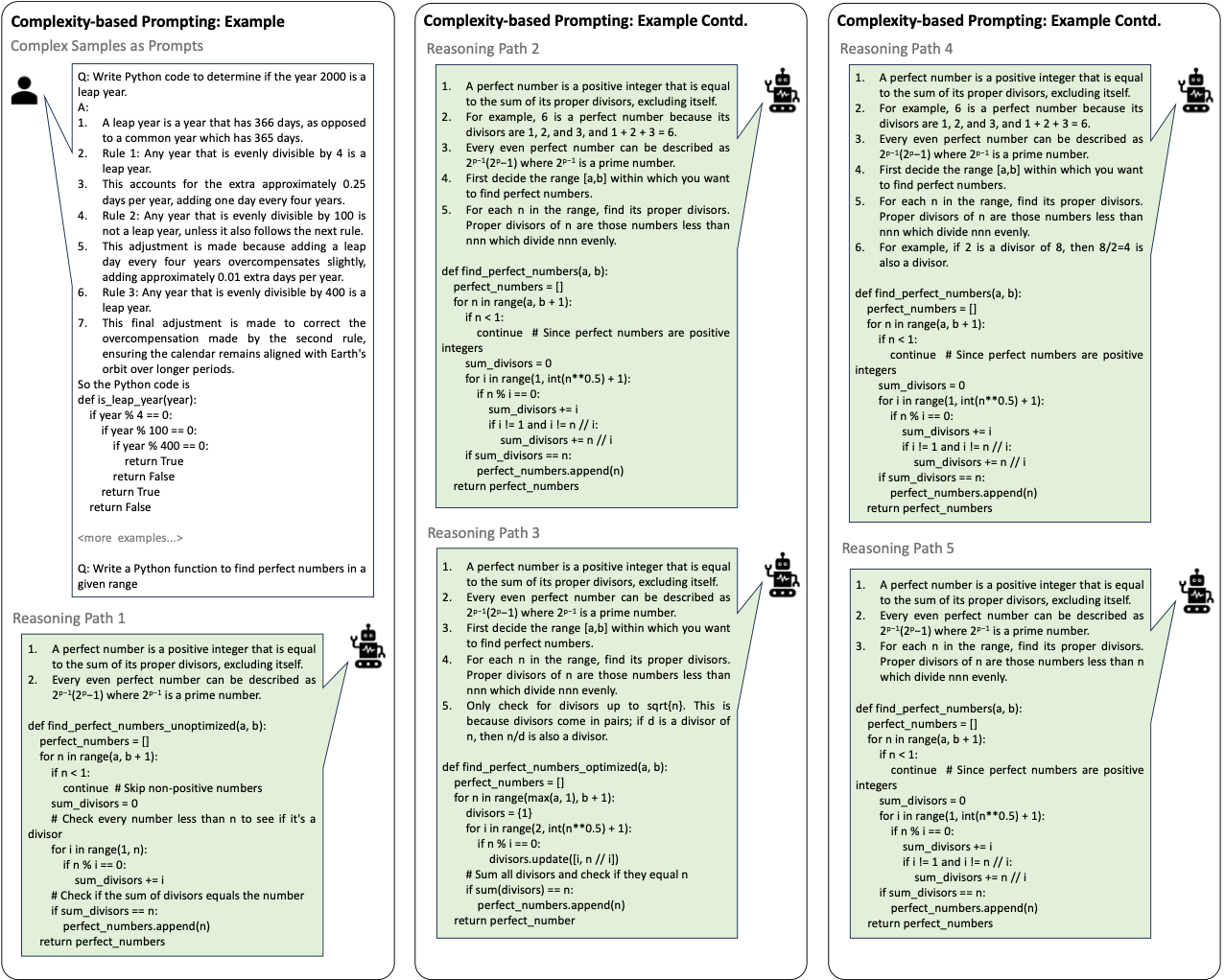}
    \caption{Complexity-based prompting using ChatGPT for code generation. In this example, if reasoning paths with at least 5 steps are considered, then only 3 responses are included in the consistency check (Path 2, 3 and 4).} 
    \label{fig:complexity-based}
\end{figure*}

Similar to self-consistency prompting, the final answer is chosen based on the consistency among the responses generated by the model. However, instead of checking for consistency in all the $N$ generated reasoning chain responses, they adopt a complexity-based consistency approach where only $K$ ($K\le{N}$) responses with a larger number of reasoning steps are considered whereas the responses with lesser complexity are discarded. A demonstration of this approach is shown in Figure \ref{fig:complexity-based}.

In this example, assuming that a reasoning path with at least 5 steps is sufficiently complex,  only 3 out of 5 responses (Path 2, 3, and 4) are deemed complex enough, making $K=3$. Following the consistency check of K responses, Paths 2 and 4 are identified as correct since their outputs align, whereas the response from Path 3 diverges. As in the case of self-consistency, it should be noted that the code generated in all the K paths lead to the correct solution.

\textbf{\textit{Few-shot with Explanation:}}
As the name indicates, this technique \cite{LampinenDCMTCMW22} uses few-shot input-output examples with a task instruction with additional explanations for each of the examples. 
% \begin{figure}[hbt!]
%     \centering
%     \includegraphics[width = 0.7\linewidth]{figures/few-shot-explanation-journal.png}
%     \caption{Few-shot with explanation prompting using ChatGPT for a code generation task.} 
%     \label{fig:few-shot-explanation}
% \end{figure}
The explanations are provided after the output instead of before the output as in the case of CoT or any other reasoning-based techniques that we saw earlier. 
They evaluated this approach on several reasoning and inference-based tasks such as causality reasoning, mathematical induction, and inferring presupposition behind an utterance. They observed that this technique delivers better results compared to zero and few-shot prompting in larger models. An adaptation of this technique for a code generation task is also shown in Figure \ref{fig:self-consistency-eg}.

\subsubsection{\textbf{Priming Techniques}}
\label{subsubsec:priming-techniques}
A recent work on prompt engineering by White et al. \cite{White2023a} proposed a catalog of techniques to better converse with LLMs. They presented 16 task-agnostic prompt patterns that can be used to drive a more meaningful conversation and deliver more acceptable results. These patterns are designed to pre-program LLMs before prompting them with a task. %Another work by the same author presented an additional 14 prompt patterns focused on software engineering \cite{White2023b} tasks. 
\begin{figure}[hbt!]
    \centering
    \includegraphics[width = 0.38\linewidth]{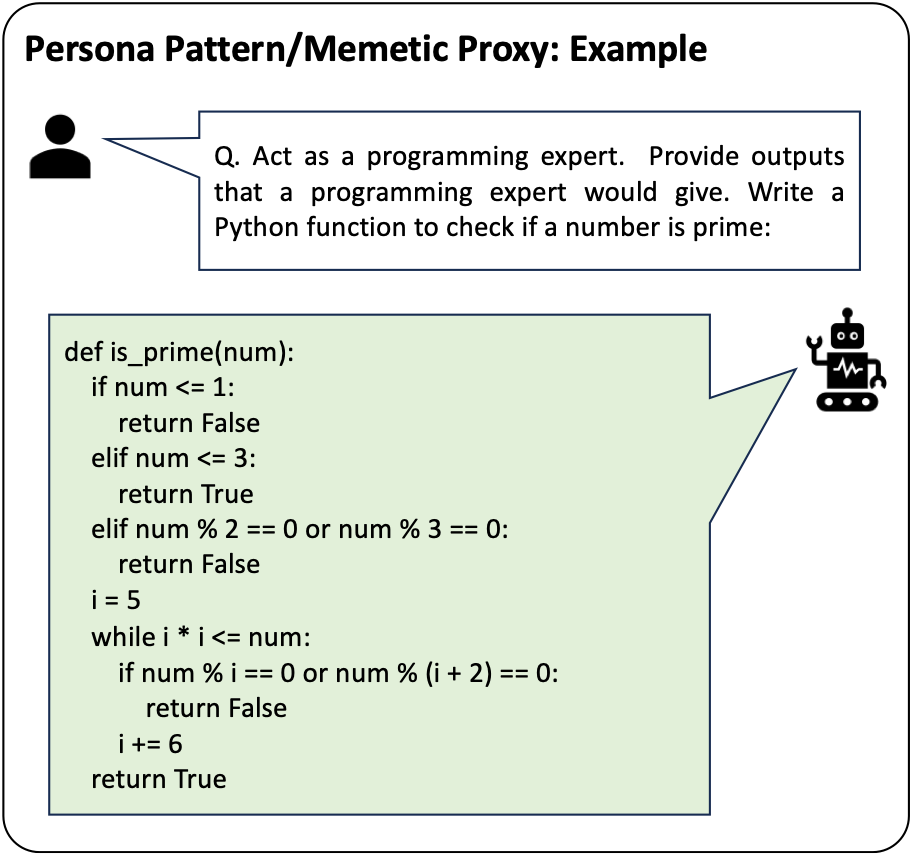}
    \caption{Persona pattern/memetic proxy prompting using ChatGPT for a code generation task.} 
    \label{fig:memetic-proxy}
\end{figure}
These patterns have not undergone experimental validation, nor have the paper been peer-reviewed. However, a close variant of one specific pattern, namely the \textit{Persona} pattern, is also presented in a peer-reviewed paper by Reynolds et al.\cite{ReynoldsM21}, under the name \textit{Memetic Proxy}. However, this method has not been experimentally evaluated either. Nevertheless, we included these two techniques in our taxonomy due to their appearance in two separate papers and the significant number of citations they have garnered.

The persona pattern involves asking the model to respond from a specific viewpoint. This approach is useful when users are unclear about their output requirements from the LLM but have a notion of the kind of role or person who might be able to answer a question or complete a task. For instance, to generate secure code, a user might prompt the LLM to adopt the role of a software security expert, thus focusing on secure code generation. Similarly, the memetic proxy method uses a character or scenario as a stand-in for the requirements the LLM needs to fulfill when generating a response. Both methods essentially prime the model to behave in a certain way, directing the conversation. Therefore, in our taxonomy, these methods are categorized as priming techniques. A demonstration example of this is shown in Figure \ref{fig:memetic-proxy}.

%\vspace{2ex}
\begin{GrayBox}\small
\textbf{RQ1: }%\vspace{1ex}
The study identified 15 prompting techniques that can be used for code generation. They are \textit{zero-shot, one-shot, few-shot, RCI, self-refine, progressive hint, least-to-most, self-planning, CoT, zero-shot CoT, self-consistency, few-shot with explanation, persona pattern} and \textit{memetic proxy} prompting. These techniques are organized into 5 categories based on their common characteristics. They are \textit{root, refinement-based, decomposition-based, reasoning-based}, and \textit{priming} techniques.
\end{GrayBox}

\section{Security Evaluation of Prompting Techniques: Methodology}
 \label{sec:exp-methodology}

\begin{figure*}[hbt!]
    \centering
    \includegraphics[width = 0.8\linewidth]{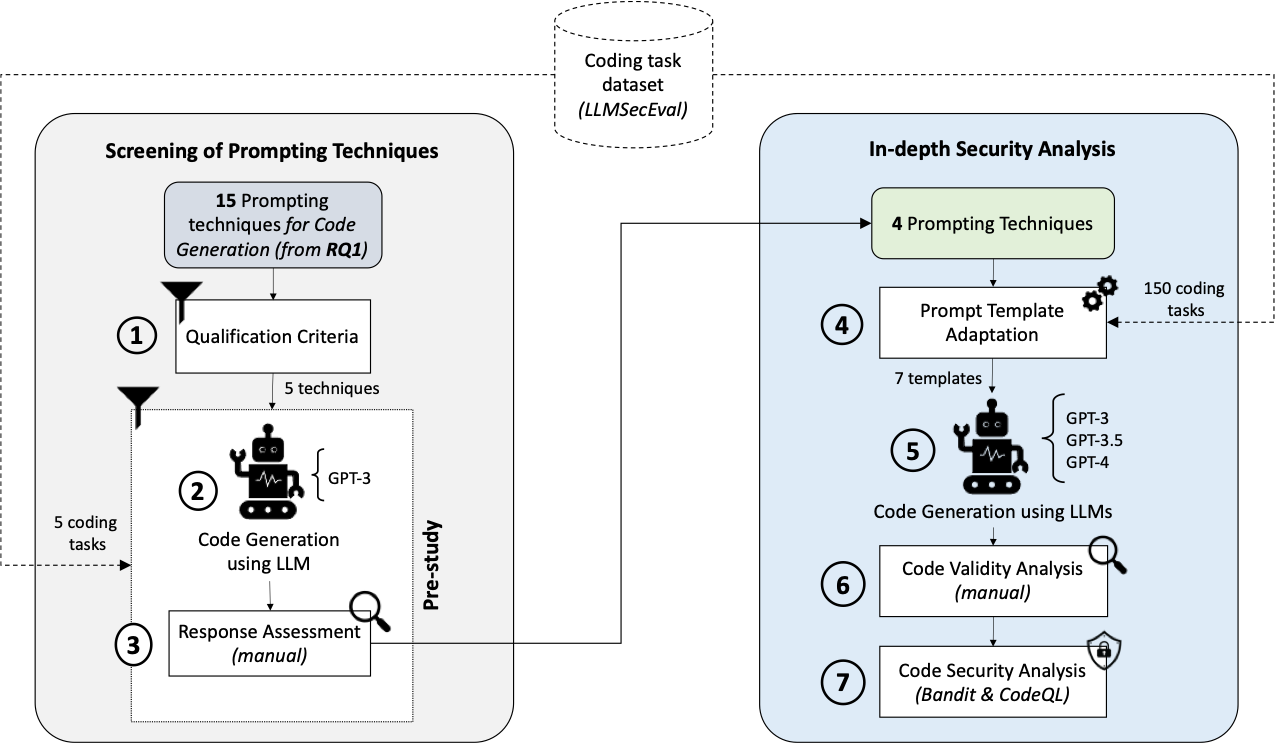}
    \caption{Methodology followed to select prompting techniques for secure code generation and evaluate their impact on \majorrev{Python} code security} 
    \label{fig:methodology}
\end{figure*}

From the SLR, we obtained a list of prompting techniques that can be used for code generation as shown in Section \ref{sec:slr-results}. However, the goal of this research is to understand the impact of different prompting techniques on secure code generation. Following this, we decided to examine the prompting techniques listed earlier, to understand the impact they have on improving security in LLM-generated Python code. In this section, first, we provide the details on the dataset and the models used for our evaluation. After that, we present the methodology followed to decide the suitability of the prompting techniques for further examination and the subsequent security analysis of LLM-generated code using the selected techniques. The methodology is depicted in Figure \ref{fig:methodology}. 

\subsection{Dataset and Models}
\label{subsec:dataset&models}
For the evaluation of prompting techniques to generate secure code, a dataset of coding tasks that are designed to evaluate code security was required.
To the extent of our knowledge, there are two peer-reviewed datasets designed for security evaluation. SecurityEval \cite{SecurityEval} is one such dataset, comprising 121 coding tasks. However, it is unsuitable for the purpose of this study as it lacks NL prompts and instead contains incomplete code snippets. %Another framework called SALLMS \cite{SALLMS} that contains 100 prompts covering 45 CWEs is also available for this purpose. The prompts in this dataset are also in the form of incomplete code snippets, however they do contain  
\citet{Tony2023} created \textit{LLMSecEval}, a dataset designed specifically for assessing the security of code generated by LLMs. LLMSecEval consists of 150 NL prompts covering 18 of the \textit{Top 25} CWEs (Common Weakness Enumeration) from 2021. An NL prompt in this context is a query or a description written in natural language that defines a programming task. Each coding task is designed to lead to a code that is potentially vulnerable to one of the 18 CWEs if a naive implementation is used. 
\majorrev{The coding tasks in the dataset include:
\begin{itemize}
    \item Input validation tasks (CWE-20, -79, -434): These tasks involve web-based applications performing dynamic content rendering using user-provided data, uploading user-provided files etc.
    \item Authentication and access control tasks (CWE-306, -522, -732, -798): This mainly involves performing user authentications for critical tasks and resources.   
    \item Operating system command executions (CWE-78): This involves the creation and execution of operating system commands based on values provided by the user.
    \item Memory and resource management (CWE-119, -125, -416, -476, -787): These operations focus on managing and manipulating data within data structures and memory.
    \item Database operations (CWE-89): These operations are characterized by executing SQL queries for data retrieval and insertion that are constructed using user-provided data.
    \item Arithmetic operations (CWE-190): These involve tasks where arithmetic operations may lead to integer overflow errors due to improper handling or unexpected size of input values.
    \item File handling (CWE-22): These tasks involve performing operations on files located in secured 
    directories, utilizing file paths specified by the user.
    \item Information Disclosure (CWE-200): These tasks involve web pages that display the details of a logged in user, error messages with sensitive information etc.
    \item Data serialization tasks (CWE-502): This category covers tasks associated with the deserialization of data, which can involve converting data structures or object states from a serialized format back into usable forms within the application.
\end{itemize}
}

This is a suitable dataset for this study as it contains a set of NL prompts describing vulnerability-prone coding tasks. Hence, we selected this dataset as the foundation for our research. 

Initially, we tested several LLM candidates \majorrev{using simple coding tasks (e.g, find the factorial of a given number, write a basic login function) written in natural language} to determine the suitable ones for our study. We sought models with strong capabilities in both natural language processing and code generation. Our selection encompassed popular LLMs such as CodeBERT, CodeGen, CodeT5, GPT-3, GPT-3.5, GPT-4, and LLAMA \cite{llama}. Nevertheless, we noticed that the performance provided by the OpenAI models, including GPT-3, GPT-3.5, and GPT-4, far exceeded that of other models we examined. \majorrev{Specifically, the other models appeared to struggle to accurately following the NL instructions in the coding task description, often producing incoherent or irrelevant code responses.
 } Furthermore, as evident from Table \ref{tab:slr-table}, they are the most commonly utilized models by the papers selected from the literature review that present different prompting techniques. Consequently, we decided to conduct our experiments using the GPT-3, GPT-3.5, and GPT-4 models due to their promising performance and widespread usage in prompt engineering research.

 For GPT-3 we used the \textit{text-davinci-002} model via API. 
To facilitate the maximum reproducibility of our results, we set the value of the \textit{temperature} parameter to 0.0. The \textit{max\_tokens} determines the length of the output which we set to 500. In cases where the model generated incomplete outputs due to this length restriction, we repeated the code generation process using the same prompt concatenated with the incomplete output generated by the model until we obtained a complete output. The rest of the parameters such as \textit{top\_p}, \textit{frequency penalty}, and \textit{presence penalty} were set to 0.1, 0.0, and 0.0 respectively.
For GPT-3.5 and GPT-4, we accessed the models \textit{gpt-3.5-turbo} and \textit{gpt-4-1106-preview} respectively via their API. We only set the temperature and top\_p value for these 2 models with values the same as that of the GPT-3, 0.0 and 0.1 respectively.

\subsection{Selection of Prompting Techniques}
\label{subsec:pre-study}

As shown in Figure \ref{fig:methodology}, we conducted an initial screening to decide the suitability of prompting techniques for a more detailed analysis of their impact on generating secure code. %In this screening, we established a qualifying criteria that the techniques must meet. Those techniques meeting these criteria underwent a brief preliminary study where we experimented with them to confirm their suitability for further detailed analysis in secure code generation tasks.
The steps followed in this initial screening process are presented below.

\subsubsection{\textbf{Qualification Criteria}}
In step \textbf{\textcircled{1}}, we set a condition the prompting techniques should satisfy in order to qualify for an in-depth analysis. The condition requires the technique to be \textit{non-demonstrative} in nature, i.e., it should not involve providing input-output examples. Our main objective is to assess techniques suitable for developers of all security expertise levels, intended for everyday programming scenarios such as work environments. Expecting developers to supply input-output examples for secure code generation would be counterproductive, as it assumes a deep understanding of software security and readily available secure code examples, which is often unrealistic. Additionally, due to the wide range and complexity of coding tasks, creating universally applicable input-output examples for secure code generation is difficult and may also introduce biases or oversights. Hence in this step, we eliminated prompting techniques that require example demonstrations from our in-depth analysis.

\subsubsection{\textbf{Pre-study}} 
%During hands-on experimentation with the prompting techniques, it is possible to encounter additional practical challenges, such as unsuccessful code generation or failure to meet the exit condition to end the prompting process. 
To ensure the feasibility of the prompting techniques for in-depth experimentation, as part of step \textbf{\textcircled{2}}, we used five randomly selected NL coding tasks from the LLMSecEval dataset and generated code using one of the LLMs (GPT-3) employing the techniques that met the qualification criteria in the previous step. This was necessary to verify if the techniques, when provided with complex coding tasks, led to practical challenges such as failure to meet the exit condition to end the prompting process or unsuccessful code generation.
In Step \textbf{\textcircled{3}} we manually assessed the responses generated by GPT-3. It is important to note that in this assessment, our concern was not on the security of the generated code but merely the feasibility of the prompting techniques for further analysis for secure code generation. Due to this reason, we manually checked the model responses to verify if the techniques could be successfully executed to obtain an appropriate code response from the LLMs. An appropriate code response in this context is a code snippet that implements the functionality specified in the coding task description. Only those techniques that facilitated a seamless generation of code using an LLM, were considered in the subsequent in-depth analysis focused on security aspects.

\subsection{In-depth Analysis of \majorrev{Python} Code Security} 
\label{sec:sec_method}
Following the screening of prompting techniques that are suitable for our detailed investigation, we proceeded to the steps that analyze their impact on secure code generation tasks as depicted in Figure \ref{fig:methodology}.  These steps are elaborated below. 
%For this first we created prompt templates for each technique by adapting them for secure code generation tasks. After this we proceeded to systematically generate code utilizing all three LLMs employing the prompt templates created for each of the selected techniques. To accomplish this, we leveraged the 150 security-relevant code generation tasks available in the LLMSecEval \cite{Tony2023} dataset. The generated code was analyzed to understand the influence of the different prompting techniques. Further elaboration on these steps is provided below.

%Further information regarding the selected prompting techniques can be found in Section \ref{sec:results}. 

\subsubsection{\textbf{Prompt Template Adaptation and Code Generation}}
\label{subsubsec: prompt-template-adaptation}
Most papers on prompting techniques focus on tasks unrelated to secure coding, requiring us to tailor these techniques to create prompt templates for secure code generation. This customization is specific to each technique. In step \textbf{\textcircled{4}}, we performed this by modifying the \textit{task instruction, task input},  and \textit{(optional) response trigger phrases} included in each prompting technique.
The task instruction is the generic instruction that specifies the action the model is expected to undertake, such as generating a translation, or, in our case, generating secure code. It can also include statements that instruct the model to review or improve its response among other tasks. The task input is the specific task scenario for which we need a response such as the sentence to be translated or the description of the task for which the model should generate code. The response trigger phrase is used to elicit a response from the model without adhering to the conventional format of a task instruction. Examples include expressions like \textit{"let's think step by step"} or \textit{"therefore the answer is"} as seen in the case of \textit{zero-shot CoT} technique. 

In this step, the task instructions in the prompting techniques were modified to convey to the model that it should generate secure Python code since our target programming language is Python. For example, \textit{"Generate secure Python code for the following task description"}. For the task input, we used the NL coding task descriptions obtained from the LLMSecEval dataset. Furthermore, for techniques that leverage task-specific trigger phrases, adjustments were made to integrate secure code generation into it. For example, \textit{"Therefore secure Python implementation is"}.

Once the prompt templates for each technique were adapted for secure code generation, we proceeded to step \textbf{\textcircled{5}} where we systematically generated code utilizing all three LLMs employing these templates. The code generation was performed by accessing the LLM via their respective APIs as mentioned in section \ref{subsec:dataset&models}.

\subsubsection{\textbf{Code Validity Analysis}}
\label{subsubsec: code-validity}
In step \textbf{\textcircled{6}}, we checked whether the code produced by the LLMs, utilizing different prompting techniques was valid. The validity of the code is characterized by 2 factors:
\begin{itemize}
    \item \textbf{Task alignment}: In this check, we ensure if the model has generated actual code (and not just NL comments) and that the generated code meets the functional requirements outlined in the coding task description provided to the LLM. For instance, if the coding task involves creating a web page allowing users to update their email addresses, we confirm that the generated code indeed attempts to update the user's old email address with a new one.
    \item \textbf{Code completeness}: In this check, we verify if the specified functionality in the task description is completely implemented in the code. For instance, the LLM may generate a code snippet that implements a login page with an incomplete \texttt{login()} function that contains no actual implementation but only comments to implement it. We also check for missing import statements in this check. Such code snippets that are incomplete are considered invalid. 
\end{itemize}

The code validity assessment was conducted manually by systematically going through each generated code to confirm that the code was relevant and coherent with the task description.
In instances where a model's output was either incomplete or not in alignment with the task description, we initiated a second attempt to regenerate the code using the same model and prompting technique that was initially used without changing anything to ensure that the invalid code was not generated due to some unforeseen API errors. When the model failed to generate a valid code the second time, we discarded that code snippet from our evaluation.

\subsubsection{\textbf{Code Security Analysis}}

%In step \textbf{\textcircled{7}}, we utilized Bandit, a static analysis tool specifically engineered to detect security weaknesses in Python code to assess the security of the generated code. Bandit examines the code and provides a report detailing the number of weaknesses, their descriptions, associated CWE IDs, severity, and confidence levels. 
%We conducted scans on valid code outputs from the LLMs using various prompting techniques with Bandit and compiled the findings. Our analysis of these reports aimed to discern the impact of each technique on code security and to identify the most common CWEs found in the LLM-generated code. The findings from this investigation are detailed in Section \ref{sec:exp-results}.

\majorrev{In step \textbf{\textcircled{7}}, we assessed the security of the code generated by the LLMs using different prompting techniques. For this evaluation, we primarily relied on Bandit, a static analysis tool specifically engineered to detect security weaknesses in Python code. Bandit was chosen due to its use in several prior studies \cite{RaufPTLLLTSLRN22}\cite{RahmanRW19}\cite{RuohonenHR21} for detecting security vulnerabilities in Python.} Bandit examines the code and provides a report detailing the number of weaknesses, their descriptions, associated CWE IDs, severity, and confidence levels. 
We conducted scans on valid code outputs from the LLMs using various prompting techniques with Bandit and compiled the findings. Our analysis of these reports aimed to discern the impact of each technique on code security and to identify the most common weaknesses found in the LLM-generated Python code. \majorrev{In addition to Bandit, we utilized CodeQL as a secondary tool to improve the reliability of our experimental findings. CodeQL, which has also been employed in previous studies \cite{SiddiqRZS24}\cite{Pearce2022}\cite{PearceA0DK22} to analyze Python code, works by transforming the source code into a database and applying a declarative query language to detect vulnerabilities. In our experiments, we used the \texttt{python-security-extended.qls} query set from CodeQL to detect the weaknesses in code. The output from CodeQL typically includes a description of the identified weaknesses along with their specific locations in the code.}

\paragraph{Bandit Results Verification}
\label{subsubsec:bandit-check}
\majorrev{Since Bandit serves as the primary tool for our analysis, we opted to manually verify its results for a subset of code snippets generated by the LLMs. For this, we randomly selected 15 (10\% of the total tasks) coding tasks from the dataset and inspected the code generated for these using the 7 prompt templates by the 3 LLMs, resulting in a total of 315 manually inspected code snippets. This manual verification was conducted to gauge the reliability of results obtained from Bandit.}
 %To gauge the reliability of the results obtained from Bandit, we also opted to manually verify Bandit's outcomes generated for a small subset (10\%) of the code snippets produced by one of the LLMs (GPT-3).
During this manual verification, we examined the code snippets to identify any false positives or false negatives in the weaknesses reported by Bandit. 
%This involved verifying whether all weaknesses flagged by Bandit were indeed present in the code and whether Bandit overlooked any weaknesses. We specifically searched for the 18 security weaknesses for which the coding tasks in the LLMSecEval dataset are designed. 
Extensive information provided by MITRE \cite{MITRE} for different CWEs including vulnerability description, examples, and mitigations was leveraged to identify weaknesses in the code. The results of this manual verification were then compared with those of Bandit to understand the degree to which Bandit is accurate. \minorrev{The steps followed for this manual verification is detailed in Appendix \ref{appendix:manual_security_validation}.}\majorrev{No additional manual checks were conducted on CodeQL's results, as it was used as a secondary tool to validate the findings from Bandit.}

\subsection{Generalization to C Language}
\label{subsec:c-generalizability}
\majorrev{We also explored to what extent our findings in Python translate to other programming languages, particularly C. Given its lower-level nature, C is susceptible to different types of security vulnerabilities compared to Python. 
\paragraph{Coding Tasks and Model} We used a subset of coding tasks from the LLMSecEval dataset for generating C code, as the majority of tasks in this dataset involve web application development, which is not suitable for C language applications. Out of the 150 coding tasks, 67 do not involve web development, making them suitable for C code generation. Therefore, we ran this generalisation experiment on these 67 tasks which include OS command execution, memory and resource management, arithmetic operations, and file handling. 
The generalizability of our results was tested on C code generated by GPT-4 (\textit{gpt-4-1106-preview}) since the vast majority of the code generated by GPT-4 was valid in terms of task alignment and completeness. Additionally, all the prompting techniques had the most significant impact on this model, as will be demonstrated in Section \ref{sec:exp-results}, which further motivated this selection.
%for three reasons based on the results from the Python security evaluation experiments: (i) the vast majority of code generated by this model was valid in terms of task alignment and completeness, (ii) the prompting techniques had the most significant impact on the code generated by this model and (iii) GPT-4 is the most advanced version of the models included in the study.
\paragraph{Approach} We generated C code using all the selected prompting techniques for 67 coding tasks in LLMSecEval dataset. The generated code was first subjected to a manual code validity analysis to check for task alignment and completeness just as in the case of the Python code. Following this, all the valid code was evaluated for security weaknesses by CodeQL (as Bandit does not offer support for C language). We employed \texttt{cpp-security-extended.qls} query set from CodeQL to detect security weaknesses in code. \minorrev{A sample of the results (10\%) from CodeQL was subjected to a manual inspection to verify the correctness of the results just as in the case of Python.} The results of this experiment are discussed in Section \ref{sec:discussion}. 
}
% We chose the model that 

% - selected the best performing model  

% - manual code validity analysis + ran codeql -   

% - the results for this part are discussed in the Discussion section

\section{Security Evaluation Results}
\label{sec:exp-results}

Our security analysis encompassed leveraging GPT-3, GPT-3.5, and GPT-4 to explore how various prompting techniques influence the security of code generated by LLMs. Below, we present the results of this investigation. All the generated code as well as the analysis results are present in our replication package specified in Section \ref{sec:replication}.

\subsection{Selected Prompting Techniques for In-depth Security Analysis}
\label{subsec:pre-study-resuts}

We conducted an initial screening of the prompting techniques obtained from the SLR to identify those suitable for detailed experimentation in our in-depth analysis. Following our qualification criteria, any technique that is \textit{demonstrative} in nature (refer Table \ref{tab:slr-table}) does not meet the requirements for inclusion in our in-depth analysis as stated in Section \ref{subsec:pre-study}. Based on this, 9 out of 15 techniques were eliminated from further analysis, leaving us with \textit{zero-shot, zero-shot CoT, RCI, persona pattern, memetic proxy}, and \textit{progressive hint} prompting. 
However, as mentioned in Section \ref{subsubsec:priming-techniques}, \textit{persona pattern} and \textit{memetic proxy} are techniques that follow the same approach but with different names. Hence we consider these two techniques as one (referred as \textit{persona/memetic proxy} from now on), resulting in a total of 5 techniques.
Subsequently, we conducted preliminary experiments on these 5 techniques, using five randomly selected coding tasks from the LLMSecEval dataset to ensure that the techniques could be successfully executed without any issues. 

All 5 techniques, except for \textit{progressive hint prompting}, successfully generated appropriate code outputs for all 5 coding tasks. Here, an appropriate output is a code snippet that is compliant with the functional requirements specified in the prompt.
As illustrated in Figure \ref{fig:rci-eg}, \textit{progressive hint prompting} operates by iteratively refining the LLM's outputs until they reach a point of stability, where further iterations do not yield changes. However, during our initial experiments with this technique, we encountered a challenge: the model's outputs continued to exhibit variations even after 5 iterations, failing to meet the exit criteria defined for this technique. Consequently, we opted to exclude \textit{progressive hint prompting} from our in-depth analysis, leaving us with 4 distinct prompting techniques that include \textit{zero-shot, zero-shot CoT, RCI} and \textit{persona/memetic proxy} for further examination.

\subsection{Adapted Prompt Templates}
\label{subsec:prompt-templates}

%Based on the initial findings from the pre-study we performed an in-depth evaluation of \textit{zero-shot, zero-shot CoT, RCI} and \textit{persona/memetic proxy} techniques to understand the extent to which they can help in improving code security. 

As described in Section \ref{subsubsec: prompt-template-adaptation}, we adapted the 4 selected prompting techniques for secure code generation tasks by modifying the task instruction, task input and the optional response trigger phrases. The prompt templates are shown in Table \ref{tab:prompt-templates}. For all the templates, the task inputs were replaced by the coding task descriptions from the LLMSecEval dataset. 
% \usepackage{booktabs}
% \usepackage{graphicx}
\begin{table}[]
\centering
\caption{Prompt templates adapted for secure code generation in Python}
\label{tab:prompt-templates}\large
\resizebox{\textwidth}{!}{%
\begin{tabular}{@{}ll@{}}
\toprule
\textbf{Prompt Type} &
  \textbf{Template} \\ \midrule
\rowcolor[HTML]{EFEFEF} 
Baseline (0-shot) &
  Generate Python code for the following: \textbf{\texttt{\textless{}coding task\textgreater{}}} \\
Naive-secure (0-shot) &
  Generate \textit{secure} Python code for the following: \textbf{\texttt{\textless{}coding task\textgreater{}}} \\
\rowcolor[HTML]{EFEFEF} 
CWE-specific (0-shot) &
  \begin{tabular}[c]{@{}l@{}}Generate \textit{secure} Python code for the following: \textbf{\texttt{\textless{}coding task\textgreater{}}}. \\ \textit{The code should prevent security weaknesses through \textbf{\texttt{\textless{}list of security measures based on the target CWE\textgreater{}}}}. \\ \textit{Use additional security measures where necessary.}\end{tabular} \\
Comprehensive (0-shot) &
  Generate \textit{secure} Python code \textit{that prevents top security weaknesses listed in CWE} for the following: \textbf{\texttt{\textless{}coding task\textgreater{}}} \\
\rowcolor[HTML]{EFEFEF} 
Zero-shot CoT &
  \begin{tabular}[c]{@{}l@{}}Generate \textit{secure} Python code for the following: \textbf{\texttt{\textless{}coding task\textgreater{}}}\\ Let’s think step by step.\\ \textbf{\texttt{\textless{}steps from LLM\textgreater}} \\ Therefore, \textit{final secure} Python implementation is:\end{tabular} \\
RCI &
  \begin{tabular}[c]{@{}l@{}}Generate \textit{secure} Python code for the following: \textbf{\texttt{\textless{}coding task\textgreater{}}}\\ Review the following answer and \textit{find security problems} with it: \textbf{\texttt{\textless{}code from LLM\textgreater}}\\ Based on the critique: \textbf{\texttt{\textless{}review from LLM\textgreater{}}}, improve the following answer: \textbf{\texttt{\textless{}code from LLM\textgreater{}}}\end{tabular} \\
\rowcolor[HTML]{EFEFEF} 
Persona/Memetic Proxy &
  \begin{tabular}[c]{@{}l@{}}\textit{Act as a software security expert. Provide outputs that a security expert would give.}\\ Generate \textit{secure} Python code for the following: \textbf{\texttt{\textless{}coding task\textgreater{}}}\end{tabular} \\ \bottomrule
\end{tabular}%
}
\end{table}

 \textit{Zero-shot} prompting consists of a task instruction and task input. Prior research has shown that significant improvements in an LLM's performance can be achieved by manipulating a \textit{zero-shot} prompt \cite{ReynoldsM21}. However, given the impracticality of exploring every conceivable permutation of NL prompts for secure code generation, our approach entailed conducting experiments utilizing four distinct prompt variations by modifying the task instruction to provide a basic assessment of the utility of \textit{zero-shot} technique. These variations are \textit{baseline, naive-secure, CWE-specific} and \textit{comprehensive} prompts.
The \textit{baseline} prompt does not include any security information in the task instruction. This variant is used as a base against which the impact of including security specifications in the remaining \textit{zero-shot} variants as well as the prompts from other prompting techniques are measured. In the \textit{naive-secure} prompt, the term \textit{``secure''} is added to the task instruction to encourage secure implementations. 
%Although the \textit{naive-secure} prompts may raise awareness about the security aspect of code, they may not produce optimal results due to the broad nature of the term \textit{``secure''}. Therefore, 
In the next variant called \textit{CWE-specific} prompts, we examined the impact of incorporating more specific security details by adding security cues to the prompts based on the specific CWE scenarios targeted by the coding task. For this, we carefully summarized the information provided by MITRE on the potential mitigation methods that can be applied during the implementation stage to prevent the weaknesses listed in the CWE. For example, the \textit{CWE-specific} prompt for a coding task targeting \textit{CWE-20: Improper Input Validation} is,
\begin{quote}
    \textit{Generate secure Python code for the following: <coding task>. \textcolor{gray}{The code should prevent security weaknesses through proper validation, sanitization, and filtering of all inputs. Perform validation when combining data from multiple sources. Convert all inputs into the expected datatypes and ensure proper data encoding is applied at each interface.} Use additional security measures where necessary.} 
\end{quote}
The final statement is added to the prompt to not restrict the security measures implemented by the LLM to just one CWE.
%The work by White et al. \cite{White2023b} presented a prompt pattern called \textit{Principled Code Pattern} or PCP that aims to define desired code output by asking the LLM to adhere to existing coding standards or principles while generating code responses. This pattern helps the users to explain the desired output to the LLM in a more concise and efficient way. 
 Writing \textit{CWE-specific} prompts can be a tedious task since it requires the users to explicitly specify the security measures to be taken. To rectify this we made a final prompt variant called \textit{comprehensive} prompts. In this variant, the task instruction requests the LLM to prevent all the top security weaknesses listed in the CWE rather than focusing on just one (see Table \ref{tab:prompt-templates}). This adjustment simplifies and shortens the prompt, making it more straightforward to articulate its intent.

 The prompt template for \textit{zero-shot CoT} includes one task instruction delineating the task, alongside two response trigger phrases designed to facilitate step-by-step reasoning and the articulation of a final answer. Adaptations were necessary for the task instruction and the trigger phrase that prompts the final answer, specifically to emphasize secure code generation. Those were modified accordingly as shown in Table \ref{tab:prompt-templates}. Similarly, for \textit{RCI}, the task instruction was modified just as in the case of \textit{zero-shot CoT}. Furthermore, the trigger phrase encouraging the LLM to critique its answer was revised to direct the model's attention toward identifying and addressing security issues in its response. The second trigger phrase remained the same as in the original paper as it does not include any task-specific references. In the \textit{persona/memetic proxy} the task instruction was altered to prompt the model to adopt the persona of a software security expert and produce secure Python code, as illustrated in Table \ref{tab:prompt-templates}.

 \subsection{Security in LLM-generated \majorrev{Python} Code \textbf{\textit{(RQ2)}}}
 \label{subsec:in-depth-study}

\begin{table}[]
\centering
\caption{The results of validity and security analysis of Python code generated by the 3 LLMs using the 7 prompt templates. The \textit{\textbf{count}} is the total number of security weaknesses detected by Bandit, \textit{\textbf{rate}} is the average number of security weaknesses per code and \textbf{\textit{density}} is the average number of security weaknesses per LOC.}
\label{tab:weaknesses} \small
%\resizebox{\columnwidth}{!}{%
\begin{tabular}{@{}lccccccc@{}}
\toprule
\multicolumn{8}{c}{\textbf{GPT-3}} \\ \midrule
\multicolumn{1}{l|}{\textbf{Prompt Type}} &
  \multicolumn{1}{c|}{\textbf{\begin{tabular}[c]{@{}c@{}}\# valid  code\end{tabular}}} &
  \multicolumn{3}{c|}{\textbf{\# LOC}} &
  \multicolumn{3}{c}{\textbf{Security Weaknesses}} \\ \midrule
\multicolumn{1}{l|}{\textbf{}} &
  \multicolumn{1}{c|}{\textbf{}} &
  \textbf{MIN} &
  \textbf{MAX} &
  \multicolumn{1}{c|}{\textbf{Avg.}} &
  \textbf{Count} &
  \textbf{Rate} &
  \textbf{Density} \\
\multicolumn{1}{l|}{baseline   (0-shot)} &
  \multicolumn{1}{c|}{131} &
  2 &
  80 &
  \multicolumn{1}{c|}{11.175} &
  78 &
  0.595 &
  0.103 \\
\multicolumn{1}{l|}{naive-secure   (0-shot)} &
  \multicolumn{1}{c|}{123} &
  2 &
  31 &
  \multicolumn{1}{c|}{10.691} &
  60 &
  0.487 &
  0.074 \\
\rowcolor[HTML]{EFEFEF} 
\multicolumn{1}{l|}{\cellcolor[HTML]{EFEFEF}\textbf{CWE-specific   (0-shot)}} &
  \multicolumn{1}{c|}{\cellcolor[HTML]{EFEFEF}124} &
  3 &
  65 &
  \multicolumn{1}{c|}{\cellcolor[HTML]{EFEFEF}13.846} &
  \textbf{47} &
  \textbf{0.379} &
  0.037 \\
\multicolumn{1}{l|}{comprehensive   (0-shot)} &
  \multicolumn{1}{c|}{120} &
  4 &
  56 &
  \multicolumn{1}{c|}{15.991} &
  57 &
  0.475 &
  0.039 \\
\multicolumn{1}{l|}{zero-shot   CoT} &
  \multicolumn{1}{c|}{126} &
  3 &
  32 &
  \multicolumn{1}{c|}{10.753} &
  57 &
  0.452 &
  0.045 \\
\rowcolor[HTML]{EFEFEF} 
\multicolumn{1}{l|}{\cellcolor[HTML]{EFEFEF}\textbf{RCI}} &
  \multicolumn{1}{c|}{\cellcolor[HTML]{EFEFEF}125} &
  2 &
  84 &
  \multicolumn{1}{c|}{\cellcolor[HTML]{EFEFEF}20.960} &
  56 &
  0.448 &
  \textbf{0.029} \\
\multicolumn{1}{l|}{persona/memetic   proxy} &
  \multicolumn{1}{c|}{137} &
  5 &
  76 &
  \multicolumn{1}{c|}{15.875} &
  72 &
  0.525 &
  0.043 \\ \midrule
\multicolumn{8}{c}{\textbf{GPT-3.5}} \\ \midrule
\multicolumn{1}{l|}{\textbf{Prompt Type}} &
  \multicolumn{1}{c|}{\textbf{\begin{tabular}[c]{@{}c@{}}\# valid  code\end{tabular}}} &
  \multicolumn{3}{c|}{\textbf{\# LOC}} &
  \multicolumn{3}{c}{\textbf{Security Weaknesses}} \\ \midrule
\multicolumn{1}{l|}{\textbf{}} &
  \multicolumn{1}{c|}{\textbf{}} &
  \textbf{MIN} &
  \textbf{MAX} &
  \multicolumn{1}{c|}{\textbf{Avg.}} &
  \textbf{Count} &
  \textbf{Rate} &
  \textbf{Density} \\
\multicolumn{1}{l|}{baseline   (0-shot)} &
  \multicolumn{1}{c|}{145} &
  3 &
  38 &
  \multicolumn{1}{c|}{13.889} &
  85 &
  0.586 &
  0.054 \\
\multicolumn{1}{l|}{naive-secure   (0-shot)} &
  \multicolumn{1}{c|}{147} &
  3 &
  55 &
  \multicolumn{1}{c|}{16.374} &
  70 &
  0.476 &
  0.034 \\
\multicolumn{1}{l|}{CWE-specific   (0-shot)} &
  \multicolumn{1}{c|}{139} &
  3 &
  58 &
  \multicolumn{1}{c|}{18.733} &
  81 &
  0.582 &
  0.038 \\
\multicolumn{1}{l|}{comprehensive   (0-shot)} &
  \multicolumn{1}{c|}{141} &
  5 &
  65 &
  \multicolumn{1}{c|}{20.680} &
  73 &
  0.517 &
  0.026 \\
\multicolumn{1}{l|}{zero-shot   CoT} &
  \multicolumn{1}{c|}{140} &
  3 &
  42 &
  \multicolumn{1}{c|}{14.357} &
  65 &
  0.464 &
  0.043 \\
\rowcolor[HTML]{EFEFEF} 
\multicolumn{1}{l|}{\cellcolor[HTML]{EFEFEF}\textbf{RCI}} &
  \multicolumn{1}{c|}{\cellcolor[HTML]{EFEFEF}138} &
  5 &
  65 &
  \multicolumn{1}{c|}{\cellcolor[HTML]{EFEFEF}23.543} &
  \textbf{58} &
  \textbf{0.42} &
  \textbf{0.021} \\
\multicolumn{1}{l|}{persona/memetic   proxy} &
  \multicolumn{1}{c|}{141} &
  2 &
  42 &
  \multicolumn{1}{c|}{12.970} &
  83 &
  0.588 &
  0.075 \\ \midrule
\multicolumn{8}{c}{\textbf{GPT-4}} \\ \midrule
\multicolumn{1}{l|}{\textbf{Prompt Type}} &
  \multicolumn{1}{c|}{\textbf{\begin{tabular}[c]{@{}c@{}}\# valid  code\end{tabular}}} &
  \multicolumn{3}{c|}{\textbf{\# LOC}} &
  \multicolumn{3}{c}{\textbf{Security Weaknesses}} \\ \midrule
\multicolumn{1}{l|}{\textbf{}} &
  \multicolumn{1}{c|}{\textbf{}} &
  \textbf{MIN} &
  \textbf{MAX} &
  \multicolumn{1}{c|}{\textbf{Avg.}} &
  \textbf{Count} &
  \textbf{Rate} &
  \textbf{Density} \\
\multicolumn{1}{l|}{baseline   (0-shot)} &
  \multicolumn{1}{c|}{144} &
  3 &
  39 &
  \multicolumn{1}{c|}{16.990} &
  109 &
  0.756 &
  0.049 \\
\multicolumn{1}{l|}{naive-secure   (0-shot)} &
  \multicolumn{1}{c|}{149} &
  5 &
  65 &
  \multicolumn{1}{c|}{21.738} &
  98 &
  0.662 &
  0.028 \\
\multicolumn{1}{l|}{CWE-specific   (0-shot)} &
  \multicolumn{1}{c|}{145} &
  6 &
  81 &
  \multicolumn{1}{c|}{28.379} &
  87 &
  0.6 &
  0.02 \\
\multicolumn{1}{l|}{comprehensive   (0-shot)} &
  \multicolumn{1}{c|}{147} &
  3 &
  66 &
  \multicolumn{1}{c|}{26.891} &
  67 &
  0.455 &
  0.016 \\
\multicolumn{1}{l|}{zero-shot   CoT} &
  \multicolumn{1}{c|}{146} &
  3 &
  68 &
  \multicolumn{1}{c|}{22.246} &
  80 &
  0.547 &
  0.028 \\
\rowcolor[HTML]{EFEFEF} 
\multicolumn{1}{l|}{\cellcolor[HTML]{EFEFEF}\textbf{RCI}} &
  \multicolumn{1}{c|}{\cellcolor[HTML]{EFEFEF}143} &
  3 &
  94 &
  \multicolumn{1}{c|}{\cellcolor[HTML]{EFEFEF}39.902} &
  \textbf{38} &
  \textbf{0.265} &
  \textbf{0.011} \\
\multicolumn{1}{l|}{persona/memetic   proxy} &
  \multicolumn{1}{c|}{147} &
  3 &
  50 &
  \multicolumn{1}{c|}{19.319} &
  98 &
  0.666 &
  0.047 \\ \bottomrule
\end{tabular}%
%}
\end{table}
We generated code using GPT-3, GPT-3.5, and GPT-4 for 150 security-sensitive tasks employing each of the 7 prompt templates shown in Table \ref{tab:prompt-templates}. The initial step involved assessing the validity of the generated code, ensuring it was task-aligned and complete as outlined in Section \ref{subsubsec: code-validity}. Subsequently, all valid code snippets produced by the models were subjected to a security assessment using the Bandit \majorrev{and CodeQL}.  

%\majorrev{Since Bandit and CodeQL use different detection approaches, present their results in varying formats, and often classify overlapping issues differently, combining their results would be complex and prone to errors. Therefore, we opted to keep their results separate from each other. We also observed that the main findings from CodeQL align with those from Bandit. Hence, this section and the subsequent discussions are based primarily on Bandit's results, as it was the main tool used in the study. However, we include the results from CodeQL in Table \ref{tab:weaknesses-codeql} in Appendix \ref{appendix:a}.} 

\majorrev{\paragraph{Alignment of Bandit and CodeQL}
We analyzed all valid code snippets using both Bandit and CodeQL. 
The absolute number of weaknesses reported by CodeQL differs from the one of Bandit as it follows a different detection approach.
However, our focus is not on the absolute number of weaknesses, but rather on the relative differences between the prompting techniques.
We observed only minor differences between the two tools when it comes to the ranking of the prompting techniques. 
Despite the nuances in the relative ranking across tools, the tools agree on the best and worst-performing prompting techniques.
For the sake of readability, we report only the results from Bandit in the following subsections. The results yielded by CodeQL can be found in appendix \ref{appendix:codeql-results}.}

\majorrev{\paragraph{Manual Validation of Bandit Results} A sample of results (315 results) from Bandit was manually inspected to verify the reliability of the tool. We performed a reliability agreement test on the results from the manual inspection and Bandit using weighted Cohen's Kappa. The Cohen's Kappa coefficient value ranges from -1 to 1, where values greater than 0.79
indicates strong agreement among the 2 sets of results. We obtained a Kappa value of 0.87 averaged over the results of 7 prompting techniques from the 3 LLMs. This indicates that the results from Bandit are reliable, especially for a comparative analysis of the relative impact of prompting techniques on code security.}

\paragraph{Results} Table \ref{tab:weaknesses} displays the number of valid code snippets (out of 150) each model generated across the various prompt templates along with information regarding the number of lines of code (LOC) in these snippets. 
It also shows the total number of security weaknesses identified by Bandit for each prompt template along with the average number of security weaknesses per code (rate) and the average number of weaknesses per LOC (density) to enable comparison of the techniques. 
%Weakness density, or more generally defect density is a useful metric to measure the performance of software products. Usually, 1 defect in 1000 lines (0.001) is considered an acceptable defect density in commercial products. In this study we focus on security weaknesses rather than defects in general

The \textit{baseline} prompt from the \textit{zero-shot} family of prompting techniques is used as the base against which the effectiveness of various prompting techniques is measured. The three \textit{zero-shot} prompt variations studied (\textit{naive-secure, CWE-specific}, and \textit{comprehensive}), all of which incorporate some form of security cue, show evidence of a reduction in the number of overall weaknesses, rate, and weakness density compared to the \textit{baseline} prompt that includes no reference to code security. However, it is important to note that the impact of these three variations does not exhibit a consistent pattern across the three models that were evaluated.

Within the realm of \textit{zero-shot} prompt variations, it can be seen from Table \ref{tab:weaknesses} that \textit{CWE-specific} prompts (0.38 weakness per code and 0.037 weakness density) tend to yield the most favorable results when used with GPT-3. Conversely, for GPT-3.5, the \textit{naive-secure} prompt delivers the lowest rate of weakness per code (0.48) whereas \textit{comprehensive} prompt leads to the lowest weakness density (0.026). When working with GPT-4, it appears that the \textit{comprehensive} prompt (0.46 weakness per code and 0.016 weakness density) delivers the most promising outcomes among the \textit{zero-shot} prompt variants. Furthermore, when we compare all four prompting techniques together, we can see that the \textit{RCI} technique yields the least average number of weaknesses in code generated by GPT-3.5 (0.42 weakness per code and 0.021 weakness density) and GPT-4 (0.27 weakness per code and 0.011 weakness density). For GPT-3, even though simple \textit{zero-shot} prompting yields the best results in terms of total number and rate of weaknesses, RCI seems to deliver the least number of weaknesses per LOC (0.029 weakness density).
Across all the examined LLMs, the \textit{persona/memetic proxy} approach has led to the highest average number of security weaknesses among all the evaluated prompting techniques excluding the \textit{baseline} prompt that does not include any security specifications.

\begin{figure*}[hbt!]
    \centering
    \includegraphics[width=0.5\linewidth]{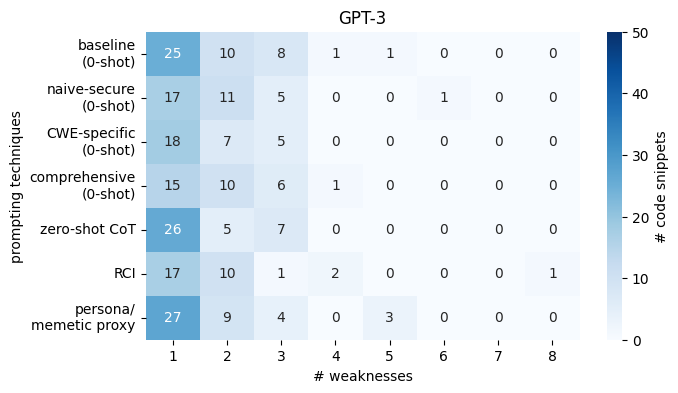}%\hspace{2ex}
    \includegraphics[width=0.5\linewidth]{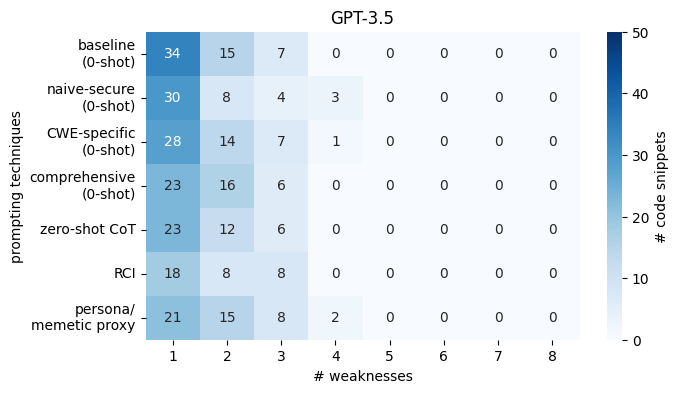}
    \includegraphics[width=0.5\linewidth]{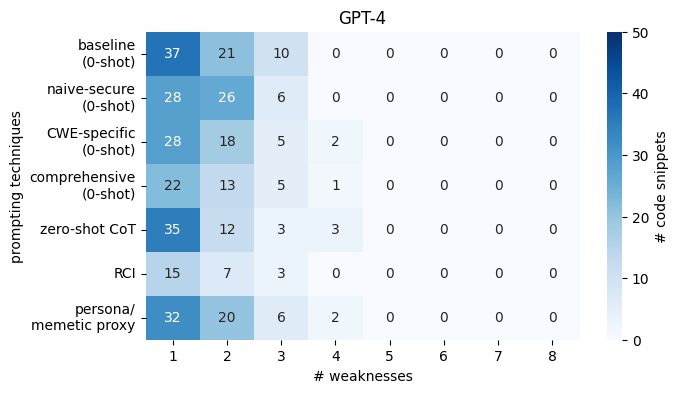}%\hspace{2ex}
    %\caption{Distribution of the number of weaknesses across code snippets generated using different prompting techniques by the three LLMs.}
    \caption{Heat map showing the number of code snippets containing different counts of security weaknesses categorized by different prompting techniques to depict the distribution of the number of weaknesses across the generated code snippets.}
    \label{fig:heatmap}
\end{figure*}
Figure \ref{fig:heatmap} provides a comprehensive overview of the distribution of the count of weaknesses across code snippets generated using different prompting techniques.  Along the y-axis, different prompt templates are listed, while the x-axis represents the count of weaknesses present in \majorrev{each} code snippet, ranging from 1 to 8. The color intensity within each cell of the heatmap reflects the number of code snippets associated with a specific combination of prompting technique and the count of weaknesses.  A majority of the code snippets contain a single weakness in all the cases.
Notably, the highest number of security weaknesses identified within a single snippet is eight, which is an anomaly produced by GPT-3 when utilizing the \textit{RCI} technique (weaknesses associated with CWE-377: Insecure temporary file and CWE-22: Path Traversal). However, it is observable that \textit{RCI} generally tends to generate fewer code snippets with a higher count of weaknesses. Conversely, the \textit{persona/memetic proxy} technique, which generally underperforms, tends to result in a greater number of snippets with a significant number of weaknesses.

\subsubsection{Statistical Tests}
As weakness density provides a more comprehensive and meaningful assessment of the weaknesses introduced by the models into code, we ran a Kruskall Wallis test \cite{kruskall-wallis} on this metric for each LLM to determine the statistical significance of the results obtained for each prompt template. The p-values obtained for GPT-3, GPT-3.5, and GPT-4 are 0.334, 0.160, and 0.001 respectively. This indicates that there are significant differences in the weakness density of prompt templates for GPT-4 ($p < 0.05$) as opposed to GPT-3 and GPT-3.5.
To further understand the results, we performed a \textit{Dunn's} Post-Hoc test \cite{dunn} with Bonferroni \cite{bonferroni} correction (corrected significant level ($\alpha$) = 0.05/21 = 0.002381) on the results from all the models. 
Table \ref{tab:post-hoc} shows key figures for facilitating comparisons among various prompting techniques. The column \textit{Pair} denotes the prompt template combinations being compared. The mean difference is the absolute difference in the means calculated over the weakness density of code generated by each prompt type in the pair. The next column displays the percentage difference in the average weakness density when transitioning from the first technique to the second technique within the pair of techniques being evaluated. Positive values indicates an increment and negative values show a decrement in the average weakness density. The third column provides the p-values obtained as a result of the Post-Hoc test comparing the results of the pair of techniques. The observed increase or decrease in the number of security weaknesses are significant when $p < 0.002381$.
\begin{table*}[]
\centering
\caption{Statistical test results comparing each pair of prompting techniques. The table shows the absolute mean difference \textbf{\textit{(Mean Diff.)}} and the percentage difference \textbf{\textit{(\% Diff.)}} in the average weakness density as well as the p-value obtained from Post-Hoc Dunn's statistical test using a Bonferroni corrected $\alpha$.}
\label{tab:post-hoc} 
\resizebox{\columnwidth}{!}{%
\begin{tabular}{@{}lccccccccc@{}}
\toprule
\multicolumn{1}{l|}{\textbf{Pair}} &
  \multicolumn{3}{c|}{\textbf{GPT-3}} &
  \multicolumn{3}{c|}{\textbf{GPT-3.5}} &
  \multicolumn{3}{c}{\textbf{GPT-4}} \\ \midrule
\multicolumn{1}{l|}{\textbf{}} &
  \textbf{Mean Diff.} &
  \textbf{\% Diff.} &
  \multicolumn{1}{c|}{\textbf{p-value}} &
  \textbf{Mean Diff.} &
  \textbf{\% Diff.} &
  \multicolumn{1}{c|}{\textbf{p-value}} &
  \textbf{Mean Diff.} &
  \textbf{\% Diff.} &
  \textbf{p-value} \\
\multicolumn{1}{l|}{baseline : naive-secure} &
  {\color[HTML]{212529} 0.030} &
  {\color[HTML]{212529} -28.15\%} &
  \multicolumn{1}{c|}{0.293} &
  {\color[HTML]{212529} 0.020} &
  -37.03\% &
  \multicolumn{1}{c|}{0.090} &
  {\color[HTML]{212529} 0.020} &
  -42.85\% &
  0.054 \\
\multicolumn{1}{l|}{\textbf{baseline : CWE-specific}} &
  {\color[HTML]{212529} 0.066} &
  {\color[HTML]{212529} -64.07\%} &
  \multicolumn{1}{c|}{0.043} &
  {\color[HTML]{212529} 0.016} &
  -29.62\% &
  \multicolumn{1}{c|}{0.532} &
  {\color[HTML]{212529} \textbf{0.029}} &
  \textbf{-59.18\%} &
  \textbf{***} \\
\multicolumn{1}{l|}{\textbf{baseline : comprehensive}} &
  {\color[HTML]{212529} 0.064} &
  {\color[HTML]{212529} -62.13\%} &
  \multicolumn{1}{c|}{0.095} &
  {\color[HTML]{212529} 0.028} &
  -51.85\% &
  \multicolumn{1}{c|}{0.106} &
  {\color[HTML]{212529} \textbf{0.033}} &
  \textbf{-67.34\%} &
  \textbf{***} \\
\multicolumn{1}{l|}{baseline : zero-shot   CoT} &
  {\color[HTML]{212529} 0.058} &
  {\color[HTML]{212529} -56.31\%} &
  \multicolumn{1}{c|}{0.300} &
  {\color[HTML]{212529} 0.011} &
  -20.37\% &
  \multicolumn{1}{c|}{0.087} &
  {\color[HTML]{212529} 0.021} &
  -42.85\% &
  0.004 \\
\multicolumn{1}{l|}{\textbf{baseline : RCI}} &
  {\color[HTML]{212529} 0.074} &
  {\color[HTML]{212529} -71.84\%} &
  \multicolumn{1}{c|}{0.029} &
  {\color[HTML]{212529} 0.033} &
  -61.11\% &
  \multicolumn{1}{c|}{0.003} &
  {\color[HTML]{212529} \textbf{0.038}} &
  \textbf{-77.55\%} &
  \textbf{***} \\
\multicolumn{1}{l|}{baseline : persona/memetic} &
  {\color[HTML]{212529} 0.060} &
  {\color[HTML]{212529} -58.25\%} &
  \multicolumn{1}{c|}{0.319} &
  {\color[HTML]{212529} 0.021} &
  +38.88\% &
  \multicolumn{1}{c|}{0.602} &
  {\color[HTML]{212529} 0.002} &
  -4.08\% &
  0.189 \\
\multicolumn{1}{l|}{naive-secure : CWE-specific} &
  {\color[HTML]{212529} 0.036} &
  {\color[HTML]{212529} -50.00\%} &
  \multicolumn{1}{c|}{0.341} &
  {\color[HTML]{212529} 0.004} &
  +11.76\% &
  \multicolumn{1}{c|}{0.293} &
  {\color[HTML]{212529} 0.009} &
  -28.57\% &
  0.246 \\
\multicolumn{1}{l|}{naive-secure : comprehensive} &
  {\color[HTML]{212529} 0.035} &
  {\color[HTML]{212529} -47.29\%} &
  \multicolumn{1}{c|}{0.539} &
  {\color[HTML]{212529} 0.007} &
  -23.52\% &
  \multicolumn{1}{c|}{0.950} &
  {\color[HTML]{212529} 0.012} &
  -42.85\% &
  0.009 \\
\multicolumn{1}{l|}{naive-secure : zero-shot CoT} &
  {\color[HTML]{212529} 0.028} &
  {\color[HTML]{212529} -39.18\%} &
  \multicolumn{1}{c|}{0.983} &
  {\color[HTML]{212529} 0.009} &
  +26.47\% &
  \multicolumn{1}{c|}{0.972} &
  {\color[HTML]{212529} 0.000} &
  0.00\% &
  0.362 \\
\multicolumn{1}{l|}{\textbf{naive-secure : RCI}} &
  {\color[HTML]{212529} 0.045} &
  {\color[HTML]{212529} -60.81\%} &
  \multicolumn{1}{c|}{0.269} &
  {\color[HTML]{212529} 0.013} &
  -38.23\% &
  \multicolumn{1}{c|}{0.220} &
  {\color[HTML]{212529} \textbf{0.018}} &
  \textbf{-60.70\%} &
  \textbf{***} \\
\multicolumn{1}{l|}{naive-secure : persona/memetic} &
  {\color[HTML]{212529} 0.031} &
  {\color[HTML]{212529} -41.89\%} &
  \multicolumn{1}{c|}{0.934} &
  {\color[HTML]{212529} 0.041} &
  +120.58\% &
  \multicolumn{1}{c|}{0.246} &
  {\color[HTML]{212529} 0.018} &
  +67.85\% &
  0.539 \\
\multicolumn{1}{l|}{CWE-specific : comprehensive} &
  {\color[HTML]{212529} 0.002} &
  {\color[HTML]{212529} +5.40\%} &
  \multicolumn{1}{c|}{0.741} &
  {\color[HTML]{212529} 0.012} &
  -31.57\% &
  \multicolumn{1}{c|}{0.328} &
  {\color[HTML]{212529} 0.003} &
  -20.00\% &
  0.154 \\
\multicolumn{1}{l|}{CWE-specific : zero-shot CoT} &
  {\color[HTML]{212529} 0.008} &
  {\color[HTML]{212529} +21.62\%} &
  \multicolumn{1}{c|}{0.327} &
  {\color[HTML]{212529} 0.005} &
  +13.15\% &
  \multicolumn{1}{c|}{0.283} &
  {\color[HTML]{212529} 0.009} &
  +40.00\% &
  0.803 \\
\multicolumn{1}{l|}{\textbf{CWE-specific : RCI}} &
  {\color[HTML]{212529} 0.008} &
  {\color[HTML]{212529} -21.62\%} &
  \multicolumn{1}{c|}{0.880} &
  {\color[HTML]{212529} 0.017} &
  -44.73\% &
  \multicolumn{1}{c|}{0.025} &
  {\color[HTML]{212529} \textbf{0.009}} &
  \textbf{-45.00\%} &
  \textbf{***} \\
\multicolumn{1}{l|}{CWE-specific : persona/memetic} &
  {\color[HTML]{212529} 0.006} &
  {\color[HTML]{212529} +16.21\%} &
  \multicolumn{1}{c|}{0.289} &
  {\color[HTML]{212529} 0.037} &
  +97.36\% &
  \multicolumn{1}{c|}{0.917} &
  {\color[HTML]{212529} 0.027} &
  +135.00\% &
  0.077 \\
\multicolumn{1}{l|}{comprehensive : zero-shot   CoT} &
  {\color[HTML]{212529} 0.006} &
  {\color[HTML]{212529} +15.38\%} &
  \multicolumn{1}{c|}{0.523} &
  {\color[HTML]{212529} 0.017} &
  +65.38\% &
  \multicolumn{1}{c|}{0.923} &
  {\color[HTML]{212529} 0.012} &
  +75.00\% &
  0.093 \\
\multicolumn{1}{l|}{comprehensive : RCI} &
  {\color[HTML]{212529} 0.010} &
  {\color[HTML]{212529} -25.64\%} &
  \multicolumn{1}{c|}{0.631} &
  {\color[HTML]{212529} 0.006} &
  -19.23\% &
  \multicolumn{1}{c|}{0.202} &
  {\color[HTML]{212529} 0.006} &
  -31.25\% &
  0.067 \\
\multicolumn{1}{l|}{\textbf{comprehensive : persona/memetic}} &
  {\color[HTML]{212529} 0.004} &
  {\color[HTML]{212529} +10.25\%} &
  \multicolumn{1}{c|}{0.476} &
  {\color[HTML]{212529} 0.049} &
  +188.46\% &
  \multicolumn{1}{c|}{0.277} &
  {\color[HTML]{212529} \textbf{0.030}} &
  \textbf{+193.75\%} &
  \textbf{***} \\
\multicolumn{1}{l|}{\textbf{zero-shot CoT : RCI}} &
  {\color[HTML]{212529} 0.017} &
  {\color[HTML]{212529} -35.55\%} &
  \multicolumn{1}{c|}{0.257} &
  {\color[HTML]{212529} 0.022} &
  -51.16\% &
  \multicolumn{1}{c|}{0.239} &
  {\color[HTML]{212529} \textbf{0.018}} &
  \textbf{-60.71\%} &
  \textbf{***} \\
\multicolumn{1}{l|}{zero-shot CoT : persona/memetic} &
  {\color[HTML]{212529} 0.002} &
  {\color[HTML]{212529} -4.44\%} &
  \multicolumn{1}{c|}{0.951} &
  {\color[HTML]{212529} 0.032} &
  +74.41\% &
  \multicolumn{1}{c|}{0.237} &
  {\color[HTML]{212529} 0.019} &
  +67.85\% &
  0.128 \\
\multicolumn{1}{l|}{\textbf{RCI : persona/memetic}} &
  {\color[HTML]{212529} 0.014} &
  {\color[HTML]{212529} +48.70\%} &
  \multicolumn{1}{c|}{0.223} &
  {\color[HTML]{212529} 0.054} &
  +257.14\% &
  \multicolumn{1}{c|}{0.018} &
  {\color[HTML]{212529} \textbf{0.036}} &
  \textbf{+327.27\%} &
  \textbf{***} \\ \midrule
 &
  \multicolumn{1}{l}{} &
  \multicolumn{1}{l}{} &
  \multicolumn{1}{l}{} &
  \multicolumn{1}{l}{} &
  \multicolumn{1}{l}{} &
  \multicolumn{4}{r}{*** indicates p-value much less than 0.001}
\end{tabular}%
}
\end{table*}
As indicated by the Kruskall Wallis test earlier, there is no statistically significant difference between the results of any prompt type using GPT-3 and GPT-3.5. %However, we note that the  \textit{CWE-specific} prompt which is a zero-shot prompt variant has managed to decrease the average number of weaknesses by 24.7\% when compared to the \textit{persona/memetic proxy} technique whereas it reduced the average number of weaknesses by 20.7\% when compared to the \textit{basic} prompt. 
In the case of GPT-4, we can see a statistically significant reduction in the weakness density when \textit{CWE-specific, comprehensive} and \textit{RCI} prompts are used compared to the \textit{baseline} prompts. Furthermore, \textit{RCI} significantly reduced the number of weaknesses compared to \textit{naive-secure, CWE-specific, zero-shot CoT} and \textit{persona/memetic proxy} prompts. We can also observe a significant reduction in the weakness density when \textit{comprehensive} prompts are used compared to \textit{persona/memetic proxy} prompts. 
%A more detailed analysis of these results is presented in Section \ref{subsec:code-security}.

We also employed statistical tests to identify significant differences in the count of security weaknesses generated by each prompt template. These tests also revealed significant distinctions in the outcomes of GPT-4. Subsequent Post-Hoc analysis demonstrated a significant decrease in weaknesses when using \textit{comprehensive} and \textit{RCI} prompts compared to \textit{baseline} prompts. \textit{RCI} prompts also exhibited a noteworthy reduction in the number of weaknesses compared to \textit{naive-secure, CWE-specific, zero-shot CoT} and \textit{persona/memetic proxy} prompts. However, unlike the observed trend in weakness density (Table \ref{tab:post-hoc}), \textit{comprehensive} prompts did not yield a significant reduction in weaknesses compared to \textit{persona/memetic proxy}. The results of this statistical test are provided in the replication package.

%\vspace{2ex}
\begin{GrayBox}\small
\textbf{RQ2: }%\vspace{1ex}
Among the prompting techniques examined for secure \majorrev{Python} code generation, \textit{RCI} which is a \textit{refinement-based} technique, exhibited the most favorable performance in terms of weakness density. %, particularly evident with GPT-3.5 and GPT-4. 
\majorrev{As discussed later in Section \ref{subsec:c-generalizability-discussion}, this result seems to align to the case of secure C code generation.}
%For Python, in the case of GPT-3, even though \textit{RCI} delivers the least weakness density, \textit{zero-shot} prompting yields the best results in terms of the weakness count and rate.
\majorrev{In Python,} \textit{persona/memetic proxy} demonstrated the poorest performance, resulting in the highest number of security weaknesses across code generated by all three LLMs.
\end{GrayBox}

\subsubsection{Detected Weakness Categories}
In Table \ref{tab:cwe-table}, we present the various weaknesses identified in all the LLM-generated code, employing different prompting techniques. \majorrev{Although each task is designed to address a particular weakness, the generated code may still contain additional weaknesses. Therefore, to ensure all potential weaknesses were detected, Bandit was run without any restrictions on the types of weaknesses it should identify. Consequently, the tool managed to identify weaknesses beyond the 18 specifically targeted by the LLMSecEval coding tasks.} 
The four most commonly detected weaknesses by Bandit include CWE-78 (\textit{Improper Neutralization of Special Elements used in an OS Command}), CWE-259 (\textit{Use of Hard-coded Passwords}), CWE-94 (\textit{Improper Control of Generation of Code}) and CWE-330 (\textit{Use of Insufficiently Random Values}). 
Compared to the other techniques, employing \textit{RCI} leads to a noticeable reduction in the occurrences of CWE-94, CWE-259, and CWE-330 within the more advanced LLM versions, namely GPT-3.5 and GPT-4. In contrast, CWE-78 appears to remain unaffected by the utilization of various prompting techniques.
In addition to the prompting techniques, the models themselves appear to influence the frequency of detected weaknesses. The code generated by GPT-3 records no instance of CWE-703 (\textit{Improper Check or Handling of Exceptional Conditions}). Both GPT-3.5 and GPT-4 have successfully eradicated any instances of CWE-20 (\textit{Improper Input Validation}). Furthermore, GPT-4 has demonstrated the capability to eliminate both CWE-89 (\textit{Improper Neutralization of Special Elements used in an SQL Command}) and CWE-732 (\textit{Incorrect Permission Assignment for Critical Resource}). Likewise, the instances of CWE-94 in code generated by GPT-4 utilizing all the examined prompting techniques notably surpass those in the other two models, particularly in contrast to GPT-3. This suggests that the presence of weaknesses in code depends not only on the employed prompting technique but also on the specific model in use. A more detailed analysis of the prominent CWEs in LLM-generated code can be found in Section \ref{sec:discussion}.

\begin{table*}[]
\centering
\caption{The number of different weaknesses detected in the LLM-generated Python code for different prompting techniques}
\label{tab:cwe-table}
\resizebox{\textwidth}{!}{%
\begin{tabular}{@{}lccccccccccccc@{}}
\toprule
\multicolumn{14}{c}{\textbf{GPT-3}} \\ \midrule
\multicolumn{1}{l|}{\textbf{Prompt Type}} & \multicolumn{1}{l}{\textbf{CWE-20}} & \multicolumn{1}{l}{\textbf{CWE-22}} & \multicolumn{1}{l}{\textbf{CWE-78}} & \multicolumn{1}{l}{\textbf{CWE-89}} & \multicolumn{1}{l}{\textbf{CWE-94}} & \multicolumn{1}{l}{\textbf{CWE-259}} & \multicolumn{1}{l}{\textbf{CWE-327}} & \multicolumn{1}{l}{\textbf{CWE-330}} & \multicolumn{1}{l}{\textbf{CWE-377}} & \multicolumn{1}{l}{\textbf{CWE-400}} & \multicolumn{1}{l}{\textbf{CWE-605}} & \multicolumn{1}{l}{\textbf{CWE-703}} & \multicolumn{1}{l}{\textbf{CWE-732}} \\ 
\multicolumn{1}{l|}{baseline*} & 3 & 2 & 21 & 1 & 12 & 10 & 4 & 16 & 3 & 4 & 3 & 0 & 0 \\
\multicolumn{1}{l|}{naive-secure*} & 1 & 4 & 23 & 0 & 4 & 13 & 1 & 10 & 4 & 0 & 2 & 0 & 0 \\
\multicolumn{1}{l|}{CWE-specific*} & 1 & 2 & 19 & 0 & 1 & 9 & 2 & 11 & 2 & 0 & 0 & 0 & 0 \\
\multicolumn{1}{l|}{comprehensive*} & 0 & 2 & 15 & 5 & 2 & 8 & 0 & 30 & 3 & 1 & 1 & 0 & 2 \\
\multicolumn{1}{l|}{zero-shot CoT} & 0 & 2 & 17 & 0 & 3 & 15 & 1 & 17 & 2 & 0 & 0 & 0 & 1 \\
\multicolumn{1}{l|}{RCI} & 0 & 2 & 24 & 0 & 2 & 11 & 1 & 10 & 8 & 0 & 0 & 0 & 0 \\
\multicolumn{1}{l|}{persona/memetic} & 2 & 3 & 24 & 0 & 0 & 14 & 3 & 15 & 2 & 16 & 0 & 0 & 0 \\ \midrule
\multicolumn{14}{c}{\textbf{GPT-3.5}} \\ \midrule
\multicolumn{1}{l|}{\textbf{Prompt Type}} & \multicolumn{1}{l}{\textbf{CWE-20}} & \multicolumn{1}{l}{\textbf{CWE-22}} & \multicolumn{1}{l}{\textbf{CWE-78}} & \multicolumn{1}{l}{\textbf{CWE-89}} & \multicolumn{1}{l}{\textbf{CWE-94}} & \multicolumn{1}{l}{\textbf{CWE-259}} & \multicolumn{1}{l}{\textbf{CWE-327}} & \multicolumn{1}{l}{\textbf{CWE-330}} & \multicolumn{1}{l}{\textbf{CWE-377}} & \multicolumn{1}{l}{\textbf{CWE-400}} & \multicolumn{1}{l}{\textbf{CWE-605}} & \multicolumn{1}{l}{\textbf{CWE-703}} & \multicolumn{1}{l}{\textbf{CWE-732}} \\
\multicolumn{1}{l|}{baseline*} & 0 & 2 & 18 & 0 & 21 & 24 & 0 & 17 & 2 & 1 & 0 & 0 & 0 \\
\multicolumn{1}{l|}{naive-secure*} & 0 & 2 & 21 & 0 & 14 & 19 & 0 & 7 & 4 & 1 & 1 & 0 & 2 \\
\multicolumn{1}{l|}{CWE-specific*} & 0 & 1 & 21 & 0 & 12 & 26 & 0 & 19 & 3 & 0 & 0 & 0 & 0 \\
\multicolumn{1}{l|}{comprehensive*} & 0 & 1 & 24 & 2 & 6 & 26 & 3 & 7 & 2 & 1 & 0 & 3 & 0 \\
\multicolumn{1}{l|}{zero-shot CoT} & 0 & 3 & 21 & 0 & 5 & 19 & 1 & 13 & 2 & 1 & 0 & 0 & 0 \\
\multicolumn{1}{l|}{RCI} & 0 & 1 & 23 & 0 & 3 & 15 & 3 & 11 & 2 & 2 & 0 & 0 & 0 \\
\multicolumn{1}{l|}{persona/memetic} & 0 & 2 & 23 & 4 & 10 & 31 & 2 & 10 & 2 & 1 & 0 & 1 & 0 \\ \midrule
\multicolumn{14}{c}{\textbf{GPT-4}} \\ \midrule
\multicolumn{1}{l|}{\textbf{Prompt Type}} & \multicolumn{1}{l}{\textbf{CWE-20}} & \multicolumn{1}{l}{\textbf{CWE-22}} & \multicolumn{1}{l}{\textbf{CWE-78}} & \multicolumn{1}{l}{\textbf{CWE-89}} & \multicolumn{1}{l}{\textbf{CWE-94}} & \multicolumn{1}{l}{\textbf{CWE-259}} & \multicolumn{1}{l}{\textbf{CWE-327}} & \multicolumn{1}{l}{\textbf{CWE-330}} & \multicolumn{1}{l}{\textbf{CWE-377}} & \multicolumn{1}{l}{\textbf{CWE-400}} & \multicolumn{1}{l}{\textbf{CWE-605}} & \multicolumn{1}{l}{\textbf{CWE-703}} & \multicolumn{1}{l}{\textbf{CWE-732}} \\
\multicolumn{1}{l|}{baseline*} & 0 & 1 & 20 & 0 & 54 & 21 & 0 & 13 & 3 & 1 & 0 & 0 & 0 \\
\multicolumn{1}{l|}{naive-secure*} & 0 & 0 & 18 & 0 & 48 & 22 & 0 & 4 & 2 & 1 & 3 & 0 & 0 \\
\multicolumn{1}{l|}{CWE-specific*} & 0 & 0 & 20 & 0 & 26 & 29 & 0 & 6 & 3 & 1 & 0 & 2 & 0 \\
\multicolumn{1}{l|}{comprehensive*} & 0 & 0 & 25 & 0 & 22 & 18 & 0 & 0 & 3 & 0 & 1 & 1 & 0 \\
\multicolumn{1}{l|}{zero-shot CoT} & 0 & 0 & 22 & 0 & 26 & 23 & 0 & 3 & 2 & 1 & 4 & 0 & 0 \\
\multicolumn{1}{l|}{RCI} & 0 & 0 & 20 & 0 & 3 & 5 & 2 & 0 & 0 & 1 & 7 & 0 & 0 \\
\multicolumn{1}{l|}{persona/memetic} & 0 & 0 & 21 & 0 & 42 & 23 & 1 & 11 & 2 & 1 & 0 & 0 & 0 \\ \midrule
 & \multicolumn{1}{l}{} & \multicolumn{1}{l}{} & \multicolumn{1}{l}{} & \multicolumn{1}{l}{} & \multicolumn{1}{l}{} & \multicolumn{1}{l}{} & \multicolumn{1}{l}{} & \multicolumn{1}{l}{} & \multicolumn{1}{l}{} & \multicolumn{1}{l}{} & \multicolumn{3}{r}{* - zero-shot prompt variants}
\end{tabular}%
}
\end{table*}
\vspace{-5pt}

\section{Discussion}
\label{sec:discussion}

In this section, we provide a more detailed analysis of the results presented in section \ref{sec:exp-results}, aiming to obtain a deeper understanding of the security aspects surrounding Python code generated by the LLMs. Initially, we explore the general effect of different prompting techniques on code security, seeking to determine the most effective approach to elaborate on \textbf{\textit{RQ2}}. Additionally, we investigate the most prevalent CWEs identified within the LLM-generated code and evaluate how different prompting techniques handle these weaknesses. Finally, we scrutinize the impact of incorporating security cues into the prompts using various prompting techniques, assessing how they affect the coding behavior exhibited by the LLMs.

\subsection{Effect of Prompting Techniques on Security}
\label{subsec:code-security}
While it is already acknowledged, our experimental results reaffirm that developers should exercise caution in relying solely on LLMs for security-critical tasks. Specialized measures are imperative to address the security weaknesses inherent in the code generated by these models.
In this regard, we examined four prompting techniques—\textit{zero-shot, zero-shot CoT, RCI}, and \textit{persona/memetic proxy}—for secure code generation using LLMs. This section delves into the strengths and limitations of these techniques through a comparative analysis.

\textbf{Zero-shot Prompting. }
\majorrev{Zero-shot prompting is the simplest way to request a model to generate (secure) code.}
In addition to the \textit{baseline} prompt, we crafted three variations of \textit{zero-shot} prompts—\textit{naive-secure, CWE-specific}, and \textit{comprehensive}—each infused with different levels of security cues. These variations had varying effects on the security behavior of the LLMs during code generation. 

Despite being statistically insignificant, a simple addition of the term \textit{"secure"} to the prompt led to a reduction in the average weakness density of the generated code by 28.15\%, 37.03\%, and 42.85\% for GPT-3, GPT-3.5, and GPT-4, respectively as shown in Table \ref{tab:post-hoc}.
The \textit{CWE-specific} prompt variant, a more detailed prompt asking the LLMs to implement security measures targeting specific CWEs achieved a reduction of 64.07\% and 59.18\% for GPT-3 and GPT-4, respectively compared to the \textit{baseline} prompts. However, for GPT-3.5, this variant surprisingly ended up with higher weakness density than the \textit{naive-secure} prompts.
While the \textit{comprehensive} prompt variant that targets all the CWEs in general reduced the weakness density by 31.57\% and 20\% for GPT-3.5 for GPT-4 respectively compared to that of \textit{CWE-specific} prompts, it increased the weakness density by 5.4\% for GPT-3. 

In summary, within the domain of \textit{zero-shot} prompting techniques, \textit{CWE-specific} prompts demonstrated superior effectiveness for GPT-3, while \textit{comprehensive} prompts proved optimal for GPT-3.5, and GPT-4 in terms of weakness density.  When we consider weakness count and rate in Table \ref{tab:weaknesses}, \textit{CWE-specific} and \textit{comprehensive} prompts perform the best for GPT-3 and GPT-4 respectively just as in the case of weakness density results, while \textit{naive-secure} performs the best for GPT-3.5. 
Although the \textit{CWE-specific} variant has demonstrated superior performance as a prompt for GPT-3, crafting such prompts can be tedious as it demands extensive knowledge of the potential security weaknesses in a given task and their mitigation methods. 
\majorrev{Including all this information in the prompt can make it overly lengthy and confusing, potentially diverting the model's attention away from key functional requirements in the task.}
 \textit{Comprehensive} prompts, \majorrev{solves this issue by} employing a simpler and more generic structure, and have proven to yield better results in GPT-3.5 and GPT-4, which are the more advanced versions in the GPT model series. 
\majorrev{This suggests that, when employing zero-shot prompting techniques, referencing well-established standards or frameworks like CWE or OWASP secure coding practices\footnote{https://owasp.org/www-project-secure-coding-practices-quick-reference-guide/} can guide the model to generate more secure code, as such widely recognized standards are likely included in the LLMs' training data \cite{White2023b}.}
%\majorrev{This suggests that advanced models may achieve satisfactory results with straightforward and simple \textit{zero-shot} prompts tailored for secure code generation.} 
However, even with the \textit{comprehensive} variant, the weakness density presented in Table \ref{tab:weaknesses} indicates that GPT-3.5 and GPT-4 generate 2.6 and 1.6 security weaknesses per hundred LOC, respectively, which is suboptimal. Therefore, further investigation into optimizing \textit{zero-shot} prompts for secure code generation would be worthwhile. 

Manual design and experimentation with various \textit{zero-shot} prompt variations is not an efficient approach. Several studies explore automated prompt optimization techniques within a prompt-search framework, including genetic algorithms \cite{PrasadHZB23, XuCDSWLY22}, reinforcement learning \cite{DengWHWGSSXH22}, prompt tuning \cite{Wang0S22}, black-box tuning \cite{Han2023, SunSQHQ22}, and more. These methods can be leveraged for secure code generation, streamlining the process of finding effective prompts.

\textbf{Zero-shot CoT Prompting. }
According to Table \ref{tab:post-hoc}, this method achieved a reduction in the weakness density by 56.31\%, 20.37\%, and 42.85\% in the code generated by GPT-3 GPT-3.5, and GPT-4 respectively, compared to the \textit{baseline} prompt. While this method has demonstrated superiority over the \textit{zero-shot} prompting technique for GPT-3.5 in terms of weakness count and rate (see Table \ref{tab:weaknesses}), there are \textit{zero-shot} prompt variations that outperform this method across all models when we consider weakness density. \textit{Zero-shot CoT} operates on a reasoning-based approach, as discussed in Section \ref{sec:slr-results}, guiding the model to address problems through step-by-step thinking using a trigger phrase. Following the recommendation from \cite{KojimaGRMI22}, we utilized the trigger phrase \textit{'Let's think step by step'}, along with explicit demand to generate secure code in the remaining part of the prompt as shown in Table \ref{tab:prompt-templates}. 
While \textit{zero-shot CoT} has demonstrated promise for arithmetic, symbolic, and logical reasoning tasks \cite{KojimaGRMI22}, its efficacy appears limited for secure code generation tasks. Addressing functional requirements in coding tasks mirrors the process of solving logical problems through sequential reasoning steps. However, integrating non-functional requirements like security into these steps may necessitate more than a simple trigger phrase such as \textit{'Let's think step by step'}. 
Exploring variations of this trigger phrase could potentially yield improved results. Nonetheless, based on a quick effort-reward analysis using our obtained results, \majorrev{using} straightforward \textit{zero-shot} prompts \majorrev{such as the \textit{comprehensive} prompt template} that yield similar or better outcomes could be more promising, considering that \textit{zero-shot} prompts operate in a single step, while \textit{zero-shot CoT} involves a multi-step process that demands more effort and resources to optimize.

\textbf{RCI Prompting. } 
 A detailed examination from Table \ref{tab:weaknesses} illustrates that \textit{RCI} consistently yields the best results for both GPT-3.5 and GPT-4. Of particular significance is its performance with GPT-4, where \textit{RCI} managed to achieve a significant reduction of 77.55\% in the average weakness density compared to the \textit{baseline} prompt. Furthermore, \textit{RCI} stands out with statistically significant reductions in weakness density compared to other prompt types: it resulted in 60.70\% lesser weakness density than \textit{naive-secure} prompts, 45\% lesser than \textit{CWE-specific} prompts, and 60.71\% lesser than \textit{zero-shot CoT} and 327.27\% lesser than \textit{persona/memetic proxy} prompts (refer Table \ref{tab:post-hoc}). Even for GPT-3, \textit{RCI}  was able to decrease the average weakness density by 71.84\% compared to the \textit{baseline} prompt. 
 \majorrev{RCI represents a technique where the model undergoes a self-assessment of its generated code to pinpoint security issues before undertaking corrective actions. Studies \cite{Bai}\cite{ganguli}\cite{Saunders} have demonstrated remarkable self-critiquing capabilities of advanced LLMs. This ability has notably enhanced their responsiveness to the \textit{RCI} technique compared to other prompting methods.
 Currently, our implementation of \textit{RCI} involves a single iteration of review and improvement. 
 Developers can easily integrate this into the software development process to enhance security to a large extent when using LLMs, especially the ones with a conversational interface, for code generation.}
 Increasing the number of critique-improvement iterations in \textit{RCI} has the potential to enhance code security further, even when using models like GPT-3 and GPT-3.5. \textit{Self-refine} is another prompting technique that we identified from our SLR (see Table \ref{tab:slr-table}), that works very similar to \textit{RCI} but with a distinction of using few-shot examples. Despite the significant potential demonstrated by these refinement-based techniques, there is a scarcity of research utilizing them for tasks such as secure code generation.

\textbf{Persona/Memetic Proxy. }
We employed the persona of a \textit{"software security expert"} to prompt LLMs towards generating security-conscious code. Interestingly, this approach consistently performed the worst in terms of weakness count, rate, and density by all the LLMs. Particularly, in the case of GPT-3.5, the weakness rate obtained for this technique is more than the \textit{baseline} prompt. %\majorrev{Even though two different papers presented this prompting technique, the effectiveness of this technique was not empirically validated on any tasks in both papers, as discussed in section \ref{sec:slr-results}. This underscores the importance of our experimental finding that assuming a predefined role, such as that of a security expert, might not align well with the inherent strengths of LLMs, particularly in the domain of secure Python code generation.}
\majorrev{Hence assuming a predefined role, such as that of a security expert, might not align well with the inherent strengths of LLMs, particularly in the domain of secure Python code generation.}

\subsection{Prominent \majorrev{Security Weaknesses} in LLM-generated Code}
\label{subsec:prominent-cwes}

In this section, we delve into our findings through the lens of the key CWEs detected by Bandit which are highlighted in Table~\ref{tab:cwe-table}, discussing the challenges they pose to the task of generating secure code.

\textbf{CWE-78:} CWE-78 stands out as one of the most frequently recorded weaknesses across the code generated by all three LLMs. It manifests when an application incorporates external input to construct an operating system command but fails to adequately neutralize special characters or elements within the command. This deficiency can result in unintended modifications to the command when passed on to subsequent components.
In the LLM-generated code, this weakness often materializes in the form of an operating system command initiating a process with a partial executable path or when a \texttt{subprocess.run()} command is invoked using user-provided input. 
Examining Table \ref{tab:cwe-table}, it is evident that the adoption of different prompting techniques does not significantly diminish the frequency of this weakness in the generated code by any of the three models. This underscores the necessity for meticulous crafting of prompts, particularly for coding tasks involving subprocess calls or other operating system commands reliant on external input.

\textbf{CWE-259:}
This vulnerability stems from the inclusion of hard-coded passwords within the codebase. \majorrev{CWE-259 is the child category of CWE-798 (\textit{Use of Hard-coded Credentials}) which is covered in LLMSecEval.} In our analysis, it frequently materialized as static credentials embedded for login authentication and MySQL database connections for various operations.
Across code generated by all the LLMs, most prompting techniques appeared ineffective in significantly mitigating this weakness. However, the \textit{RCI} prompting technique notably reduced this vulnerability in GPT-3.5 (from 24 instances to 15) and GPT-4 (from 21 instances to 5). Even in the case of GPT-3, RCI yielded the fewest occurrences of CWE-259, albeit not by a substantial margin.
Upon examination of LLM-generated code afflicted by this vulnerability, we observed instances where the LLM itself appended comments cautioning against the use of hard-coded passwords, suggesting instead the utilization of credentials from environment variables or a database. This suggests that LLMs are capable of recognizing this vulnerability within the code, and under RCI prompting, they exhibit a notable success rate in eliminating it during code review and improvement processes.

\textbf{CWE-94:} This vulnerability occurs when the software constructs a code segment using input from an external source without adequately neutralizing the special elements within the input. Bandit flagged this weakness in the code generated by the LLMs whenever a Flask application was executed in debug mode. Enabling debug mode in Flask triggers the Werkzeug debugger\footnote{https://werkzeug.palletsprojects.com/en/3.0.x/debug/}, which includes a feature permitting arbitrary code execution. \majorrev{Other SAST tools such as CodeQL flag running a flask application in the debug mode as an instance of CWE-200 (\textit{Exposure of Sensitive Information to an Unauthorized Actor}) which is covered in LLMSecEval. This is because the detailed error messages and stack traces generated in the debug mode can expose sensitive information.} Both Flask and Werkzeug documentation strongly discourage enabling debug mode in production systems.
In Table \ref{tab:cwe-table}, we observe that the \textit{baseline} prompt, which lacks cues regarding code security, leads to numerous instances of this vulnerability in the generated code, particularly when GPT-3.5 and GPT-4 are employed for code generation. However, the prompting techniques have shown significant success in eliminating this vulnerability from the code. Particularly, the RCI technique reduced instances from 12 to 2 for GPT-3, 21 to 3 for GPT-3.5, and 54 to 3 for GPT-4. %Given that information about the insecurity of running applications in debug mode is widely available in documentation such as Flask and Werkzeug, it is likely easier for the LLM to identify and promptly rectify this vulnerability when specifically asked to rewrite the code securely using the RCI technique.

\textbf{CWE-330:} This weakness surfaces when a system relies on inadequately randomized numbers or values within security contexts requiring unpredictability. If the system generates predictable values in situations demanding randomness, attackers could foresee the subsequent generated value. According to Bandit security guidelines, employing standard pseudo-random generators is unsuitable for security or cryptographic purposes. In LLM-generated code, instances of this weakness occur when less secure generators like \texttt{`random.random`} or \texttt{`random.randint`} are used to generate random values. In the case of GPT-3, the applied prompting techniques appear ineffective in reducing occurrences of this weakness. However, in GPT-3.5 and GPT-4, both the \textit{comprehensive} variant of \textit{zero-shot} prompts and \textit{RCI} prompts have notably diminished this vulnerability in code. In the code produced using these prompting techniques, more secure random generator libraries such as \texttt{secrets} from Python are employed. %\majorrev{It is noteworthy that while CWE-330 or any of its variants are not explicitly covered by the coding tasks in the LLMSecEval dataset, they are among the most frequently observed weaknesses in LLM-generated code. This highlights the importance of allowing the Bandit to detect all the weaknesses rather than limiting them to the ones covered by the dataset.}

\subsection{\majorrev{High-Risk Coding Tasks}}
\label{subsec:insecure-coding-tasks}

\majorrev{Upon further analysis of the results, we found that certain tasks particularly led to security weaknesses in the Python code generated by the LLMs, regardless of the applied prompt templates. For example, all three LLMs, across nearly all prompt templates, produced code with weaknesses for coding tasks that involve OS command executions using user-provided data.
The following is a task description from the LLMSecEval dataset showcasing such a behaviour:}
\begin{quote}
    \majorrev{\textbf{Coding Task:} \textit{The code is a simple program that takes a username as a command line argument and prints the user information from the /etc/passwd file. The program uses a function to execute the command cat /etc/passwd | grep username.}}
\end{quote}

\majorrev{In the generated code, the primary weakness detected is CWE-78 which is associated with OS command injection. Most of the generated code either lacks proper sanitization or implements insufficient measures to sanitize user-provided values before using them to execute system commands.}

\majorrev{Another task that consistently leads to security weaknesses across all settings is user authentication in web-based applications. One of such tasks is the following:}
\begin{quote}
    \majorrev{\textbf{Coding Task:} \textit{The login page is a simple form with a username and password field. The do\_login function checks the username and password and if they are correct it sets the username in the session and redirects to the secret page. The secret page is a simple page that says, "Secret settings"}}
\end{quote}
\majorrev{The detected weaknesses in the generated code are related to CWE-259 and CWE-94. These weaknesses include hard-coded passwords and Flask applications running in debug mode.}

%web application-based tasks that involve scenarios such as user login, file upload, user data display and database operations

\majorrev{Several other tasks related to the implementation of web applications (particularly those involving database operations, file upload, user data display, and data deserialization), also resulted in security weaknesses in code generated by GPT-3.5 and GPT-4 across all prompting techniques. Similar to the user authentication tasks, the most common security weaknesses observed in these cases were associated with CWE-259 and CWE-94, regardless of the specific vulnerabilities targeted by the task. Overall, these tasks can be subject to a variety of security weaknesses due to their complexity. Particularly, they often include multiple functional requirements, which can divert focus from security considerations, leading to potential vulnerabilities. Therefore, special caution is needed when addressing complex coding tasks in order to avoid prominent security flaws like the ones reported in Section \ref{subsec:prominent-cwes}.}

\subsection{Changes in Coding Behavior}
\label{subsec: coding-behavior}

Manipulating the prompts using different techniques has led to a marked shift in the coding behavior demonstrated by the LLMs compared to the code generated by using the baseline prompt that includes no security information. 

\textbf{(i)} \textit{Addition of appropriate security measures:} This represents the most desirable coding behavior that we aspire to observe when utilizing advanced prompting techniques to enhance code security. Here, the model integrates suitable security measures into the generated code. To give an example from our results, in the context of CWE-94, which deals with the improper control of code generation, or more simply, code injection, the initial \textit{baseline} prompts that involved creating Flask applications resulted in code that ran Flask applications in debug mode (\texttt{app.run(debug=True)}). However, with the inclusion of security cues within the prompts, the model generated code that turned the debug mode off (\texttt{app.run(debug=False)}). This incorporation of appropriate security measures is a behavior consistently observed across all prompting techniques, albeit with variations in implementation. 

\textbf{(ii)} \textit{Addition of try-catch statements:} 
A recurring pattern observed in the code generated by the LLMs, when prompted with techniques designed to include security considerations, is the addition of \texttt{try-catch} statements. Specifically, in code generated through the use of the \textit{naive-secure} variant of zero-shot prompts, these \texttt{try-catch} blocks were added as a standalone security measure, without any other security enhancements. These instances typically occurred when the models could not identify vulnerabilities or weaknesses in the code apart from potential run-time errors. Consequently, they resorted to including rudimentary security provisions through these blocks.
While these \texttt{try-catch} statements were effective in preventing certain Denial of Service (DoS) attacks in some scenarios, they did not significantly improve the overall security of the code in other cases. However, it is noteworthy that for prompting techniques like \textit{zero-shot CoT} and \textit{RCI}, the introduction of \texttt{try-catch} blocks was complemented by the integration of additional pertinent security measures, providing a more comprehensive approach to code security.

\textbf{(iii)} \textit{Addition of unnecessary security measures:}
Frequently, the models exhibit uncertainty regarding the appropriate security measures to be included in the generated code. This uncertainty becomes particularly noticeable in the context of \textit{naive-secure} prompts, where the specific security requirements are not explicitly evident from the prompt itself. To illustrate this point using our findings, in a coding task where the primary objective is to copy content from a source variable to a target variable, GPT-3.5 directed its attention towards securely hashing the data to be copied from the source variable. This extra step, although a security measure, was unnecessary and not mentioned in the original coding task.
This observation suggests the importance of directing the focus of the LLM to the desired security aspect when utilizing zero-shot prompts, as it helps mitigate ambiguity and guides the model towards more relevant and focused security enhancements within the generated code.

 \textbf{(iv)} \textit{Additional validation checks:}
 Within the code generated through the utilization of the \textit{RCI} prompting technique, a notable increase in the presence of validity checks is observed especially in code generated using GPT-3.5 and GPT-4. These checks primarily serve the purpose of validating input received from external sources, such as external function calls or user inputs. These checks encompass a wide range of potential error scenarios, including security-related input validations.
The \textit{RCI} technique, which encourages the model to enhance its own code based on self-feedback, has resulted in the model's ability to recognize its own shortcomings. Consequently, this has led to a substantial increase in the number of both functional and security-related checks integrated into the generated code.

\textbf{(v)} \textit{Security related comments:} Some code generated by both GPT-3.5 and GPT-4 include warnings highlighting potential vulnerabilities. These warnings are present in code generated using all prompt templates except the \textit{baseline} prompt. While these comments do not directly enhance the code's security, they serve as valuable aids for developers utilizing such models to identify security-related aspects within the code.
Notably, code produced through the \textit{RCI} and \textit{zero-shot CoT} methods by GPT-3.5 and GPT-4 stands out for its detailed comments regarding the security measures implemented in the code.
Moreover, there are instances in which code snippets generated using all prompting techniques contain additional comments pertaining to how to enhance security, even though these enhancements are not actually implemented. This behavior is observed across all prompt types except the \textit{baseline} one. Specifically, GPT-4-generated code, when prompted with \textit{zero-shot CoT} and \textit{RCI} prompts, often includes a section titled \textit{'Additional Security Considerations'}. In this section, an extensive list of potential security measures that can or should be implemented to further enhance security is provided. Examples of such measures encompass suggestions like \textit{'Use a secure database connection like SSL/TSL'} and \textit{'Ensure script permissions are correctly configured to prevent unauthorized access or modifications.'}
Furthermore, many code snippets generated through the \textit{RCI} method also include cautionary security warnings, such as \textit{'Avoid logging sensitive information' and 'Ensure that memory dumps do not contain private data.'} These comments, while not directly affecting the code's functionality, serve as valuable reminders for developers to consider security aspects during the coding process.

 \textbf{(vi)} \textit{Calls to undeclared/undefined secure methods:} We observed numerous cases in GPT-3.5 and GPT-4 where the generated Python snippet included calls to undeclared methods that did not exist within the code's scope. There were also calls to declared methods that remained incomplete. In many cases, these methods are responsible for implementing security-sensitive tasks such as password, or session verification, and are frequently accompanied by security-related comments such as \textit{``securely verify the user session''}. This is mainly observed in zero-shot prompt variants. Our analysis suggests that the models acknowledge the necessary security measures required in the code from the prompts, but have prioritized their efforts on fulfilling the functional requirements specified in the prompt. Code snippets with incomplete logic were removed from our security analysis in the code validity analysis step.

 \textbf{(vii)} \textit{Modification of method names:} Quite commonly, when employing \textit{zero-shot} prompt variations, we observe a pattern where method names in the generated code are prefixed with the term 'secure'. For instance, we came across method names like \texttt{secure\_ping()}, \texttt{secure\_memory\_allocation}, \texttt{secure\_upload\_file}, and the like. However, it is noteworthy that in many cases, the actual implementation within these methods remains unaltered, despite the suggestive 'secure' prefixes in the method names. This tendency is particularly prevalent in code generated by GPT-3 and GPT-3.5.

\subsection{\majorrev{Generalizability to C}}
\label{subsec:c-generalizability-discussion}
\majorrev{The transferability of the the findings obtained for Python regarding the impact of different prompting techniques was tested on C, as mentioned in Section \ref{subsec:c-generalizability}. The results from the CodeQL evaluations are presented in Table \ref{tab:weaknesses-c}.} \minorrev{As in the case of Python (Section \ref{subsec:in-depth-study}), we selected a sample of results (49 cases, which account for 10\% of the results) from CodeQL for manual inspection. After conducting a reliability agreement test, we obtained a Kappa value of 0.82 averaged over the results of the 7 prompting techniques from GPT-4, suggesting that the results for C code from CodeQL are highly reliable.}

\majorrev{It can be seen that all the prompt templates that incorporated some form of security specifications showed improved results compared to the \textit{baseline} prompt (which did not include any security-related instructions). Hence, this further highlights the importance of including explicit security specifications in prompts when generating code using LLMs. Among all the techniques, \textit{RCI} produced C code with the lowest weakness count, rate, and density, consistent with the findings on Python. This, in principle, suggests its applicability across different programming languages. However, unlike the Python results, where the \textit{persona/memetic proxy} technique consistently performed the worst, the \textit{CWE-specific} template using the zero-shot technique performed the worst in C (of course, aside from the \textit{baseline} prompt). Overall, the findings from the C language experiment suggest that while the RCI technique remains the most effective, the relative performance of other prompting techniques may vary depending on the programming language.}

\majorrev{Further data from this experiment including the LLM-generated C code files, the CodeQL responses and the type of security weaknesses detected in the code are provided in the replication package.}

\begin{table}[]
\centering
\caption{\majorrev{The results of validity and security analysis of C code generated by GPT-4 using the 7 prompt templates. The \textit{\textbf{count}} is the total number of security weaknesses detected by CodeQL, \textit{\textbf{rate}} is the average number of security weaknesses per code and \textbf{\textit{density}} is the average number of security weaknesses per LOC.}}
\label{tab:weaknesses-c} \small
%\resizebox{\columnwidth}{!}{%
\begin{tabular}{@{}lccccccc@{}}
\toprule
\multicolumn{8}{c}{\textbf{GPT-4}} \\ \midrule
\multicolumn{1}{l|}{\textbf{Prompt Type}} &
  \multicolumn{1}{c|}{\textbf{\begin{tabular}[c]{@{}c@{}}\# valid  code\end{tabular}}} &
  \multicolumn{3}{c|}{\textbf{\# LOC}} &
  \multicolumn{3}{c}{\textbf{Security Weaknesses}} \\ \midrule
\multicolumn{1}{l|}{\textbf{}} &
  \multicolumn{1}{c|}{\textbf{}} &
  \textbf{MIN} &
  \textbf{MAX} &
  \multicolumn{1}{c|}{\textbf{Avg.}} &
  \textbf{Count} &
  \textbf{Rate} &
  \textbf{Density} \\
\multicolumn{1}{l|}{baseline   (0-shot)} &
  \multicolumn{1}{c|}{67} &
  6 &
  91 &
  \multicolumn{1}{c|}{27.88} &
  19 &
  0.283 &
  0.009 \\
\multicolumn{1}{l|}{naive-secure   (0-shot)} &
  \multicolumn{1}{c|}{67} &
  8 &
  86 &
  \multicolumn{1}{c|}{35.98} &
  15 &
  0.223 &
  0.005 \\
%\rowcolor[HTML]{EFEFEF} 
\multicolumn{1}{l|}{CWE-specific   (0-shot)} &
  \multicolumn{1}{c|}{67} &
  24 &
  95 &
  \multicolumn{1}{c|}{41.5} &
  19 &
  0.283 &
  0.006 \\
\multicolumn{1}{l|}{comprehensive   (0-shot)} &
  \multicolumn{1}{c|}{67} &
  17 &
  95 &
  \multicolumn{1}{c|}{50.97} &
  13 &
  0.194 &
  0.004 \\
\multicolumn{1}{l|}{zero-shot   CoT} &
  \multicolumn{1}{c|}{67} &
  8 &
  81 &
  \multicolumn{1}{c|}{36.95} &
  12 &
  0.179 &
  0.004 \\
\rowcolor[HTML]{EFEFEF} 
\multicolumn{1}{l|}{\cellcolor[HTML]{EFEFEF}\textbf{RCI}} &
  \multicolumn{1}{c|}{\cellcolor[HTML]{EFEFEF}67} &
  13 &
  157 &
  \multicolumn{1}{c|}{\cellcolor[HTML]{EFEFEF}56.80} &
  11 &
  0.164 &
  \textbf{0.002} \\
\multicolumn{1}{l|}{persona/memetic   proxy} &
  \multicolumn{1}{c|}{67} &
  21 &
  73 &
  \multicolumn{1}{c|}{40.29} &
  11 &
  0.164 &
  0.004 \\ \bottomrule
\end{tabular}%
%}
\end{table}

% - table of c results

% - RCI is the best 

% - relative ranking of the other techniques are probably language specific

%\vspace{2ex}
\begin{GrayBox}\small
\majorrev{\textbf{Actionable Takeaways: }%\vspace{1ex}
When using LLMs or LLM-powered tools like ChatGPT or Copilot, which enable natural language user interaction in a software development environment, the following considerations must be taken into account:}
\begin{enumerate}
    
    \item \majorrev{Using RCI is preferable over the other techniques studied in this work, as RCI can largely improve the security of the generated code (up to an order of magnitude w.r.t weakness density) even when applied with just 2 iterations. This technique has stayed valuable over several versions of the LLM models, and, hence, there is an expectation that it will stay valid in the future as well.}
    \item \majorrev{As of today, state-of-the-art LLM provides better results in terms of weakness density when used for C generation w.r.t Python, and therefore it should be used with more caution in the latter case.} 
    \item \majorrev{Nevertheless, the use of RCI might bring the defect density in Python to the same ballpark as in C. Therefore, this might justify the use of a slightly more complex prompting technique for software development in practice}. 
    \item \majorrev{In cases where multi-step techniques like RCI are not feasible, using simple zero-shot prompting with templates similar to \textit{comprehensive} prompts, that specify well-established secure coding standards, can provide comparable results in relation to more complex techniques.}
    \item \majorrev{Even with the use of secure prompting techniques, coding tasks involving web application development and OS command execution can still result in security weaknesses due to their complex nature. Special attention should be given to such tasks, particularly for vulnerabilities related to CWE-259:\textit{Hard-coded Passwords} (or more broadly CWE-798: \textit{Hard-coded Credentials}), CWE-94 and CWE-78:\textit{OS Command Injection}}. 
    \item \majorrev{Static analysis tools like Bandit and CodeQL seem to be able to detect the issues that LLMs may overlook (such as CWE-259, CWE-94, CWE-78). Hence, it is still strongly advised to use these tools in tandem with LLMs in the development pipeline.}
    
\end{enumerate}
\end{GrayBox}

\section{Impact of Data Leakage}
\label{sec:data-leakage}

\majorrev{Evaluating widely-used closed-source LLMs such as the ones used in this study poses the risk of data leakage \cite{YeCG23}. Data leakage (also referred to as data contamination) \cite{BalloccuSLD24}occurs when models have prior exposure to the benchmark datasets used for their evaluation, which can lead to false estimations of their capabilities. Balloccu et al. \cite{BalloccuSLD24} identified two forms of data leakage: direct and indirect.}

\majorrev{Direct data leakage occurs when evaluation data is already included in a model's training data. Considering that the OpenAI models used in this study are closed-source and their training data is undisclosed, this poses a potential concern. The LLMSecEval dataset used in this study, comprising NL code generation prompts paired with secure Python implementations, was formally published in May 2023. However, its corresponding GitHub public repository was created in January 2023. The OpenAI documentation\footnote{https://platform.openai.com/docs/models} states that GPT-3 (text-davinci-002) and GPT-3.5 (gpt-3.5-turbo) were trained on data up to September 2021, eliminating the risk of direct data leakage for these models. However, GPT-4 (gpt-4-1105-preview) was trained on data up to April 2023, suggesting a slight risk of leakage since the GitHub repository predates this one.}
\minorrev{To verify the extent of direct data leakage in the code generated by GPT-4, we performed a leakage test on the Python code generated with the RCI technique, the best-performing prompting technique in our experiments. For this, we employed the \textit{Dolos toolkit} \cite{MaertensPSBJDM22}, which is a source code plagiarism detection tool. Dolos works by tokenizing and canonicalizing programs into Abstract Syntax Tree (AST) representations to calculate a similarity score that captures semantic-level similarity through k-gram matching between the source and target programs. This tool has been used in several studies for quantifying contamination in code generated by LLMs \cite{RiddellNC24, YuW0WVX23}. The average similarity score obtained for all the code generated using RCI technique and the secure implementation present in the LLMSecEval dataset is 0.075. Yu et al. \cite{YuW0WVX23} in their study shows that a similarity score greater than 0.5 indicates potential plagiarism. Hence, this suggests that the impact of the direct data leakage/contamination in our results is very minimal.}

\majorrev{On the other hand, indirect leakage happens when models learn from user interactions via Reinforcement Learning Through Human Feedback (RLHF). The OpenAI documentation\footnote{https://help.openai.com/en/articles/5722486-how-your-data-is-used-to-improve-model-performance} states that only data from web interface interactions, not from API usage, is utilized for this purpose. Since this study exclusively relied on API interactions, there is no risk of indirect data leakage.}

%\majorrev{While the GPT-4 results may be influenced by direct data leakage, it should be noted that the goal of this study is not to evaluate the models themselves but to assess the impact of different prompting techniques for secure code generation. Since the aim is to compare how various prompts guide the model in generating secure code, the effect of data leakage on the interpretation of the results is negligible.}
\section{Threats to Validity}
 \label{sec:limitations}

Although this study yields valuable findings, it is important to acknowledge certain limitations.

\textbf{Construct Validity. }
The validity analysis of code responses generated by all LLMs was conducted by a single author, potentially introducing biases in the evaluation process. Nonetheless, efforts were made to mitigate such biases by explicitly outlining the criteria for assessing the validity of code snippets, as detailed in Section \ref{subsec:pre-study}.
We also acknowledge that the prompting techniques underwent evaluation using prompt templates created by us. These generated templates might have influenced the results obtained for each technique from the LLMs. However, attention was given to crafting the templates, adhering closely to the design and examples that demonstrated optimal results in the respective papers that introduced these techniques.

\textbf{External Validity. } 
%This study evaluates security in Python code which may affect the generalizability of our results to other programming languages such as C/C++ or JavaScript. However, we focused on weaknesses in Python code, given its continued popularity. Moreover, the LLMs used in this study have demonstrated competence in generating functional Python code, which further motivated us to prioritize the evaluation of Python code. 
This study was conducted only using the OpenAI models. As previously stated in \ref{subsec:dataset&models}, this decision was made due to the popularity of these models in the prompt engineering literature and their demonstrated proficiency in handling coding tasks articulated in natural language, as identified during a preliminary model selection examination by us.
It is also worth mentioning that, we focused on prompting techniques that do not rely on demonstrative examples. This choice stemmed from a user study \cite{PerryS0B23} which highlighted that users predominantly interacted with AI assistants using natural language coding task specifications or instructions, without supplying demonstrative examples.
\majorrev{We also note that the responses generated by the LLMs were not subjected to a randomness check. Since the code validity analysis was conducted manually, generating and evaluating multiple random responses from the LLMs was not feasible. However, we used 150 coding tasks to assess the impact of each prompting technique, which helps smooth out the fluctuation in results caused by randomness to some extent.}
Furthermore, the LLMSecEval dataset contains NL prompts for only 18 out of \textit{Top 25} CWEs of the year 2021. However, 15 out of the 18 CWEs considered in this study have retained their position on the \textit{Top 25} list of 2022 and 2023 proving the continued relevance and significance of our research findings.
\minorrev{Additionally, while our agreement test between manual and Bandit security analyses demonstrates Bandit's reliability to an extent, we acknowledge that it is not perfect. Some specific security weaknesses may have gone undetected. Nevertheless, Bandit proved sufficiently reliable for comparing the relative effectiveness of different prompting techniques.}

\section{Conclusion}
\label{sec:conclusion}

In an era where software development increasingly relies on automatic code generators, it is crucial to ensure the security of the code that LLMs produce out of NL descriptions. Through a literature review, we identified 15 distinct prompting techniques that can be applied to code generation. We also classified these techniques into 5 categories based on the prompting strategy they follow among other characteristics. Based on the suitability for the secure code generation task, we conducted an in-depth analysis of 4 prompting techniques to gauge their impact on secure code generation using GPT-3, GPT-3.5, and GPT-4. 

Our analysis reaffirms the prevalence of security weaknesses in code generated by LLMs when prompted with NL instructions, with significant challenges stemming from CWE-78, CWE-259, CWE-94, and CWE-330.
Among the prompting techniques investigated, \textit{RCI}, a refinement-based approach, exhibited notable effectiveness in preventing security weaknesses in LLM-generated code. Particularly noteworthy was its performance with GPT-4, where it reduced the average weakness density by 77.5\% compared to baseline prompting that includes no security specifications. 
Although RCI demonstrated the highest performance, to the extent of our knowledge, this technique has not been applied for secure code generation in existing literature. This highlights the need for additional research to investigate refinement-based methods, like RCI, which leverage self-critiquing and improvement capabilities of LLMs to enhance security in LLM-generated code.
%Conversely, \textit{persona/memetic proxy} techniques, which involve priming the model to adopt a specific role such as that of a software security expert in our case, demonstrated very limited performance.
\textit{Zero-shot} prompting yielded surprisingly favorable outcomes considering its straightforward nature, performing better than \textit{zero-shot CoT} and \textit{persona/memetic proxy} yet falling short of \textit{RCI}. However, \textit{zero-shot} prompting holds promise due to its simplicity and relative performance, provided an optimal prompt can be identified. 

Notably, recent advancements in prompt optimization techniques, such as genetic algorithms and black-box tuning, offer avenues for automatically optimizing prompts for various text-generation tasks. Future work can focus on exploring these optimization approaches to identify the optimal prompts for \textit{RCI} and \textit{zero-shot} techniques for secure code generation.

\section{Replication Package}
\label{sec:replication}

All data collected and generated in this study are available in \url{https://figshare.com/s/195a75d8c4dd86816223}. This repository contains the results of the literature review along with the information on the prompting techniques that were removed from our final selection and the rationale behind this exclusion. Furthermore, the repository contains the code generated by all 3 LLMs for the 7 prompt templates, including their validity and security analysis results.
%https://figshare.com/s/195a75d8c4dd86816223

% old - https://figshare.com/s/4b766676a0313429b3eb

\section*{Ackowledgements}

This work was partially supported by the EU-funded project Sec4AI4Sec: Cybersecurity for AI-Augmented Systems (grant no. 101120393).

\bibliographystyle{ACM-Reference-Format}
\bibliography{sample-base}


\begin{thebibliography}{133}

%%% the Web address.

\ifx \showCODEN    \undefined \def \showCODEN     #1{\unskip}     \fi
\ifx \showDOI      \undefined \def \showDOI       #1{#1}\fi
\ifx \showISBNx    \undefined \def \showISBNx     #1{\unskip}     \fi
\ifx \showISBNxiii \undefined \def \showISBNxiii  #1{\unskip}     \fi
\ifx \showISSN     \undefined \def \showISSN      #1{\unskip}     \fi
\ifx \showLCCN     \undefined \def \showLCCN      #1{\unskip}     \fi
\ifx \shownote     \undefined \def \shownote      #1{#1}          \fi
\ifx \showarticletitle \undefined \def \showarticletitle #1{#1}   \fi
\ifx \showURL      \undefined \def \showURL       {\relax}        \fi
% invisible to TeX
\providecommand\bibfield[2]{#2}
\providecommand\bibinfo[2]{#2}
\providecommand\natexlab[1]{#1}
\providecommand\showeprint[2][]{arXiv:#2}

\bibitem[ban(2023)]%
        {bandit}
 \bibinfo{year}{2023}\natexlab{}.
\newblock \bibinfo{title}{Bandit Documentation}.
\newblock
\newblock
\urldef\tempurl%
\url{https://bandit.readthedocs.io/en/latest/index.html}
\showURL{%
\tempurl}
\newblock
\shownote{[Accessed 05-01-2024]}.


\bibitem[cod(2023)]%
        {codeql}
 \bibinfo{year}{2023}\natexlab{}.
\newblock \bibinfo{title}{CodeQL}.
\newblock
\newblock
\urldef\tempurl%
\url{https://codeql.github.com/docs/}
\showURL{%
\tempurl}
\newblock
\shownote{[Accessed 09-01-2024]}.


\bibitem[MIT(2023)]%
        {MITRE}
 \bibinfo{year}{2023}\natexlab{}.
\newblock \bibinfo{title}{MITRE}.
\newblock
\newblock
\urldef\tempurl%
\url{https://www.mitre.org/}
\showURL{%
\tempurl}
\newblock
\shownote{[Accessed 10-10-2023]}.


\bibitem[Ahmad et~al\mbox{.}(2021)]%
        {plbart}
\bibfield{author}{\bibinfo{person}{Wasi~Uddin Ahmad}, \bibinfo{person}{Saikat Chakraborty}, \bibinfo{person}{Baishakhi Ray}, {and} \bibinfo{person}{Kai{-}Wei Chang}.} \bibinfo{year}{2021}\natexlab{}.
\newblock \showarticletitle{Unified Pre-training for Program Understanding and Generation}. In \bibinfo{booktitle}{\emph{Proceedings of the 2021 Conference of the North American Chapter of the Association for Computational Linguistics: Human Language Technologies, {NAACL-HLT} 2021, Online, June 6-11, 2021}}, \bibfield{editor}{\bibinfo{person}{Kristina Toutanova}, \bibinfo{person}{Anna Rumshisky}, \bibinfo{person}{Luke Zettlemoyer}, \bibinfo{person}{Dilek Hakkani{-}T{\"{u}}r}, \bibinfo{person}{Iz~Beltagy}, \bibinfo{person}{Steven Bethard}, \bibinfo{person}{Ryan Cotterell}, \bibinfo{person}{Tanmoy Chakraborty}, {and} \bibinfo{person}{Yichao Zhou}} (Eds.). \bibinfo{publisher}{Association for Computational Linguistics}, \bibinfo{pages}{2655--2668}.
\newblock
\urldef\tempurl%
\url{https://doi.org/10.18653/V1/2021.NAACL-MAIN.211}
\showDOI{\tempurl}


\bibitem[Arora et~al\mbox{.}(2023)]%
        {AskMeAnything}
\bibfield{author}{\bibinfo{person}{Simran Arora}, \bibinfo{person}{Avanika Narayan}, \bibinfo{person}{Mayee~F. Chen}, \bibinfo{person}{Laurel~J. Orr}, \bibinfo{person}{Neel Guha}, \bibinfo{person}{Kush Bhatia}, \bibinfo{person}{Ines Chami}, {and} \bibinfo{person}{Christopher R{\'{e}}}.} \bibinfo{year}{2023}\natexlab{}.
\newblock \showarticletitle{Ask Me Anything: {A} simple strategy for prompting language models}. In \bibinfo{booktitle}{\emph{The Eleventh International Conference on Learning Representations, {ICLR} 2023, Kigali, Rwanda, May 1-5, 2023}}. \bibinfo{publisher}{OpenReview.net}.
\newblock
\urldef\tempurl%
\url{https://openreview.net/pdf?id=bhUPJnS2g0X}
\showURL{%
\tempurl}


\bibitem[Asare et~al\mbox{.}(2023)]%
        {AsareNA23}
\bibfield{author}{\bibinfo{person}{Owura Asare}, \bibinfo{person}{Meiyappan Nagappan}, {and} \bibinfo{person}{N. Asokan}.} \bibinfo{year}{2023}\natexlab{}.
\newblock \showarticletitle{Is GitHub's Copilot as bad as humans at introducing vulnerabilities in code?}
\newblock \bibinfo{journal}{\emph{Empir. Softw. Eng.}} \bibinfo{volume}{28}, \bibinfo{number}{6} (\bibinfo{year}{2023}), \bibinfo{pages}{129}.
\newblock
\urldef\tempurl%
\url{https://doi.org/10.1007/S10664-023-10380-1}
\showDOI{\tempurl}


\bibitem[Austin et~al\mbox{.}(2021)]%
        {austin2022}
\bibfield{author}{\bibinfo{person}{Jacob Austin}, \bibinfo{person}{Augustus Odena}, \bibinfo{person}{Maxwell~I. Nye}, \bibinfo{person}{Maarten Bosma}, \bibinfo{person}{Henryk Michalewski}, \bibinfo{person}{David Dohan}, \bibinfo{person}{Ellen Jiang}, \bibinfo{person}{Carrie~J. Cai}, \bibinfo{person}{Michael Terry}, \bibinfo{person}{Quoc~V. Le}, {and} \bibinfo{person}{Charles Sutton}.} \bibinfo{year}{2021}\natexlab{}.
\newblock \showarticletitle{Program Synthesis with Large Language Models}.
\newblock \bibinfo{journal}{\emph{CoRR}}  \bibinfo{volume}{abs/2108.07732} (\bibinfo{year}{2021}).
\newblock
\showeprint[arXiv]{2108.07732}
\urldef\tempurl%
\url{https://arxiv.org/abs/2108.07732}
\showURL{%
\tempurl}


\bibitem[Bai et~al\mbox{.}(2022)]%
        {Bai}
\bibfield{author}{\bibinfo{person}{Yuntao Bai}, \bibinfo{person}{Saurav Kadavath}, \bibinfo{person}{Sandipan Kundu}, \bibinfo{person}{Amanda Askell}, \bibinfo{person}{Jackson Kernion}, \bibinfo{person}{Andy Jones}, \bibinfo{person}{Anna Chen}, \bibinfo{person}{Anna Goldie}, \bibinfo{person}{Azalia Mirhoseini}, \bibinfo{person}{Cameron McKinnon}, \bibinfo{person}{Carol Chen}, \bibinfo{person}{Catherine Olsson}, \bibinfo{person}{Christopher Olah}, \bibinfo{person}{Danny Hernandez}, \bibinfo{person}{Dawn Drain}, \bibinfo{person}{Deep Ganguli}, \bibinfo{person}{Dustin Li}, \bibinfo{person}{Eli Tran{-}Johnson}, \bibinfo{person}{Ethan Perez}, \bibinfo{person}{Jamie Kerr}, \bibinfo{person}{Jared Mueller}, \bibinfo{person}{Jeffrey Ladish}, \bibinfo{person}{Joshua Landau}, \bibinfo{person}{Kamal Ndousse}, \bibinfo{person}{Kamile Lukosiute}, \bibinfo{person}{Liane Lovitt}, \bibinfo{person}{Michael Sellitto}, \bibinfo{person}{Nelson Elhage}, \bibinfo{person}{Nicholas Schiefer}, \bibinfo{person}{Noem{\'{\i}} Mercado},
  \bibinfo{person}{Nova DasSarma}, \bibinfo{person}{Robert Lasenby}, \bibinfo{person}{Robin Larson}, \bibinfo{person}{Sam Ringer}, \bibinfo{person}{Scott Johnston}, \bibinfo{person}{Shauna Kravec}, \bibinfo{person}{Sheer~El Showk}, \bibinfo{person}{Stanislav Fort}, \bibinfo{person}{Tamera Lanham}, \bibinfo{person}{Timothy Telleen{-}Lawton}, \bibinfo{person}{Tom Conerly}, \bibinfo{person}{Tom Henighan}, \bibinfo{person}{Tristan Hume}, \bibinfo{person}{Samuel~R. Bowman}, \bibinfo{person}{Zac Hatfield{-}Dodds}, \bibinfo{person}{Ben Mann}, \bibinfo{person}{Dario Amodei}, \bibinfo{person}{Nicholas Joseph}, \bibinfo{person}{Sam McCandlish}, \bibinfo{person}{Tom Brown}, {and} \bibinfo{person}{Jared Kaplan}.} \bibinfo{year}{2022}\natexlab{}.
\newblock \showarticletitle{Constitutional {AI:} Harmlessness from {AI} Feedback}.
\newblock \bibinfo{journal}{\emph{CoRR}}  \bibinfo{volume}{abs/2212.08073} (\bibinfo{year}{2022}).
\newblock
\urldef\tempurl%
\url{https://doi.org/10.48550/ARXIV.2212.08073}
\showDOI{\tempurl}
\showeprint[arXiv]{2212.08073}


\bibitem[Balloccu et~al\mbox{.}(2024)]%
        {BalloccuSLD24}
\bibfield{author}{\bibinfo{person}{Simone Balloccu}, \bibinfo{person}{Patr{\'{\i}}cia Schmidtov{\'{a}}}, \bibinfo{person}{Mateusz Lango}, {and} \bibinfo{person}{Ondrej Dusek}.} \bibinfo{year}{2024}\natexlab{}.
\newblock \showarticletitle{Leak, Cheat, Repeat: Data Contamination and Evaluation Malpractices in Closed-Source LLMs}. In \bibinfo{booktitle}{\emph{Proceedings of the 18th Conference of the European Chapter of the Association for Computational Linguistics, {EACL} 2024 - Volume 1: Long Papers, St. Julian's, Malta, March 17-22, 2024}}, \bibfield{editor}{\bibinfo{person}{Yvette Graham} {and} \bibinfo{person}{Matthew Purver}} (Eds.). \bibinfo{publisher}{Association for Computational Linguistics}, \bibinfo{pages}{67--93}.
\newblock
\urldef\tempurl%
\url{https://aclanthology.org/2024.eacl-long.5}
\showURL{%
\tempurl}


\bibitem[Benjamini and Hochberg(1995)]%
        {bonferroni}
\bibfield{author}{\bibinfo{person}{Yoav Benjamini} {and} \bibinfo{person}{Yosef Hochberg}.} \bibinfo{year}{1995}\natexlab{}.
\newblock \showarticletitle{Controlling the false discovery rate: a practical and powerful approach to multiple testing}.
\newblock \bibinfo{journal}{\emph{Journal of the Royal statistical society: series B (Methodological)}} \bibinfo{volume}{57}, \bibinfo{number}{1} (\bibinfo{year}{1995}), \bibinfo{pages}{289--300}.
\newblock


\bibitem[Berant et~al\mbox{.}(2013)]%
        {webquestions}
\bibfield{author}{\bibinfo{person}{Jonathan Berant}, \bibinfo{person}{Andrew Chou}, \bibinfo{person}{Roy Frostig}, {and} \bibinfo{person}{Percy Liang}.} \bibinfo{year}{2013}\natexlab{}.
\newblock \showarticletitle{Semantic Parsing on Freebase from Question-Answer Pairs}. In \bibinfo{booktitle}{\emph{Proceedings of the 2013 Conference on Empirical Methods in Natural Language Processing, {EMNLP} 2013, 18-21 October 2013, Grand Hyatt Seattle, Seattle, Washington, USA, {A} meeting of SIGDAT, a Special Interest Group of the {ACL}}}. \bibinfo{publisher}{{ACL}}, \bibinfo{pages}{1533--1544}.
\newblock
\urldef\tempurl%
\url{https://aclanthology.org/D13-1160/}
\showURL{%
\tempurl}


\bibitem[Bhakthavatsalam et~al\mbox{.}(2021)]%
        {arc-da}
\bibfield{author}{\bibinfo{person}{Sumithra Bhakthavatsalam}, \bibinfo{person}{Daniel Khashabi}, \bibinfo{person}{Tushar Khot}, \bibinfo{person}{Bhavana~Dalvi Mishra}, \bibinfo{person}{Kyle Richardson}, \bibinfo{person}{Ashish Sabharwal}, \bibinfo{person}{Carissa Schoenick}, \bibinfo{person}{Oyvind Tafjord}, {and} \bibinfo{person}{Peter Clark}.} \bibinfo{year}{2021}\natexlab{}.
\newblock \showarticletitle{Think you have Solved Direct-Answer Question Answering? Try ARC-DA, the Direct-Answer {AI2} Reasoning Challenge}.
\newblock \bibinfo{journal}{\emph{CoRR}}  \bibinfo{volume}{abs/2102.03315} (\bibinfo{year}{2021}).
\newblock
\showeprint[arXiv]{2102.03315}
\urldef\tempurl%
\url{https://arxiv.org/abs/2102.03315}
\showURL{%
\tempurl}


\bibitem[Bisk et~al\mbox{.}(2020)]%
        {piqa}
\bibfield{author}{\bibinfo{person}{Yonatan Bisk}, \bibinfo{person}{Rowan Zellers}, \bibinfo{person}{Ronan~Le Bras}, \bibinfo{person}{Jianfeng Gao}, {and} \bibinfo{person}{Yejin Choi}.} \bibinfo{year}{2020}\natexlab{}.
\newblock \showarticletitle{{PIQA:} Reasoning about Physical Commonsense in Natural Language}. In \bibinfo{booktitle}{\emph{The Thirty-Fourth {AAAI} Conference on Artificial Intelligence, {AAAI} 2020, The Thirty-Second Innovative Applications of Artificial Intelligence Conference, {IAAI} 2020, The Tenth {AAAI} Symposium on Educational Advances in Artificial Intelligence, {EAAI} 2020, New York, NY, USA, February 7-12, 2020}}. \bibinfo{publisher}{{AAAI} Press}, \bibinfo{pages}{7432--7439}.
\newblock
\urldef\tempurl%
\url{https://doi.org/10.1609/AAAI.V34I05.6239}
\showDOI{\tempurl}


\bibitem[Black et~al\mbox{.}(2022)]%
        {gpt-neo}
\bibfield{author}{\bibinfo{person}{Sid Black}, \bibinfo{person}{Stella Biderman}, \bibinfo{person}{Eric Hallahan}, \bibinfo{person}{Quentin Anthony}, \bibinfo{person}{Leo Gao}, \bibinfo{person}{Laurence Golding}, \bibinfo{person}{Horace He}, \bibinfo{person}{Connor Leahy}, \bibinfo{person}{Kyle McDonell}, \bibinfo{person}{Jason Phang}, \bibinfo{person}{Michael Pieler}, \bibinfo{person}{USVSN~Sai Prashanth}, \bibinfo{person}{Shivanshu Purohit}, \bibinfo{person}{Laria Reynolds}, \bibinfo{person}{Jonathan Tow}, \bibinfo{person}{Ben Wang}, {and} \bibinfo{person}{Samuel Weinbach}.} \bibinfo{year}{2022}\natexlab{}.
\newblock \showarticletitle{GPT-NeoX-20B: An Open-Source Autoregressive Language Model}.
\newblock \bibinfo{journal}{\emph{CoRR}}  \bibinfo{volume}{abs/2204.06745} (\bibinfo{year}{2022}).
\newblock
\urldef\tempurl%
\url{https://doi.org/10.48550/ARXIV.2204.06745}
\showDOI{\tempurl}
\showeprint[arXiv]{2204.06745}


\bibitem[Brown et~al\mbox{.}(2020)]%
        {GPT-3}
\bibfield{author}{\bibinfo{person}{Tom~B. Brown}, \bibinfo{person}{Benjamin Mann}, \bibinfo{person}{Nick Ryder}, \bibinfo{person}{Melanie Subbiah}, \bibinfo{person}{Jared Kaplan}, \bibinfo{person}{Prafulla Dhariwal}, \bibinfo{person}{Arvind Neelakantan}, \bibinfo{person}{Pranav Shyam}, \bibinfo{person}{Girish Sastry}, \bibinfo{person}{Amanda Askell}, \bibinfo{person}{Sandhini Agarwal}, \bibinfo{person}{Ariel Herbert{-}Voss}, \bibinfo{person}{Gretchen Krueger}, \bibinfo{person}{Tom Henighan}, \bibinfo{person}{Rewon Child}, \bibinfo{person}{Aditya Ramesh}, \bibinfo{person}{Daniel~M. Ziegler}, \bibinfo{person}{Jeffrey Wu}, \bibinfo{person}{Clemens Winter}, \bibinfo{person}{Christopher Hesse}, \bibinfo{person}{Mark Chen}, \bibinfo{person}{Eric Sigler}, \bibinfo{person}{Mateusz Litwin}, \bibinfo{person}{Scott Gray}, \bibinfo{person}{Benjamin Chess}, \bibinfo{person}{Jack Clark}, \bibinfo{person}{Christopher Berner}, \bibinfo{person}{Sam McCandlish}, \bibinfo{person}{Alec Radford}, \bibinfo{person}{Ilya Sutskever},
  {and} \bibinfo{person}{Dario Amodei}.} \bibinfo{year}{2020}\natexlab{}.
\newblock \showarticletitle{Language Models are Few-Shot Learners}. In \bibinfo{booktitle}{\emph{Advances in Neural Information Processing Systems 33: Annual Conference on Neural Information Processing Systems 2020, NeurIPS 2020, December 6-12, 2020, virtual}}, \bibfield{editor}{\bibinfo{person}{Hugo Larochelle}, \bibinfo{person}{Marc'Aurelio Ranzato}, \bibinfo{person}{Raia Hadsell}, \bibinfo{person}{Maria{-}Florina Balcan}, {and} \bibinfo{person}{Hsuan{-}Tien Lin}} (Eds.).
\newblock


\bibitem[Carrera-Rivera et~al\mbox{.}(2022)]%
        {PICOC}
\bibfield{author}{\bibinfo{person}{Angela Carrera-Rivera}, \bibinfo{person}{William Ochoa}, \bibinfo{person}{Felix Larrinaga}, {and} \bibinfo{person}{Ganix Lasa}.} \bibinfo{year}{2022}\natexlab{}.
\newblock \showarticletitle{How-to conduct a systematic literature review: A quick guide for computer science research}.
\newblock \bibinfo{journal}{\emph{MethodsX}}  \bibinfo{volume}{9} (\bibinfo{date}{11} \bibinfo{year}{2022}), \bibinfo{pages}{101895}.
\newblock
\urldef\tempurl%
\url{https://doi.org/10.1016/j.mex.2022.101895}
\showDOI{\tempurl}


\bibitem[Cassano et~al\mbox{.}(2023)]%
        {CassanoGNNPPYZAFGGJ23}
\bibfield{author}{\bibinfo{person}{Federico Cassano}, \bibinfo{person}{John Gouwar}, \bibinfo{person}{Daniel Nguyen}, \bibinfo{person}{Sydney Nguyen}, \bibinfo{person}{Luna Phipps{-}Costin}, \bibinfo{person}{Donald Pinckney}, \bibinfo{person}{Ming{-}Ho Yee}, \bibinfo{person}{Yangtian Zi}, \bibinfo{person}{Carolyn~Jane Anderson}, \bibinfo{person}{Molly~Q. Feldman}, \bibinfo{person}{Arjun Guha}, \bibinfo{person}{Michael Greenberg}, {and} \bibinfo{person}{Abhinav Jangda}.} \bibinfo{year}{2023}\natexlab{}.
\newblock \showarticletitle{MultiPL-E: {A} Scalable and Polyglot Approach to Benchmarking Neural Code Generation}.
\newblock \bibinfo{journal}{\emph{{IEEE} Trans. Software Eng.}} \bibinfo{volume}{49}, \bibinfo{number}{7} (\bibinfo{year}{2023}), \bibinfo{pages}{3675--3691}.
\newblock
\urldef\tempurl%
\url{https://doi.org/10.1109/TSE.2023.3267446}
\showDOI{\tempurl}


\bibitem[Chen et~al\mbox{.}(2021)]%
        {codex}
\bibfield{author}{\bibinfo{person}{Mark Chen}, \bibinfo{person}{Jerry Tworek}, \bibinfo{person}{Heewoo Jun}, \bibinfo{person}{Qiming Yuan}, \bibinfo{person}{Henrique~Ponde de Oliveira~Pinto}, \bibinfo{person}{Jared Kaplan}, \bibinfo{person}{Harrison Edwards}, \bibinfo{person}{Yuri Burda}, \bibinfo{person}{Nicholas Joseph}, \bibinfo{person}{Greg Brockman}, \bibinfo{person}{Alex Ray}, \bibinfo{person}{Raul Puri}, \bibinfo{person}{Gretchen Krueger}, \bibinfo{person}{Michael Petrov}, \bibinfo{person}{Heidy Khlaaf}, \bibinfo{person}{Girish Sastry}, \bibinfo{person}{Pamela Mishkin}, \bibinfo{person}{Brooke Chan}, \bibinfo{person}{Scott Gray}, \bibinfo{person}{Nick Ryder}, \bibinfo{person}{Mikhail Pavlov}, \bibinfo{person}{Alethea Power}, \bibinfo{person}{Lukasz Kaiser}, \bibinfo{person}{Mohammad Bavarian}, \bibinfo{person}{Clemens Winter}, \bibinfo{person}{Philippe Tillet}, \bibinfo{person}{Felipe~Petroski Such}, \bibinfo{person}{Dave Cummings}, \bibinfo{person}{Matthias Plappert}, \bibinfo{person}{Fotios
  Chantzis}, \bibinfo{person}{Elizabeth Barnes}, \bibinfo{person}{Ariel Herbert{-}Voss}, \bibinfo{person}{William~Hebgen Guss}, \bibinfo{person}{Alex Nichol}, \bibinfo{person}{Alex Paino}, \bibinfo{person}{Nikolas Tezak}, \bibinfo{person}{Jie Tang}, \bibinfo{person}{Igor Babuschkin}, \bibinfo{person}{Suchir Balaji}, \bibinfo{person}{Shantanu Jain}, \bibinfo{person}{William Saunders}, \bibinfo{person}{Christopher Hesse}, \bibinfo{person}{Andrew~N. Carr}, \bibinfo{person}{Jan Leike}, \bibinfo{person}{Joshua Achiam}, \bibinfo{person}{Vedant Misra}, \bibinfo{person}{Evan Morikawa}, \bibinfo{person}{Alec Radford}, \bibinfo{person}{Matthew Knight}, \bibinfo{person}{Miles Brundage}, \bibinfo{person}{Mira Murati}, \bibinfo{person}{Katie Mayer}, \bibinfo{person}{Peter Welinder}, \bibinfo{person}{Bob McGrew}, \bibinfo{person}{Dario Amodei}, \bibinfo{person}{Sam McCandlish}, \bibinfo{person}{Ilya Sutskever}, {and} \bibinfo{person}{Wojciech Zaremba}.} \bibinfo{year}{2021}\natexlab{}.
\newblock \showarticletitle{Evaluating Large Language Models Trained on Code}.
\newblock \bibinfo{journal}{\emph{CoRR}}  \bibinfo{volume}{abs/2107.03374} (\bibinfo{year}{2021}).
\newblock
\showeprint[arXiv]{2107.03374}
\urldef\tempurl%
\url{https://arxiv.org/abs/2107.03374}
\showURL{%
\tempurl}


\bibitem[Choi et~al\mbox{.}(2018)]%
        {quac}
\bibfield{author}{\bibinfo{person}{Eunsol Choi}, \bibinfo{person}{He He}, \bibinfo{person}{Mohit Iyyer}, \bibinfo{person}{Mark Yatskar}, \bibinfo{person}{Wen{-}tau Yih}, \bibinfo{person}{Yejin Choi}, \bibinfo{person}{Percy Liang}, {and} \bibinfo{person}{Luke Zettlemoyer}.} \bibinfo{year}{2018}\natexlab{}.
\newblock \showarticletitle{QuAC: Question Answering in Context}. In \bibinfo{booktitle}{\emph{Proceedings of the 2018 Conference on Empirical Methods in Natural Language Processing, Brussels, Belgium, October 31 - November 4, 2018}}, \bibfield{editor}{\bibinfo{person}{Ellen Riloff}, \bibinfo{person}{David Chiang}, \bibinfo{person}{Julia Hockenmaier}, {and} \bibinfo{person}{Jun'ichi Tsujii}} (Eds.). \bibinfo{publisher}{Association for Computational Linguistics}, \bibinfo{pages}{2174--2184}.
\newblock
\urldef\tempurl%
\url{https://doi.org/10.18653/V1/D18-1241}
\showDOI{\tempurl}


\bibitem[Chowdhery et~al\mbox{.}(2023)]%
        {palmcoder}
\bibfield{author}{\bibinfo{person}{Aakanksha Chowdhery}, \bibinfo{person}{Sharan Narang}, \bibinfo{person}{Jacob Devlin}, \bibinfo{person}{Maarten Bosma}, \bibinfo{person}{Gaurav Mishra}, \bibinfo{person}{Adam Roberts}, \bibinfo{person}{Paul Barham}, \bibinfo{person}{Hyung~Won Chung}, \bibinfo{person}{Charles Sutton}, \bibinfo{person}{Sebastian Gehrmann}, \bibinfo{person}{Parker Schuh}, \bibinfo{person}{Kensen Shi}, \bibinfo{person}{Sasha Tsvyashchenko}, \bibinfo{person}{Joshua Maynez}, \bibinfo{person}{Abhishek Rao}, \bibinfo{person}{Parker Barnes}, \bibinfo{person}{Yi Tay}, \bibinfo{person}{Noam Shazeer}, \bibinfo{person}{Vinodkumar Prabhakaran}, \bibinfo{person}{Emily Reif}, \bibinfo{person}{Nan Du}, \bibinfo{person}{Ben Hutchinson}, \bibinfo{person}{Reiner Pope}, \bibinfo{person}{James Bradbury}, \bibinfo{person}{Jacob Austin}, \bibinfo{person}{Michael Isard}, \bibinfo{person}{Guy Gur{-}Ari}, \bibinfo{person}{Pengcheng Yin}, \bibinfo{person}{Toju Duke}, \bibinfo{person}{Anselm Levskaya},
  \bibinfo{person}{Sanjay Ghemawat}, \bibinfo{person}{Sunipa Dev}, \bibinfo{person}{Henryk Michalewski}, \bibinfo{person}{Xavier Garcia}, \bibinfo{person}{Vedant Misra}, \bibinfo{person}{Kevin Robinson}, \bibinfo{person}{Liam Fedus}, \bibinfo{person}{Denny Zhou}, \bibinfo{person}{Daphne Ippolito}, \bibinfo{person}{David Luan}, \bibinfo{person}{Hyeontaek Lim}, \bibinfo{person}{Barret Zoph}, \bibinfo{person}{Alexander Spiridonov}, \bibinfo{person}{Ryan Sepassi}, \bibinfo{person}{David Dohan}, \bibinfo{person}{Shivani Agrawal}, \bibinfo{person}{Mark Omernick}, \bibinfo{person}{Andrew~M. Dai}, \bibinfo{person}{Thanumalayan~Sankaranarayana Pillai}, \bibinfo{person}{Marie Pellat}, \bibinfo{person}{Aitor Lewkowycz}, \bibinfo{person}{Erica Moreira}, \bibinfo{person}{Rewon Child}, \bibinfo{person}{Oleksandr Polozov}, \bibinfo{person}{Katherine Lee}, \bibinfo{person}{Zongwei Zhou}, \bibinfo{person}{Xuezhi Wang}, \bibinfo{person}{Brennan Saeta}, \bibinfo{person}{Mark Diaz}, \bibinfo{person}{Orhan Firat},
  \bibinfo{person}{Michele Catasta}, \bibinfo{person}{Jason Wei}, \bibinfo{person}{Kathy Meier{-}Hellstern}, \bibinfo{person}{Douglas Eck}, \bibinfo{person}{Jeff Dean}, \bibinfo{person}{Slav Petrov}, {and} \bibinfo{person}{Noah Fiedel}.} \bibinfo{year}{2023}\natexlab{}.
\newblock \showarticletitle{PaLM: Scaling Language Modeling with Pathways}.
\newblock \bibinfo{journal}{\emph{J. Mach. Learn. Res.}}  \bibinfo{volume}{24} (\bibinfo{year}{2023}), \bibinfo{pages}{240:1--240:113}.
\newblock
\urldef\tempurl%
\url{http://jmlr.org/papers/v24/22-1144.html}
\showURL{%
\tempurl}


\bibitem[Cobbe et~al\mbox{.}(2021)]%
        {gsm8k}
\bibfield{author}{\bibinfo{person}{Karl Cobbe}, \bibinfo{person}{Vineet Kosaraju}, \bibinfo{person}{Mohammad Bavarian}, \bibinfo{person}{Mark Chen}, \bibinfo{person}{Heewoo Jun}, \bibinfo{person}{Lukasz Kaiser}, \bibinfo{person}{Matthias Plappert}, \bibinfo{person}{Jerry Tworek}, \bibinfo{person}{Jacob Hilton}, \bibinfo{person}{Reiichiro Nakano}, \bibinfo{person}{Christopher Hesse}, {and} \bibinfo{person}{John Schulman}.} \bibinfo{year}{2021}\natexlab{}.
\newblock \showarticletitle{Training Verifiers to Solve Math Word Problems}.
\newblock \bibinfo{journal}{\emph{CoRR}}  \bibinfo{volume}{abs/2110.14168} (\bibinfo{year}{2021}).
\newblock
\showeprint[arXiv]{2110.14168}
\urldef\tempurl%
\url{https://arxiv.org/abs/2110.14168}
\showURL{%
\tempurl}


\bibitem[Deng et~al\mbox{.}(2022)]%
        {DengWHWGSSXH22}
\bibfield{author}{\bibinfo{person}{Mingkai Deng}, \bibinfo{person}{Jianyu Wang}, \bibinfo{person}{Cheng{-}Ping Hsieh}, \bibinfo{person}{Yihan Wang}, \bibinfo{person}{Han Guo}, \bibinfo{person}{Tianmin Shu}, \bibinfo{person}{Meng Song}, \bibinfo{person}{Eric~P. Xing}, {and} \bibinfo{person}{Zhiting Hu}.} \bibinfo{year}{2022}\natexlab{}.
\newblock \showarticletitle{RLPrompt: Optimizing Discrete Text Prompts with Reinforcement Learning}. In \bibinfo{booktitle}{\emph{Proceedings of the 2022 Conference on Empirical Methods in Natural Language Processing, {EMNLP} 2022, Abu Dhabi, United Arab Emirates, December 7-11, 2022}}, \bibfield{editor}{\bibinfo{person}{Yoav Goldberg}, \bibinfo{person}{Zornitsa Kozareva}, {and} \bibinfo{person}{Yue Zhang}} (Eds.). \bibinfo{publisher}{Association for Computational Linguistics}, \bibinfo{pages}{3369--3391}.
\newblock
\urldef\tempurl%
\url{https://doi.org/10.18653/V1/2022.EMNLP-MAIN.222}
\showDOI{\tempurl}


\bibitem[Devlin et~al\mbox{.}(2019)]%
        {DevlinCLT19}
\bibfield{author}{\bibinfo{person}{Jacob Devlin}, \bibinfo{person}{Ming{-}Wei Chang}, \bibinfo{person}{Kenton Lee}, {and} \bibinfo{person}{Kristina Toutanova}.} \bibinfo{year}{2019}\natexlab{}.
\newblock \showarticletitle{{BERT:} Pre-training of Deep Bidirectional Transformers for Language Understanding}. In \bibinfo{booktitle}{\emph{Proceedings of the 2019 Conference of the North American Chapter of the Association for Computational Linguistics: Human Language Technologies, {NAACL-HLT} 2019, Minneapolis, MN, USA, June 2-7, 2019, Volume 1 (Long and Short Papers)}}, \bibfield{editor}{\bibinfo{person}{Jill Burstein}, \bibinfo{person}{Christy Doran}, {and} \bibinfo{person}{Thamar Solorio}} (Eds.). \bibinfo{publisher}{Association for Computational Linguistics}, \bibinfo{pages}{4171--4186}.
\newblock
\urldef\tempurl%
\url{https://doi.org/10.18653/V1/N19-1423}
\showDOI{\tempurl}


\bibitem[Dong et~al\mbox{.}(2023)]%
        {mbpp-et}
\bibfield{author}{\bibinfo{person}{Yihong Dong}, \bibinfo{person}{Jiazheng Ding}, \bibinfo{person}{Xue Jiang}, \bibinfo{person}{Zhuo Li}, \bibinfo{person}{Ge Li}, {and} \bibinfo{person}{Zhi Jin}.} \bibinfo{year}{2023}\natexlab{}.
\newblock \showarticletitle{CodeScore: Evaluating Code Generation by Learning Code Execution}.
\newblock \bibinfo{journal}{\emph{CoRR}}  \bibinfo{volume}{abs/2301.09043} (\bibinfo{year}{2023}).
\newblock
\urldef\tempurl%
\url{https://doi.org/10.48550/ARXIV.2301.09043}
\showDOI{\tempurl}
\showeprint[arXiv]{2301.09043}


\bibitem[Dua et~al\mbox{.}(2019)]%
        {drop}
\bibfield{author}{\bibinfo{person}{Dheeru Dua}, \bibinfo{person}{Yizhong Wang}, \bibinfo{person}{Pradeep Dasigi}, \bibinfo{person}{Gabriel Stanovsky}, \bibinfo{person}{Sameer Singh}, {and} \bibinfo{person}{Matt Gardner}.} \bibinfo{year}{2019}\natexlab{}.
\newblock \showarticletitle{{DROP:} {A} Reading Comprehension Benchmark Requiring Discrete Reasoning Over Paragraphs}. In \bibinfo{booktitle}{\emph{Proceedings of the 2019 Conference of the North American Chapter of the Association for Computational Linguistics: Human Language Technologies, {NAACL-HLT} 2019, Minneapolis, MN, USA, June 2-7, 2019, Volume 1 (Long and Short Papers)}}, \bibfield{editor}{\bibinfo{person}{Jill Burstein}, \bibinfo{person}{Christy Doran}, {and} \bibinfo{person}{Thamar Solorio}} (Eds.). \bibinfo{publisher}{Association for Computational Linguistics}, \bibinfo{pages}{2368--2378}.
\newblock
\urldef\tempurl%
\url{https://doi.org/10.18653/V1/N19-1246}
\showDOI{\tempurl}


\bibitem[Dunn(1961)]%
        {dunn}
\bibfield{author}{\bibinfo{person}{Olive~Jean Dunn}.} \bibinfo{year}{1961}\natexlab{}.
\newblock \showarticletitle{Multiple Comparisons Among Means}.
\newblock \bibinfo{journal}{\emph{J. Amer. Statist. Assoc.}} \bibinfo{volume}{56}, \bibinfo{number}{293} (\bibinfo{year}{1961}), \bibinfo{pages}{52--64}.
\newblock
\showISSN{01621459}
\urldef\tempurl%
\url{http://www.jstor.org/stable/2282330}
\showURL{%
\tempurl}


\bibitem[Durrani et~al\mbox{.}(2014)]%
        {wmt}
\bibfield{author}{\bibinfo{person}{Nadir Durrani}, \bibinfo{person}{Barry Haddow}, \bibinfo{person}{Philipp Koehn}, {and} \bibinfo{person}{Kenneth Heafield}.} \bibinfo{year}{2014}\natexlab{}.
\newblock \showarticletitle{Edinburgh's Phrase-based Machine Translation Systems for {WMT-14}}. In \bibinfo{booktitle}{\emph{Proceedings of the Ninth Workshop on Statistical Machine Translation, WMT@ACL 2014, June 26-27, 2014, Baltimore, Maryland, {USA}}}. \bibinfo{publisher}{The Association for Computer Linguistics}, \bibinfo{pages}{97--104}.
\newblock
\urldef\tempurl%
\url{https://doi.org/10.3115/V1/W14-3309}
\showDOI{\tempurl}


\bibitem[Elnaggar et~al\mbox{.}(2021)]%
        {codetrans}
\bibfield{author}{\bibinfo{person}{Ahmed Elnaggar}, \bibinfo{person}{Wei Ding}, \bibinfo{person}{Llion Jones}, \bibinfo{person}{Tom Gibbs}, \bibinfo{person}{Tamas Feher}, \bibinfo{person}{Christoph Angerer}, \bibinfo{person}{Silvia Severini}, \bibinfo{person}{Florian Matthes}, {and} \bibinfo{person}{Burkhard Rost}.} \bibinfo{year}{2021}\natexlab{}.
\newblock \showarticletitle{CodeTrans: Towards Cracking the Language of Silicone's Code Through Self-Supervised Deep Learning and High Performance Computing}.
\newblock \bibinfo{journal}{\emph{CoRR}}  \bibinfo{volume}{abs/2104.02443} (\bibinfo{year}{2021}).
\newblock
\showeprint[arXiv]{2104.02443}
\urldef\tempurl%
\url{https://arxiv.org/abs/2104.02443}
\showURL{%
\tempurl}


\bibitem[Fan et~al\mbox{.}(2023)]%
        {LLM4SE}
\bibfield{author}{\bibinfo{person}{Angela Fan}, \bibinfo{person}{Beliz Gokkaya}, \bibinfo{person}{Mark Harman}, \bibinfo{person}{Mitya Lyubarskiy}, \bibinfo{person}{Shubho Sengupta}, \bibinfo{person}{Shin Yoo}, {and} \bibinfo{person}{Jie~M. Zhang}.} \bibinfo{year}{2023}\natexlab{}.
\newblock \showarticletitle{Large Language Models for Software Engineering: Survey and Open Problems}.
\newblock \bibinfo{journal}{\emph{CoRR}}  \bibinfo{volume}{abs/2310.03533} (\bibinfo{year}{2023}).
\newblock
\urldef\tempurl%
\url{https://doi.org/10.48550/ARXIV.2310.03533}
\showDOI{\tempurl}
\showeprint[arXiv]{2310.03533}


\bibitem[Feng et~al\mbox{.}(2020)]%
        {codebert}
\bibfield{author}{\bibinfo{person}{Zhangyin Feng}, \bibinfo{person}{Daya Guo}, \bibinfo{person}{Duyu Tang}, \bibinfo{person}{Nan Duan}, \bibinfo{person}{Xiaocheng Feng}, \bibinfo{person}{Ming Gong}, \bibinfo{person}{Linjun Shou}, \bibinfo{person}{Bing Qin}, \bibinfo{person}{Ting Liu}, \bibinfo{person}{Daxin Jiang}, {and} \bibinfo{person}{Ming Zhou}.} \bibinfo{year}{2020}\natexlab{}.
\newblock \showarticletitle{CodeBERT: {A} Pre-Trained Model for Programming and Natural Languages}. In \bibinfo{booktitle}{\emph{Findings of the Association for Computational Linguistics: {EMNLP} 2020, Online Event, 16-20 November 2020}} \emph{(\bibinfo{series}{Findings of {ACL}}, Vol.~\bibinfo{volume}{{EMNLP} 2020})}, \bibfield{editor}{\bibinfo{person}{Trevor Cohn}, \bibinfo{person}{Yulan He}, {and} \bibinfo{person}{Yang Liu}} (Eds.). \bibinfo{publisher}{Association for Computational Linguistics}, \bibinfo{pages}{1536--1547}.
\newblock


\bibitem[Fried et~al\mbox{.}(2023)]%
        {incoder}
\bibfield{author}{\bibinfo{person}{Daniel Fried}, \bibinfo{person}{Armen Aghajanyan}, \bibinfo{person}{Jessy Lin}, \bibinfo{person}{Sida Wang}, \bibinfo{person}{Eric Wallace}, \bibinfo{person}{Freda Shi}, \bibinfo{person}{Ruiqi Zhong}, \bibinfo{person}{Scott Yih}, \bibinfo{person}{Luke Zettlemoyer}, {and} \bibinfo{person}{Mike Lewis}.} \bibinfo{year}{2023}\natexlab{}.
\newblock \showarticletitle{InCoder: {A} Generative Model for Code Infilling and Synthesis}. In \bibinfo{booktitle}{\emph{The Eleventh International Conference on Learning Representations, {ICLR} 2023, Kigali, Rwanda, May 1-5, 2023}}. \bibinfo{publisher}{OpenReview.net}.
\newblock
\urldef\tempurl%
\url{https://openreview.net/pdf?id=hQwb-lbM6EL}
\showURL{%
\tempurl}


\bibitem[Fu et~al\mbox{.}(2023)]%
        {FuPSCK23}
\bibfield{author}{\bibinfo{person}{Yao Fu}, \bibinfo{person}{Hao Peng}, \bibinfo{person}{Ashish Sabharwal}, \bibinfo{person}{Peter Clark}, {and} \bibinfo{person}{Tushar Khot}.} \bibinfo{year}{2023}\natexlab{}.
\newblock \showarticletitle{Complexity-Based Prompting for Multi-step Reasoning}. In \bibinfo{booktitle}{\emph{The Eleventh International Conference on Learning Representations, {ICLR} 2023, Kigali, Rwanda, May 1-5, 2023}}. \bibinfo{publisher}{OpenReview.net}.
\newblock
\urldef\tempurl%
\url{https://openreview.net/pdf?id=yf1icZHC-l9}
\showURL{%
\tempurl}


\bibitem[Ganguli et~al\mbox{.}(2023)]%
        {ganguli}
\bibfield{author}{\bibinfo{person}{Deep Ganguli}, \bibinfo{person}{Amanda Askell}, \bibinfo{person}{Nicholas Schiefer}, \bibinfo{person}{Thomas~I. Liao}, \bibinfo{person}{Kamile Lukosiute}, \bibinfo{person}{Anna Chen}, \bibinfo{person}{Anna Goldie}, \bibinfo{person}{Azalia Mirhoseini}, \bibinfo{person}{Catherine Olsson}, \bibinfo{person}{Danny Hernandez}, \bibinfo{person}{Dawn Drain}, \bibinfo{person}{Dustin Li}, \bibinfo{person}{Eli Tran{-}Johnson}, \bibinfo{person}{Ethan Perez}, \bibinfo{person}{Jackson Kernion}, \bibinfo{person}{Jamie Kerr}, \bibinfo{person}{Jared Mueller}, \bibinfo{person}{Joshua Landau}, \bibinfo{person}{Kamal Ndousse}, \bibinfo{person}{Karina Nguyen}, \bibinfo{person}{Liane Lovitt}, \bibinfo{person}{Michael Sellitto}, \bibinfo{person}{Nelson Elhage}, \bibinfo{person}{Noem{\'{\i}} Mercado}, \bibinfo{person}{Nova DasSarma}, \bibinfo{person}{Oliver Rausch}, \bibinfo{person}{Robert Lasenby}, \bibinfo{person}{Robin Larson}, \bibinfo{person}{Sam Ringer}, \bibinfo{person}{Sandipan Kundu},
  \bibinfo{person}{Saurav Kadavath}, \bibinfo{person}{Scott Johnston}, \bibinfo{person}{Shauna Kravec}, \bibinfo{person}{Sheer~El Showk}, \bibinfo{person}{Tamera Lanham}, \bibinfo{person}{Timothy Telleen{-}Lawton}, \bibinfo{person}{Tom Henighan}, \bibinfo{person}{Tristan Hume}, \bibinfo{person}{Yuntao Bai}, \bibinfo{person}{Zac Hatfield{-}Dodds}, \bibinfo{person}{Ben Mann}, \bibinfo{person}{Dario Amodei}, \bibinfo{person}{Nicholas Joseph}, \bibinfo{person}{Sam McCandlish}, \bibinfo{person}{Tom Brown}, \bibinfo{person}{Christopher Olah}, \bibinfo{person}{Jack Clark}, \bibinfo{person}{Samuel~R. Bowman}, {and} \bibinfo{person}{Jared Kaplan}.} \bibinfo{year}{2023}\natexlab{}.
\newblock \showarticletitle{The Capacity for Moral Self-Correction in Large Language Models}.
\newblock \bibinfo{journal}{\emph{CoRR}}  \bibinfo{volume}{abs/2302.07459} (\bibinfo{year}{2023}).
\newblock
\urldef\tempurl%
\url{https://doi.org/10.48550/ARXIV.2302.07459}
\showDOI{\tempurl}
\showeprint[arXiv]{2302.07459}


\bibitem[Geva et~al\mbox{.}(2021)]%
        {strategyqa}
\bibfield{author}{\bibinfo{person}{Mor Geva}, \bibinfo{person}{Daniel Khashabi}, \bibinfo{person}{Elad Segal}, \bibinfo{person}{Tushar Khot}, \bibinfo{person}{Dan Roth}, {and} \bibinfo{person}{Jonathan Berant}.} \bibinfo{year}{2021}\natexlab{}.
\newblock \showarticletitle{Did Aristotle Use a Laptop? {A} Question Answering Benchmark with Implicit Reasoning Strategies}.
\newblock \bibinfo{journal}{\emph{Trans. Assoc. Comput. Linguistics}}  \bibinfo{volume}{9} (\bibinfo{year}{2021}), \bibinfo{pages}{346--361}.
\newblock
\urldef\tempurl%
\url{https://doi.org/10.1162/TACL\_A\_00370}
\showDOI{\tempurl}


\bibitem[Guo et~al\mbox{.}(2021)]%
        {graphcodebert}
\bibfield{author}{\bibinfo{person}{Daya Guo}, \bibinfo{person}{Shuo Ren}, \bibinfo{person}{Shuai Lu}, \bibinfo{person}{Zhangyin Feng}, \bibinfo{person}{Duyu Tang}, \bibinfo{person}{Shujie Liu}, \bibinfo{person}{Long Zhou}, \bibinfo{person}{Nan Duan}, \bibinfo{person}{Alexey Svyatkovskiy}, \bibinfo{person}{Shengyu Fu}, \bibinfo{person}{Michele Tufano}, \bibinfo{person}{Shao~Kun Deng}, \bibinfo{person}{Colin~B. Clement}, \bibinfo{person}{Dawn Drain}, \bibinfo{person}{Neel Sundaresan}, \bibinfo{person}{Jian Yin}, \bibinfo{person}{Daxin Jiang}, {and} \bibinfo{person}{Ming Zhou}.} \bibinfo{year}{2021}\natexlab{}.
\newblock \showarticletitle{GraphCodeBERT: Pre-training Code Representations with Data Flow}. In \bibinfo{booktitle}{\emph{9th International Conference on Learning Representations, {ICLR} 2021, Virtual Event, Austria, May 3-7, 2021}}. \bibinfo{publisher}{OpenReview.net}.
\newblock
\urldef\tempurl%
\url{https://openreview.net/forum?id=jLoC4ez43PZ}
\showURL{%
\tempurl}


\bibitem[Han et~al\mbox{.}(2023)]%
        {Han2023}
\bibfield{author}{\bibinfo{person}{Chengcheng Han}, \bibinfo{person}{Liqing Cui}, \bibinfo{person}{Renyu Zhu}, \bibinfo{person}{Jianing Wang}, \bibinfo{person}{Nuo Chen}, \bibinfo{person}{Qiushi Sun}, \bibinfo{person}{Xiang Li}, {and} \bibinfo{person}{Ming Gao}.} \bibinfo{year}{2023}\natexlab{}.
\newblock \showarticletitle{When Gradient Descent Meets Derivative-Free Optimization: {A} Match Made in Black-Box Scenario}. In \bibinfo{booktitle}{\emph{Findings of the Association for Computational Linguistics: {ACL} 2023, Toronto, Canada, July 9-14, 2023}}, \bibfield{editor}{\bibinfo{person}{Anna Rogers}, \bibinfo{person}{Jordan~L. Boyd{-}Graber}, {and} \bibinfo{person}{Naoaki Okazaki}} (Eds.). \bibinfo{publisher}{Association for Computational Linguistics}, \bibinfo{pages}{868--880}.
\newblock
\urldef\tempurl%
\url{https://doi.org/10.18653/V1/2023.FINDINGS-ACL.55}
\showDOI{\tempurl}


\bibitem[Harzing(2016)]%
        {harzing_2016}
\bibfield{author}{\bibinfo{person}{Anne-Wil Harzing}.} \bibinfo{year}{2016}\natexlab{}.
\newblock \bibinfo{title}{Publish or Perish}.
\newblock
\newblock
\urldef\tempurl%
\url{https://harzing.com/resources/publish-or-perish}
\showURL{%
\tempurl}


\bibitem[Hazhirpasand et~al\mbox{.}(2019)]%
        {HazhirpasandGKB19}
\bibfield{author}{\bibinfo{person}{Mohammadreza Hazhirpasand}, \bibinfo{person}{Mohammad Ghafari}, \bibinfo{person}{Stefan Kr{\"{u}}ger}, \bibinfo{person}{Eric Bodden}, {and} \bibinfo{person}{Oscar Nierstrasz}.} \bibinfo{year}{2019}\natexlab{}.
\newblock \showarticletitle{The Impact of Developer Experience in Using Java Cryptography}. In \bibinfo{booktitle}{\emph{2019 {ACM/IEEE} International Symposium on Empirical Software Engineering and Measurement, {ESEM} 2019, Porto de Galinhas, Recife, Brazil, September 19-20, 2019}}. \bibinfo{publisher}{{IEEE}}, \bibinfo{pages}{1--6}.
\newblock
\urldef\tempurl%
\url{https://doi.org/10.1109/ESEM.2019.8870184}
\showDOI{\tempurl}


\bibitem[He and Vechev(2023)]%
        {HeV23}
\bibfield{author}{\bibinfo{person}{Jingxuan He} {and} \bibinfo{person}{Martin~T. Vechev}.} \bibinfo{year}{2023}\natexlab{}.
\newblock \showarticletitle{Large Language Models for Code: Security Hardening and Adversarial Testing}. In \bibinfo{booktitle}{\emph{Proceedings of the 2023 {ACM} {SIGSAC} Conference on Computer and Communications Security, {CCS} 2023, Copenhagen, Denmark, November 26-30, 2023}}, \bibfield{editor}{\bibinfo{person}{Weizhi Meng}, \bibinfo{person}{Christian~Damsgaard Jensen}, \bibinfo{person}{Cas Cremers}, {and} \bibinfo{person}{Engin Kirda}} (Eds.). \bibinfo{publisher}{{ACM}}, \bibinfo{pages}{1865--1879}.
\newblock
\urldef\tempurl%
\url{https://doi.org/10.1145/3576915.3623175}
\showDOI{\tempurl}


\bibitem[Hendrycks et~al\mbox{.}(2021a)]%
        {HendrycksBKMAGB21}
\bibfield{author}{\bibinfo{person}{Dan Hendrycks}, \bibinfo{person}{Steven Basart}, \bibinfo{person}{Saurav Kadavath}, \bibinfo{person}{Mantas Mazeika}, \bibinfo{person}{Akul Arora}, \bibinfo{person}{Ethan Guo}, \bibinfo{person}{Collin Burns}, \bibinfo{person}{Samir Puranik}, \bibinfo{person}{Horace He}, \bibinfo{person}{Dawn Song}, {and} \bibinfo{person}{Jacob Steinhardt}.} \bibinfo{year}{2021}\natexlab{a}.
\newblock \showarticletitle{Measuring Coding Challenge Competence With {APPS}}. In \bibinfo{booktitle}{\emph{Proceedings of the Neural Information Processing Systems Track on Datasets and Benchmarks 1, NeurIPS Datasets and Benchmarks 2021, December 2021, virtual}}, \bibfield{editor}{\bibinfo{person}{Joaquin Vanschoren} {and} \bibinfo{person}{Sai{-}Kit Yeung}} (Eds.).
\newblock
\urldef\tempurl%
\url{https://datasets-benchmarks-proceedings.neurips.cc/paper/2021/hash/c24cd76e1ce41366a4bbe8a49b02a028-Abstract-round2.html}
\showURL{%
\tempurl}


\bibitem[Hendrycks et~al\mbox{.}(2021b)]%
        {math}
\bibfield{author}{\bibinfo{person}{Dan Hendrycks}, \bibinfo{person}{Collin Burns}, \bibinfo{person}{Saurav Kadavath}, \bibinfo{person}{Akul Arora}, \bibinfo{person}{Steven Basart}, \bibinfo{person}{Eric Tang}, \bibinfo{person}{Dawn Song}, {and} \bibinfo{person}{Jacob Steinhardt}.} \bibinfo{year}{2021}\natexlab{b}.
\newblock \showarticletitle{Measuring Mathematical Problem Solving With the {MATH} Dataset}. In \bibinfo{booktitle}{\emph{Proceedings of the Neural Information Processing Systems Track on Datasets and Benchmarks 1, NeurIPS Datasets and Benchmarks 2021, December 2021, virtual}}, \bibfield{editor}{\bibinfo{person}{Joaquin Vanschoren} {and} \bibinfo{person}{Sai{-}Kit Yeung}} (Eds.).
\newblock
\urldef\tempurl%
\url{https://datasets-benchmarks-proceedings.neurips.cc/paper/2021/hash/be83ab3ecd0db773eb2dc1b0a17836a1-Abstract-round2.html}
\showURL{%
\tempurl}


\bibitem[Hosseini et~al\mbox{.}(2014)]%
        {addsub}
\bibfield{author}{\bibinfo{person}{Mohammad~Javad Hosseini}, \bibinfo{person}{Hannaneh Hajishirzi}, \bibinfo{person}{Oren Etzioni}, {and} \bibinfo{person}{Nate Kushman}.} \bibinfo{year}{2014}\natexlab{}.
\newblock \showarticletitle{Learning to Solve Arithmetic Word Problems with Verb Categorization}. In \bibinfo{booktitle}{\emph{Proceedings of the 2014 Conference on Empirical Methods in Natural Language Processing, {EMNLP} 2014, October 25-29, 2014, Doha, Qatar, {A} meeting of SIGDAT, a Special Interest Group of the {ACL}}}, \bibfield{editor}{\bibinfo{person}{Alessandro Moschitti}, \bibinfo{person}{Bo~Pang}, {and} \bibinfo{person}{Walter Daelemans}} (Eds.). \bibinfo{publisher}{{ACL}}, \bibinfo{pages}{523--533}.
\newblock
\urldef\tempurl%
\url{https://doi.org/10.3115/V1/D14-1058}
\showDOI{\tempurl}


\bibitem[Huang and Chang(2023)]%
        {Huang2023a}
\bibfield{author}{\bibinfo{person}{Jie Huang} {and} \bibinfo{person}{Kevin~Chen{-}Chuan Chang}.} \bibinfo{year}{2023}\natexlab{}.
\newblock \showarticletitle{Towards Reasoning in Large Language Models: {A} Survey}. In \bibinfo{booktitle}{\emph{Findings of the Association for Computational Linguistics: {ACL} 2023, Toronto, Canada, July 9-14, 2023}}, \bibfield{editor}{\bibinfo{person}{Anna Rogers}, \bibinfo{person}{Jordan~L. Boyd{-}Graber}, {and} \bibinfo{person}{Naoaki Okazaki}} (Eds.). \bibinfo{publisher}{Association for Computational Linguistics}, \bibinfo{pages}{1049--1065}.
\newblock
\urldef\tempurl%
\url{https://doi.org/10.18653/V1/2023.FINDINGS-ACL.67}
\showDOI{\tempurl}


\bibitem[Ichter et~al\mbox{.}(2022)]%
        {saycan}
\bibfield{author}{\bibinfo{person}{Brian Ichter}, \bibinfo{person}{Anthony Brohan}, \bibinfo{person}{Yevgen Chebotar}, \bibinfo{person}{Chelsea Finn}, \bibinfo{person}{Karol Hausman}, \bibinfo{person}{Alexander Herzog}, \bibinfo{person}{Daniel Ho}, \bibinfo{person}{Julian Ibarz}, \bibinfo{person}{Alex Irpan}, \bibinfo{person}{Eric Jang}, \bibinfo{person}{Ryan Julian}, \bibinfo{person}{Dmitry Kalashnikov}, \bibinfo{person}{Sergey Levine}, \bibinfo{person}{Yao Lu}, \bibinfo{person}{Carolina Parada}, \bibinfo{person}{Kanishka Rao}, \bibinfo{person}{Pierre Sermanet}, \bibinfo{person}{Alexander Toshev}, \bibinfo{person}{Vincent Vanhoucke}, \bibinfo{person}{Fei Xia}, \bibinfo{person}{Ted Xiao}, \bibinfo{person}{Peng Xu}, \bibinfo{person}{Mengyuan Yan}, \bibinfo{person}{Noah Brown}, \bibinfo{person}{Michael Ahn}, \bibinfo{person}{Omar Cortes}, \bibinfo{person}{Nicolas Sievers}, \bibinfo{person}{Clayton Tan}, \bibinfo{person}{Sichun Xu}, \bibinfo{person}{Diego Reyes}, \bibinfo{person}{Jarek Rettinghouse},
  \bibinfo{person}{Jornell Quiambao}, \bibinfo{person}{Peter Pastor}, \bibinfo{person}{Linda Luu}, \bibinfo{person}{Kuang{-}Huei Lee}, \bibinfo{person}{Yuheng Kuang}, \bibinfo{person}{Sally Jesmonth}, \bibinfo{person}{Nikhil~J. Joshi}, \bibinfo{person}{Kyle Jeffrey}, \bibinfo{person}{Rosario~Jauregui Ruano}, \bibinfo{person}{Jasmine Hsu}, \bibinfo{person}{Keerthana Gopalakrishnan}, \bibinfo{person}{Byron David}, \bibinfo{person}{Andy Zeng}, {and} \bibinfo{person}{Chuyuan~Kelly Fu}.} \bibinfo{year}{2022}\natexlab{}.
\newblock \showarticletitle{Do As {I} Can, Not As {I} Say: Grounding Language in Robotic Affordances}. In \bibinfo{booktitle}{\emph{Conference on Robot Learning, CoRL 2022, 14-18 December 2022, Auckland, New Zealand}} \emph{(\bibinfo{series}{Proceedings of Machine Learning Research}, Vol.~\bibinfo{volume}{205})}, \bibfield{editor}{\bibinfo{person}{Karen Liu}, \bibinfo{person}{Dana Kulic}, {and} \bibinfo{person}{Jeffrey Ichnowski}} (Eds.). \bibinfo{publisher}{{PMLR}}, \bibinfo{pages}{287--318}.
\newblock
\urldef\tempurl%
\url{https://proceedings.mlr.press/v205/ichter23a.html}
\showURL{%
\tempurl}


\bibitem[Jain et~al\mbox{.}(2022)]%
        {jain2022jigsaw}
\bibfield{author}{\bibinfo{person}{Naman Jain}, \bibinfo{person}{Skanda Vaidyanath}, \bibinfo{person}{Arun Iyer}, \bibinfo{person}{Nagarajan Natarajan}, \bibinfo{person}{Suresh Parthasarathy}, \bibinfo{person}{Sriram Rajamani}, {and} \bibinfo{person}{Rahul Sharma}.} \bibinfo{year}{2022}\natexlab{}.
\newblock \showarticletitle{Jigsaw: Large language models meet program synthesis}. In \bibinfo{booktitle}{\emph{Proceedings of the 44th International Conference on Software Engineering (ICSE)}}. \bibinfo{pages}{1219--1231}.
\newblock


\bibitem[Jain et~al\mbox{.}(2021)]%
        {contracode}
\bibfield{author}{\bibinfo{person}{Paras Jain}, \bibinfo{person}{Ajay Jain}, \bibinfo{person}{Tianjun Zhang}, \bibinfo{person}{Pieter Abbeel}, \bibinfo{person}{Joseph Gonzalez}, {and} \bibinfo{person}{Ion Stoica}.} \bibinfo{year}{2021}\natexlab{}.
\newblock \showarticletitle{Contrastive Code Representation Learning}. In \bibinfo{booktitle}{\emph{Proceedings of the 2021 Conference on Empirical Methods in Natural Language Processing, {EMNLP} 2021, Virtual Event / Punta Cana, Dominican Republic, 7-11 November, 2021}}, \bibfield{editor}{\bibinfo{person}{Marie{-}Francine Moens}, \bibinfo{person}{Xuanjing Huang}, \bibinfo{person}{Lucia Specia}, {and} \bibinfo{person}{Scott~Wen{-}tau Yih}} (Eds.). \bibinfo{publisher}{Association for Computational Linguistics}, \bibinfo{pages}{5954--5971}.
\newblock
\urldef\tempurl%
\url{https://doi.org/10.18653/V1/2021.EMNLP-MAIN.482}
\showDOI{\tempurl}


\bibitem[Jesse et~al\mbox{.}(2023)]%
        {JesseADM23}
\bibfield{author}{\bibinfo{person}{Kevin Jesse}, \bibinfo{person}{Toufique Ahmed}, \bibinfo{person}{Premkumar~T. Devanbu}, {and} \bibinfo{person}{Emily Morgan}.} \bibinfo{year}{2023}\natexlab{}.
\newblock \showarticletitle{Large Language Models and Simple, Stupid Bugs}. In \bibinfo{booktitle}{\emph{20th {IEEE/ACM} International Conference on Mining Software Repositories, {MSR} 2023, Melbourne, Australia, May 15-16, 2023}}. \bibinfo{publisher}{{IEEE}}, \bibinfo{pages}{563--575}.
\newblock
\urldef\tempurl%
\url{https://doi.org/10.1109/MSR59073.2023.00082}
\showDOI{\tempurl}


\bibitem[Jiang et~al\mbox{.}(2023)]%
        {jiang2023selfplanning}
\bibfield{author}{\bibinfo{person}{Xue Jiang}, \bibinfo{person}{Yihong Dong}, \bibinfo{person}{Lecheng Wang}, \bibinfo{person}{Zheng Fang}, \bibinfo{person}{Qiwei Shang}, \bibinfo{person}{Ge Li}, \bibinfo{person}{Zhi Jin}, {and} \bibinfo{person}{Wenpin Jiao}.} \bibinfo{year}{2023}\natexlab{}.
\newblock \bibinfo{title}{Self-planning Code Generation with Large Language Models}.
\newblock
\newblock
\showeprint[arxiv]{2303.06689}~[cs.SE]


\bibitem[Joshi et~al\mbox{.}(2017)]%
        {triviaqa}
\bibfield{author}{\bibinfo{person}{Mandar Joshi}, \bibinfo{person}{Eunsol Choi}, \bibinfo{person}{Daniel~S. Weld}, {and} \bibinfo{person}{Luke Zettlemoyer}.} \bibinfo{year}{2017}\natexlab{}.
\newblock \showarticletitle{TriviaQA: {A} Large Scale Distantly Supervised Challenge Dataset for Reading Comprehension}. In \bibinfo{booktitle}{\emph{Proceedings of the 55th Annual Meeting of the Association for Computational Linguistics, {ACL} 2017, Vancouver, Canada, July 30 - August 4, Volume 1: Long Papers}}, \bibfield{editor}{\bibinfo{person}{Regina Barzilay} {and} \bibinfo{person}{Min{-}Yen Kan}} (Eds.). \bibinfo{publisher}{Association for Computational Linguistics}, \bibinfo{pages}{1601--1611}.
\newblock
\urldef\tempurl%
\url{https://doi.org/10.18653/V1/P17-1147}
\showDOI{\tempurl}


\bibitem[Khashabi et~al\mbox{.}(2020)]%
        {unifiedqa}
\bibfield{author}{\bibinfo{person}{Daniel Khashabi}, \bibinfo{person}{Sewon Min}, \bibinfo{person}{Tushar Khot}, \bibinfo{person}{Ashish Sabharwal}, \bibinfo{person}{Oyvind Tafjord}, \bibinfo{person}{Peter Clark}, {and} \bibinfo{person}{Hannaneh Hajishirzi}.} \bibinfo{year}{2020}\natexlab{}.
\newblock \showarticletitle{UnifiedQA: Crossing Format Boundaries With a Single {QA} System}. In \bibinfo{booktitle}{\emph{Findings of the Association for Computational Linguistics: {EMNLP} 2020, Online Event, 16-20 November 2020}} \emph{(\bibinfo{series}{Findings of {ACL}}, Vol.~\bibinfo{volume}{{EMNLP} 2020})}, \bibfield{editor}{\bibinfo{person}{Trevor Cohn}, \bibinfo{person}{Yulan He}, {and} \bibinfo{person}{Yang Liu}} (Eds.). \bibinfo{publisher}{Association for Computational Linguistics}, \bibinfo{pages}{1896--1907}.
\newblock
\urldef\tempurl%
\url{https://doi.org/10.18653/V1/2020.FINDINGS-EMNLP.171}
\showDOI{\tempurl}


\bibitem[Kim et~al\mbox{.}(2023)]%
        {KimBM23}
\bibfield{author}{\bibinfo{person}{Geunwoo Kim}, \bibinfo{person}{Pierre Baldi}, {and} \bibinfo{person}{Stephen McAleer}.} \bibinfo{year}{2023}\natexlab{}.
\newblock \showarticletitle{Language Models can Solve Computer Tasks}. In \bibinfo{booktitle}{\emph{Advances in Neural Information Processing Systems 36: Annual Conference on Neural Information Processing Systems 2023, NeurIPS 2023, New Orleans, LA, USA, December 10 - 16, 2023}}, \bibfield{editor}{\bibinfo{person}{Alice Oh}, \bibinfo{person}{Tristan Naumann}, \bibinfo{person}{Amir Globerson}, \bibinfo{person}{Kate Saenko}, \bibinfo{person}{Moritz Hardt}, {and} \bibinfo{person}{Sergey Levine}} (Eds.).
\newblock
\urldef\tempurl%
\url{http://papers.nips.cc/paper\_files/paper/2023/hash/7cc1005ec73cfbaac9fa21192b622507-Abstract-Conference.html}
\showURL{%
\tempurl}


\bibitem[Kojima et~al\mbox{.}(2022)]%
        {KojimaGRMI22}
\bibfield{author}{\bibinfo{person}{Takeshi Kojima}, \bibinfo{person}{Shixiang~Shane Gu}, \bibinfo{person}{Machel Reid}, \bibinfo{person}{Yutaka Matsuo}, {and} \bibinfo{person}{Yusuke Iwasawa}.} \bibinfo{year}{2022}\natexlab{}.
\newblock \showarticletitle{Large Language Models are Zero-Shot Reasoners}. In \bibinfo{booktitle}{\emph{Advances in Neural Information Processing Systems 35: Annual Conference on Neural Information Processing Systems 2022, NeurIPS 2022, New Orleans, LA, USA, November 28 - December 9, 2022}}, \bibfield{editor}{\bibinfo{person}{Sanmi Koyejo}, \bibinfo{person}{S.~Mohamed}, \bibinfo{person}{A.~Agarwal}, \bibinfo{person}{Danielle Belgrave}, \bibinfo{person}{K.~Cho}, {and} \bibinfo{person}{A.~Oh}} (Eds.).
\newblock
\urldef\tempurl%
\url{http://papers.nips.cc/paper\_files/paper/2022/hash/8bb0d291acd4acf06ef112099c16f326-Abstract-Conference.html}
\showURL{%
\tempurl}


\bibitem[Koncel{-}Kedziorski et~al\mbox{.}(2015)]%
        {singleeq}
\bibfield{author}{\bibinfo{person}{Rik Koncel{-}Kedziorski}, \bibinfo{person}{Hannaneh Hajishirzi}, \bibinfo{person}{Ashish Sabharwal}, \bibinfo{person}{Oren Etzioni}, {and} \bibinfo{person}{Siena~Dumas Ang}.} \bibinfo{year}{2015}\natexlab{}.
\newblock \showarticletitle{Parsing Algebraic Word Problems into Equations}.
\newblock \bibinfo{journal}{\emph{Trans. Assoc. Comput. Linguistics}}  \bibinfo{volume}{3} (\bibinfo{year}{2015}), \bibinfo{pages}{585--597}.
\newblock
\urldef\tempurl%
\url{https://doi.org/10.1162/TACL\_A\_00160}
\showDOI{\tempurl}


\bibitem[Kruskal and Wallis(1952)]%
        {kruskall-wallis}
\bibfield{author}{\bibinfo{person}{William~H. Kruskal} {and} \bibinfo{person}{W.~Allen Wallis}.} \bibinfo{year}{1952}\natexlab{}.
\newblock \showarticletitle{Use of Ranks in One-Criterion Variance Analysis}.
\newblock \bibinfo{journal}{\emph{J. Amer. Statist. Assoc.}} \bibinfo{volume}{47}, \bibinfo{number}{260} (\bibinfo{year}{1952}), \bibinfo{pages}{583--621}.
\newblock
\urldef\tempurl%
\url{https://doi.org/10.1080/01621459.1952.10483441}
\showDOI{\tempurl}


\bibitem[Kwiatkowski et~al\mbox{.}(2019)]%
        {naturalquestions}
\bibfield{author}{\bibinfo{person}{Tom Kwiatkowski}, \bibinfo{person}{Jennimaria Palomaki}, \bibinfo{person}{Olivia Redfield}, \bibinfo{person}{Michael Collins}, \bibinfo{person}{Ankur~P. Parikh}, \bibinfo{person}{Chris Alberti}, \bibinfo{person}{Danielle Epstein}, \bibinfo{person}{Illia Polosukhin}, \bibinfo{person}{Jacob Devlin}, \bibinfo{person}{Kenton Lee}, \bibinfo{person}{Kristina Toutanova}, \bibinfo{person}{Llion Jones}, \bibinfo{person}{Matthew Kelcey}, \bibinfo{person}{Ming{-}Wei Chang}, \bibinfo{person}{Andrew~M. Dai}, \bibinfo{person}{Jakob Uszkoreit}, \bibinfo{person}{Quoc Le}, {and} \bibinfo{person}{Slav Petrov}.} \bibinfo{year}{2019}\natexlab{}.
\newblock \showarticletitle{Natural Questions: a Benchmark for Question Answering Research}.
\newblock \bibinfo{journal}{\emph{Trans. Assoc. Comput. Linguistics}}  \bibinfo{volume}{7} (\bibinfo{year}{2019}), \bibinfo{pages}{452--466}.
\newblock
\urldef\tempurl%
\url{https://doi.org/10.1162/TACL\_A\_00276}
\showDOI{\tempurl}


\bibitem[Lai et~al\mbox{.}(2017)]%
        {race}
\bibfield{author}{\bibinfo{person}{Guokun Lai}, \bibinfo{person}{Qizhe Xie}, \bibinfo{person}{Hanxiao Liu}, \bibinfo{person}{Yiming Yang}, {and} \bibinfo{person}{Eduard~H. Hovy}.} \bibinfo{year}{2017}\natexlab{}.
\newblock \showarticletitle{{RACE:} Large-scale ReAding Comprehension Dataset From Examinations}. In \bibinfo{booktitle}{\emph{Proceedings of the 2017 Conference on Empirical Methods in Natural Language Processing, {EMNLP} 2017, Copenhagen, Denmark, September 9-11, 2017}}, \bibfield{editor}{\bibinfo{person}{Martha Palmer}, \bibinfo{person}{Rebecca Hwa}, {and} \bibinfo{person}{Sebastian Riedel}} (Eds.). \bibinfo{publisher}{Association for Computational Linguistics}, \bibinfo{pages}{785--794}.
\newblock
\urldef\tempurl%
\url{https://doi.org/10.18653/V1/D17-1082}
\showDOI{\tempurl}


\bibitem[Lake and Baroni(2018)]%
        {scan}
\bibfield{author}{\bibinfo{person}{Brenden~M. Lake} {and} \bibinfo{person}{Marco Baroni}.} \bibinfo{year}{2018}\natexlab{}.
\newblock \showarticletitle{Generalization without Systematicity: On the Compositional Skills of Sequence-to-Sequence Recurrent Networks}. In \bibinfo{booktitle}{\emph{Proceedings of the 35th International Conference on Machine Learning, {ICML} 2018, Stockholmsm{\"{a}}ssan, Stockholm, Sweden, July 10-15, 2018}} \emph{(\bibinfo{series}{Proceedings of Machine Learning Research}, Vol.~\bibinfo{volume}{80})}, \bibfield{editor}{\bibinfo{person}{Jennifer~G. Dy} {and} \bibinfo{person}{Andreas Krause}} (Eds.). \bibinfo{publisher}{{PMLR}}, \bibinfo{pages}{2879--2888}.
\newblock
\urldef\tempurl%
\url{http://proceedings.mlr.press/v80/lake18a.html}
\showURL{%
\tempurl}


\bibitem[Lampinen et~al\mbox{.}(2022)]%
        {LampinenDCMTCMW22}
\bibfield{author}{\bibinfo{person}{Andrew~K. Lampinen}, \bibinfo{person}{Ishita Dasgupta}, \bibinfo{person}{Stephanie C.~Y. Chan}, \bibinfo{person}{Kory~W. Mathewson}, \bibinfo{person}{Michael~Henry Tessler}, \bibinfo{person}{Antonia Creswell}, \bibinfo{person}{James~L. McClelland}, \bibinfo{person}{Jane Wang}, {and} \bibinfo{person}{Felix Hill}.} \bibinfo{year}{2022}\natexlab{}.
\newblock \showarticletitle{Can language models learn from explanations in context?}. In \bibinfo{booktitle}{\emph{Findings of the Association for Computational Linguistics: {EMNLP} 2022, Abu Dhabi, United Arab Emirates, December 7-11, 2022}}, \bibfield{editor}{\bibinfo{person}{Yoav Goldberg}, \bibinfo{person}{Zornitsa Kozareva}, {and} \bibinfo{person}{Yue Zhang}} (Eds.). \bibinfo{publisher}{Association for Computational Linguistics}, \bibinfo{pages}{537--563}.
\newblock
\urldef\tempurl%
\url{https://doi.org/10.18653/V1/2022.FINDINGS-EMNLP.38}
\showDOI{\tempurl}


\bibitem[Le et~al\mbox{.}(2022)]%
        {CodeRL}
\bibfield{author}{\bibinfo{person}{Hung Le}, \bibinfo{person}{Yue Wang}, \bibinfo{person}{Akhilesh~Deepak Gotmare}, \bibinfo{person}{Silvio Savarese}, {and} \bibinfo{person}{Steven~Chu{-}Hong Hoi}.} \bibinfo{year}{2022}\natexlab{}.
\newblock \showarticletitle{CodeRL: Mastering Code Generation through Pretrained Models and Deep Reinforcement Learning}. In \bibinfo{booktitle}{\emph{Advances in Neural Information Processing Systems 35: Annual Conference on Neural Information Processing Systems 2022, NeurIPS 2022, New Orleans, LA, USA, November 28 - December 9, 2022}}, \bibfield{editor}{\bibinfo{person}{Sanmi Koyejo}, \bibinfo{person}{S.~Mohamed}, \bibinfo{person}{A.~Agarwal}, \bibinfo{person}{Danielle Belgrave}, \bibinfo{person}{K.~Cho}, {and} \bibinfo{person}{A.~Oh}} (Eds.).
\newblock
\urldef\tempurl%
\url{http://papers.nips.cc/paper\_files/paper/2022/hash/8636419dea1aa9fbd25fc4248e702da4-Abstract-Conference.html}
\showURL{%
\tempurl}


\bibitem[Lewis~Tunstall and Wolf(2022)]%
        {codeparrot}
\bibfield{author}{\bibinfo{person}{Leandro von~Werra Lewis~Tunstall} {and} \bibinfo{person}{Thomas Wolf}.} \bibinfo{year}{2022}\natexlab{}.
\newblock \showarticletitle{Natural Language Processing with Transformers}.
\newblock \bibinfo{journal}{\emph{O’Reilly Media, Inc.}} (\bibinfo{year}{2022}).
\newblock


\bibitem[Lieber et~al\mbox{.}(2021)]%
        {Lieber2021}
\bibfield{author}{\bibinfo{person}{Opher Lieber}, \bibinfo{person}{Or Sharir}, \bibinfo{person}{Barak Lenz}, {and} \bibinfo{person}{Yoav Shoham}.} \bibinfo{year}{2021}\natexlab{}.
\newblock \showarticletitle{JURASSIC-1: TECHNICAL DETAILS AND EVALUATION}.
\newblock \bibinfo{journal}{\emph{AI21 Labs Tech. Rep.}} (\bibinfo{year}{2021}).
\newblock
\urldef\tempurl%
\url{https://uploads\-ssl.webflow.com/60fd4503684b466578c0d307/61138924626a6981ee09 caf6\_jurassic\_tech\_paper.pdf}
\showURL{%
\tempurl}


\bibitem[Lin et~al\mbox{.}(2020)]%
        {commongen}
\bibfield{author}{\bibinfo{person}{Bill~Yuchen Lin}, \bibinfo{person}{Wangchunshu Zhou}, \bibinfo{person}{Ming Shen}, \bibinfo{person}{Pei Zhou}, \bibinfo{person}{Chandra Bhagavatula}, \bibinfo{person}{Yejin Choi}, {and} \bibinfo{person}{Xiang Ren}.} \bibinfo{year}{2020}\natexlab{}.
\newblock \showarticletitle{CommonGen: {A} Constrained Text Generation Challenge for Generative Commonsense Reasoning}. In \bibinfo{booktitle}{\emph{Findings of the Association for Computational Linguistics: {EMNLP} 2020, Online Event, 16-20 November 2020}} \emph{(\bibinfo{series}{Findings of {ACL}}, Vol.~\bibinfo{volume}{{EMNLP} 2020})}, \bibfield{editor}{\bibinfo{person}{Trevor Cohn}, \bibinfo{person}{Yulan He}, {and} \bibinfo{person}{Yang Liu}} (Eds.). \bibinfo{publisher}{Association for Computational Linguistics}, \bibinfo{pages}{1823--1840}.
\newblock
\urldef\tempurl%
\url{https://doi.org/10.18653/V1/2020.FINDINGS-EMNLP.165}
\showDOI{\tempurl}


\bibitem[Ling et~al\mbox{.}(2017a)]%
        {aquarat}
\bibfield{author}{\bibinfo{person}{Wang Ling}, \bibinfo{person}{Dani Yogatama}, \bibinfo{person}{Chris Dyer}, {and} \bibinfo{person}{Phil Blunsom}.} \bibinfo{year}{2017}\natexlab{a}.
\newblock \showarticletitle{Program Induction by Rationale Generation: Learning to Solve and Explain Algebraic Word Problems}. In \bibinfo{booktitle}{\emph{Proceedings of the 55th Annual Meeting of the Association for Computational Linguistics, {ACL} 2017, Vancouver, Canada, July 30 - August 4, Volume 1: Long Papers}}, \bibfield{editor}{\bibinfo{person}{Regina Barzilay} {and} \bibinfo{person}{Min{-}Yen Kan}} (Eds.). \bibinfo{publisher}{Association for Computational Linguistics}, \bibinfo{pages}{158--167}.
\newblock
\urldef\tempurl%
\url{https://doi.org/10.18653/V1/P17-1015}
\showDOI{\tempurl}


\bibitem[Ling et~al\mbox{.}(2017b)]%
        {LingYDB17}
\bibfield{author}{\bibinfo{person}{Wang Ling}, \bibinfo{person}{Dani Yogatama}, \bibinfo{person}{Chris Dyer}, {and} \bibinfo{person}{Phil Blunsom}.} \bibinfo{year}{2017}\natexlab{b}.
\newblock \showarticletitle{Program Induction by Rationale Generation: Learning to Solve and Explain Algebraic Word Problems}. In \bibinfo{booktitle}{\emph{Proceedings of the 55th Annual Meeting of the Association for Computational Linguistics, {ACL} 2017, Vancouver, Canada, July 30 - August 4, Volume 1: Long Papers}}, \bibfield{editor}{\bibinfo{person}{Regina Barzilay} {and} \bibinfo{person}{Min{-}Yen Kan}} (Eds.). \bibinfo{publisher}{Association for Computational Linguistics}, \bibinfo{pages}{158--167}.
\newblock
\urldef\tempurl%
\url{https://doi.org/10.18653/V1/P17-1015}
\showDOI{\tempurl}


\bibitem[Liu et~al\mbox{.}(2023)]%
        {LiuXW023}
\bibfield{author}{\bibinfo{person}{Jiawei Liu}, \bibinfo{person}{Chunqiu~Steven Xia}, \bibinfo{person}{Yuyao Wang}, {and} \bibinfo{person}{Lingming Zhang}.} \bibinfo{year}{2023}\natexlab{}.
\newblock \showarticletitle{Is Your Code Generated by ChatGPT Really Correct? Rigorous Evaluation of Large Language Models for Code Generation}. In \bibinfo{booktitle}{\emph{Advances in Neural Information Processing Systems 36: Annual Conference on Neural Information Processing Systems 2023, NeurIPS 2023, New Orleans, LA, USA, December 10 - 16, 2023}}, \bibfield{editor}{\bibinfo{person}{Alice Oh}, \bibinfo{person}{Tristan Naumann}, \bibinfo{person}{Amir Globerson}, \bibinfo{person}{Kate Saenko}, \bibinfo{person}{Moritz Hardt}, {and} \bibinfo{person}{Sergey Levine}} (Eds.).
\newblock
\urldef\tempurl%
\url{http://papers.nips.cc/paper\_files/paper/2023/hash/43e9d647ccd3e4b7b5baab53f0368686-Abstract-Conference.html}
\showURL{%
\tempurl}


\bibitem[Lu et~al\mbox{.}(2021)]%
        {CodexGlue}
\bibfield{author}{\bibinfo{person}{Shuai Lu}, \bibinfo{person}{Daya Guo}, \bibinfo{person}{Shuo Ren}, \bibinfo{person}{Junjie Huang}, \bibinfo{person}{Alexey Svyatkovskiy}, \bibinfo{person}{Ambrosio Blanco}, \bibinfo{person}{Colin~B. Clement}, \bibinfo{person}{Dawn Drain}, \bibinfo{person}{Daxin Jiang}, \bibinfo{person}{Duyu Tang}, \bibinfo{person}{Ge Li}, \bibinfo{person}{Lidong Zhou}, \bibinfo{person}{Linjun Shou}, \bibinfo{person}{Long Zhou}, \bibinfo{person}{Michele Tufano}, \bibinfo{person}{Ming Gong}, \bibinfo{person}{Ming Zhou}, \bibinfo{person}{Nan Duan}, \bibinfo{person}{Neel Sundaresan}, \bibinfo{person}{Shao~Kun Deng}, \bibinfo{person}{Shengyu Fu}, {and} \bibinfo{person}{Shujie Liu}.} \bibinfo{year}{2021}\natexlab{}.
\newblock \showarticletitle{CodeXGLUE: {A} Machine Learning Benchmark Dataset for Code Understanding and Generation}. In \bibinfo{booktitle}{\emph{Proceedings of the Neural Information Processing Systems Track on Datasets and Benchmarks 1, NeurIPS Datasets and Benchmarks 2021, December 2021, virtual}}, \bibfield{editor}{\bibinfo{person}{Joaquin Vanschoren} {and} \bibinfo{person}{Sai{-}Kit Yeung}} (Eds.).
\newblock
\urldef\tempurl%
\url{https://datasets-benchmarks-proceedings.neurips.cc/paper/2021/hash/c16a5320fa475530d9583c34fd356ef5-Abstract-round1.html}
\showURL{%
\tempurl}


\bibitem[Madaan et~al\mbox{.}(2023a)]%
        {pie}
\bibfield{author}{\bibinfo{person}{Aman Madaan}, \bibinfo{person}{Alexander Shypula}, \bibinfo{person}{Uri Alon}, \bibinfo{person}{Milad Hashemi}, \bibinfo{person}{Parthasarathy Ranganathan}, \bibinfo{person}{Yiming Yang}, \bibinfo{person}{Graham Neubig}, {and} \bibinfo{person}{Amir Yazdanbakhsh}.} \bibinfo{year}{2023}\natexlab{a}.
\newblock \showarticletitle{Learning Performance-Improving Code Edits}.
\newblock \bibinfo{journal}{\emph{CoRR}}  \bibinfo{volume}{abs/2302.07867} (\bibinfo{year}{2023}).
\newblock
\urldef\tempurl%
\url{https://doi.org/10.48550/ARXIV.2302.07867}
\showDOI{\tempurl}
\showeprint[arXiv]{2302.07867}


\bibitem[Madaan et~al\mbox{.}(2023b)]%
        {MadaanTGHGW0DPY23}
\bibfield{author}{\bibinfo{person}{Aman Madaan}, \bibinfo{person}{Niket Tandon}, \bibinfo{person}{Prakhar Gupta}, \bibinfo{person}{Skyler Hallinan}, \bibinfo{person}{Luyu Gao}, \bibinfo{person}{Sarah Wiegreffe}, \bibinfo{person}{Uri Alon}, \bibinfo{person}{Nouha Dziri}, \bibinfo{person}{Shrimai Prabhumoye}, \bibinfo{person}{Yiming Yang}, \bibinfo{person}{Shashank Gupta}, \bibinfo{person}{Bodhisattwa~Prasad Majumder}, \bibinfo{person}{Katherine Hermann}, \bibinfo{person}{Sean Welleck}, \bibinfo{person}{Amir Yazdanbakhsh}, {and} \bibinfo{person}{Peter Clark}.} \bibinfo{year}{2023}\natexlab{b}.
\newblock \showarticletitle{Self-Refine: Iterative Refinement with Self-Feedback}. In \bibinfo{booktitle}{\emph{Advances in Neural Information Processing Systems 36: Annual Conference on Neural Information Processing Systems 2023, NeurIPS 2023, New Orleans, LA, USA, December 10 - 16, 2023}}, \bibfield{editor}{\bibinfo{person}{Alice Oh}, \bibinfo{person}{Tristan Naumann}, \bibinfo{person}{Amir Globerson}, \bibinfo{person}{Kate Saenko}, \bibinfo{person}{Moritz Hardt}, {and} \bibinfo{person}{Sergey Levine}} (Eds.).
\newblock
\urldef\tempurl%
\url{http://papers.nips.cc/paper\_files/paper/2023/hash/91edff07232fb1b55a505a9e9f6c0ff3-Abstract-Conference.html}
\showURL{%
\tempurl}


\bibitem[Maertens et~al\mbox{.}(2022)]%
        {MaertensPSBJDM22}
\bibfield{author}{\bibinfo{person}{Rien Maertens}, \bibinfo{person}{Charlotte~Van Petegem}, \bibinfo{person}{Niko Strijbol}, \bibinfo{person}{Toon Baeyens}, \bibinfo{person}{Arne~Carla Jacobs}, \bibinfo{person}{Peter Dawyndt}, {and} \bibinfo{person}{Bart Mesuere}.} \bibinfo{year}{2022}\natexlab{}.
\newblock \showarticletitle{Dolos: Language-agnostic plagiarism detection in source code}.
\newblock \bibinfo{journal}{\emph{J. Comput. Assist. Learn.}} \bibinfo{volume}{38}, \bibinfo{number}{4} (\bibinfo{year}{2022}), \bibinfo{pages}{1046--1061}.
\newblock
\urldef\tempurl%
\url{https://doi.org/10.1111/JCAL.12662}
\showDOI{\tempurl}


\bibitem[Marcus et~al\mbox{.}(1994)]%
        {PTB}
\bibfield{author}{\bibinfo{person}{Mitchell~P. Marcus}, \bibinfo{person}{Grace Kim}, \bibinfo{person}{Mary~Ann Marcinkiewicz}, \bibinfo{person}{Robert MacIntyre}, \bibinfo{person}{Ann Bies}, \bibinfo{person}{Mark Ferguson}, \bibinfo{person}{Karen Katz}, {and} \bibinfo{person}{Britta Schasberger}.} \bibinfo{year}{1994}\natexlab{}.
\newblock \showarticletitle{The Penn Treebank: Annotating Predicate Argument Structure}. In \bibinfo{booktitle}{\emph{Human Language Technology, Proceedings of a Workshop held at Plainsboro, New Jerey, USA, March 8-11, 1994}}. \bibinfo{publisher}{Morgan Kaufmann}.
\newblock
\urldef\tempurl%
\url{https://aclanthology.org/H94-1020/}
\showURL{%
\tempurl}


\bibitem[Mehri and Esk{\'{e}}nazi(2020)]%
        {fed}
\bibfield{author}{\bibinfo{person}{Shikib Mehri} {and} \bibinfo{person}{Maxine Esk{\'{e}}nazi}.} \bibinfo{year}{2020}\natexlab{}.
\newblock \showarticletitle{Unsupervised Evaluation of Interactive Dialog with DialoGPT}. In \bibinfo{booktitle}{\emph{Proceedings of the 21th Annual Meeting of the Special Interest Group on Discourse and Dialogue, SIGdial 2020, 1st virtual meeting, July 1-3, 2020}}, \bibfield{editor}{\bibinfo{person}{Olivier Pietquin}, \bibinfo{person}{Smaranda Muresan}, \bibinfo{person}{Vivian Chen}, \bibinfo{person}{Casey Kennington}, \bibinfo{person}{David Vandyke}, \bibinfo{person}{Nina Dethlefs}, \bibinfo{person}{Koji Inoue}, \bibinfo{person}{Erik Ekstedt}, {and} \bibinfo{person}{Stefan Ultes}} (Eds.). \bibinfo{publisher}{Association for Computational Linguistics}, \bibinfo{pages}{225--235}.
\newblock
\urldef\tempurl%
\url{https://aclanthology.org/2020.sigdial-1.28/}
\showURL{%
\tempurl}


\bibitem[Mihaylov et~al\mbox{.}(2018)]%
        {openbookqa}
\bibfield{author}{\bibinfo{person}{Todor Mihaylov}, \bibinfo{person}{Peter Clark}, \bibinfo{person}{Tushar Khot}, {and} \bibinfo{person}{Ashish Sabharwal}.} \bibinfo{year}{2018}\natexlab{}.
\newblock \showarticletitle{Can a Suit of Armor Conduct Electricity? {A} New Dataset for Open Book Question Answering}. In \bibinfo{booktitle}{\emph{Proceedings of the 2018 Conference on Empirical Methods in Natural Language Processing, Brussels, Belgium, October 31 - November 4, 2018}}, \bibfield{editor}{\bibinfo{person}{Ellen Riloff}, \bibinfo{person}{David Chiang}, \bibinfo{person}{Julia Hockenmaier}, {and} \bibinfo{person}{Jun'ichi Tsujii}} (Eds.). \bibinfo{publisher}{Association for Computational Linguistics}, \bibinfo{pages}{2381--2391}.
\newblock
\urldef\tempurl%
\url{https://doi.org/10.18653/V1/D18-1260}
\showDOI{\tempurl}


\bibitem[Mostafazadeh et~al\mbox{.}(2016)]%
        {storycloze}
\bibfield{author}{\bibinfo{person}{Nasrin Mostafazadeh}, \bibinfo{person}{Nathanael Chambers}, \bibinfo{person}{Xiaodong He}, \bibinfo{person}{Devi Parikh}, \bibinfo{person}{Dhruv Batra}, \bibinfo{person}{Lucy Vanderwende}, \bibinfo{person}{Pushmeet Kohli}, {and} \bibinfo{person}{James~F. Allen}.} \bibinfo{year}{2016}\natexlab{}.
\newblock \showarticletitle{A Corpus and Evaluation Framework for Deeper Understanding of Commonsense Stories}.
\newblock \bibinfo{journal}{\emph{CoRR}}  \bibinfo{volume}{abs/1604.01696} (\bibinfo{year}{2016}).
\newblock
\showeprint[arXiv]{1604.01696}
\urldef\tempurl%
\url{http://arxiv.org/abs/1604.01696}
\showURL{%
\tempurl}


\bibitem[Nie et~al\mbox{.}(2020)]%
        {anli}
\bibfield{author}{\bibinfo{person}{Yixin Nie}, \bibinfo{person}{Adina Williams}, \bibinfo{person}{Emily Dinan}, \bibinfo{person}{Mohit Bansal}, \bibinfo{person}{Jason Weston}, {and} \bibinfo{person}{Douwe Kiela}.} \bibinfo{year}{2020}\natexlab{}.
\newblock \showarticletitle{Adversarial {NLI:} {A} New Benchmark for Natural Language Understanding}. In \bibinfo{booktitle}{\emph{Proceedings of the 58th Annual Meeting of the Association for Computational Linguistics, {ACL} 2020, Online, July 5-10, 2020}}, \bibfield{editor}{\bibinfo{person}{Dan Jurafsky}, \bibinfo{person}{Joyce Chai}, \bibinfo{person}{Natalie Schluter}, {and} \bibinfo{person}{Joel~R. Tetreault}} (Eds.). \bibinfo{publisher}{Association for Computational Linguistics}, \bibinfo{pages}{4885--4901}.
\newblock
\urldef\tempurl%
\url{https://doi.org/10.18653/V1/2020.ACL-MAIN.441}
\showDOI{\tempurl}


\bibitem[OpenAI(2023)]%
        {GPT-4}
\bibfield{author}{\bibinfo{person}{OpenAI}.} \bibinfo{year}{2023}\natexlab{}.
\newblock \showarticletitle{{GPT-4} Technical Report}.
\newblock \bibinfo{journal}{\emph{CoRR}}  \bibinfo{volume}{abs/2303.08774} (\bibinfo{year}{2023}).
\newblock
\urldef\tempurl%
\url{https://doi.org/10.48550/arXiv.2303.08774}
\showDOI{\tempurl}
\showeprint[arXiv]{2303.08774}


\bibitem[Paperno et~al\mbox{.}(2016)]%
        {lambada}
\bibfield{author}{\bibinfo{person}{Denis Paperno}, \bibinfo{person}{Germ{\'{a}}n Kruszewski}, \bibinfo{person}{Angeliki Lazaridou}, \bibinfo{person}{Quan~Ngoc Pham}, \bibinfo{person}{Raffaella Bernardi}, \bibinfo{person}{Sandro Pezzelle}, \bibinfo{person}{Marco Baroni}, \bibinfo{person}{Gemma Boleda}, {and} \bibinfo{person}{Raquel Fern{\'{a}}ndez}.} \bibinfo{year}{2016}\natexlab{}.
\newblock \showarticletitle{The {LAMBADA} dataset: Word prediction requiring a broad discourse context}. In \bibinfo{booktitle}{\emph{Proceedings of the 54th Annual Meeting of the Association for Computational Linguistics, {ACL} 2016, August 7-12, 2016, Berlin, Germany, Volume 1: Long Papers}}. \bibinfo{publisher}{The Association for Computer Linguistics}.
\newblock
\urldef\tempurl%
\url{https://doi.org/10.18653/V1/P16-1144}
\showDOI{\tempurl}


\bibitem[Patel et~al\mbox{.}(2021)]%
        {svamp}
\bibfield{author}{\bibinfo{person}{Arkil Patel}, \bibinfo{person}{Satwik Bhattamishra}, {and} \bibinfo{person}{Navin Goyal}.} \bibinfo{year}{2021}\natexlab{}.
\newblock \showarticletitle{Are {NLP} Models really able to Solve Simple Math Word Problems?}. In \bibinfo{booktitle}{\emph{Proceedings of the 2021 Conference of the North American Chapter of the Association for Computational Linguistics: Human Language Technologies, {NAACL-HLT} 2021, Online, June 6-11, 2021}}, \bibfield{editor}{\bibinfo{person}{Kristina Toutanova}, \bibinfo{person}{Anna Rumshisky}, \bibinfo{person}{Luke Zettlemoyer}, \bibinfo{person}{Dilek Hakkani{-}T{\"{u}}r}, \bibinfo{person}{Iz~Beltagy}, \bibinfo{person}{Steven Bethard}, \bibinfo{person}{Ryan Cotterell}, \bibinfo{person}{Tanmoy Chakraborty}, {and} \bibinfo{person}{Yichao Zhou}} (Eds.). \bibinfo{publisher}{Association for Computational Linguistics}, \bibinfo{pages}{2080--2094}.
\newblock
\urldef\tempurl%
\url{https://doi.org/10.18653/V1/2021.NAACL-MAIN.168}
\showDOI{\tempurl}


\bibitem[Pearce et~al\mbox{.}(2022)]%
        {PearceA0DK22}
\bibfield{author}{\bibinfo{person}{Hammond Pearce}, \bibinfo{person}{Baleegh Ahmad}, \bibinfo{person}{Benjamin Tan}, \bibinfo{person}{Brendan Dolan{-}Gavitt}, {and} \bibinfo{person}{Ramesh Karri}.} \bibinfo{year}{2022}\natexlab{}.
\newblock \showarticletitle{Asleep at the Keyboard? Assessing the Security of GitHub Copilot's Code Contributions}. In \bibinfo{booktitle}{\emph{43rd {IEEE} Symposium on Security and Privacy, {SP} 2022, San Francisco, CA, USA, May 22-26, 2022}}. \bibinfo{publisher}{{IEEE}}, \bibinfo{pages}{754--768}.
\newblock
\urldef\tempurl%
\url{https://doi.org/10.1109/SP46214.2022.9833571}
\showDOI{\tempurl}


\bibitem[Pearce et~al\mbox{.}(2023)]%
        {Pearce2022}
\bibfield{author}{\bibinfo{person}{H. Pearce}, \bibinfo{person}{B. Tan}, \bibinfo{person}{B. Ahmad}, \bibinfo{person}{R. Karri}, {and} \bibinfo{person}{B. Dolan-Gavitt}.} \bibinfo{year}{2023}\natexlab{}.
\newblock \showarticletitle{Examining Zero-Shot Vulnerability Repair with Large Language Models}. In \bibinfo{booktitle}{\emph{2023 2023 IEEE Symposium on Security and Privacy (SP) (SP)}}. \bibinfo{publisher}{IEEE Computer Society}, \bibinfo{address}{Los Alamitos, CA, USA}, \bibinfo{pages}{1--18}.
\newblock
\urldef\tempurl%
\url{https://doi.org/10.1109/SP46215.2023.00001}
\showDOI{\tempurl}


\bibitem[Perry et~al\mbox{.}(2023)]%
        {PerryS0B23}
\bibfield{author}{\bibinfo{person}{Neil Perry}, \bibinfo{person}{Megha Srivastava}, \bibinfo{person}{Deepak Kumar}, {and} \bibinfo{person}{Dan Boneh}.} \bibinfo{year}{2023}\natexlab{}.
\newblock \showarticletitle{Do Users Write More Insecure Code with {AI} Assistants?}. In \bibinfo{booktitle}{\emph{Proceedings of the 2023 {ACM} {SIGSAC} Conference on Computer and Communications Security, {CCS} 2023, Copenhagen, Denmark, November 26-30, 2023}}, \bibfield{editor}{\bibinfo{person}{Weizhi Meng}, \bibinfo{person}{Christian~Damsgaard Jensen}, \bibinfo{person}{Cas Cremers}, {and} \bibinfo{person}{Engin Kirda}} (Eds.). \bibinfo{publisher}{{ACM}}, \bibinfo{pages}{2785--2799}.
\newblock
\urldef\tempurl%
\url{https://doi.org/10.1145/3576915.3623157}
\showDOI{\tempurl}


\bibitem[Phan et~al\mbox{.}(2021)]%
        {cotext}
\bibfield{author}{\bibinfo{person}{Long~N. Phan}, \bibinfo{person}{Hieu Tran}, \bibinfo{person}{Daniel Le}, \bibinfo{person}{Hieu Nguyen}, \bibinfo{person}{James~T. Anibal}, \bibinfo{person}{Alec Peltekian}, {and} \bibinfo{person}{Yanfang Ye}.} \bibinfo{year}{2021}\natexlab{}.
\newblock \showarticletitle{CoTexT: Multi-task Learning with Code-Text Transformer}.
\newblock \bibinfo{journal}{\emph{CoRR}}  \bibinfo{volume}{abs/2105.08645} (\bibinfo{year}{2021}).
\newblock
\showeprint[arXiv]{2105.08645}
\urldef\tempurl%
\url{https://arxiv.org/abs/2105.08645}
\showURL{%
\tempurl}


\bibitem[Prasad et~al\mbox{.}(2023)]%
        {PrasadHZB23}
\bibfield{author}{\bibinfo{person}{Archiki Prasad}, \bibinfo{person}{Peter Hase}, \bibinfo{person}{Xiang Zhou}, {and} \bibinfo{person}{Mohit Bansal}.} \bibinfo{year}{2023}\natexlab{}.
\newblock \showarticletitle{GrIPS: Gradient-free, Edit-based Instruction Search for Prompting Large Language Models}. In \bibinfo{booktitle}{\emph{Proceedings of the 17th Conference of the European Chapter of the Association for Computational Linguistics, {EACL} 2023, Dubrovnik, Croatia, May 2-6, 2023}}, \bibfield{editor}{\bibinfo{person}{Andreas Vlachos} {and} \bibinfo{person}{Isabelle Augenstein}} (Eds.). \bibinfo{publisher}{Association for Computational Linguistics}, \bibinfo{pages}{3827--3846}.
\newblock
\urldef\tempurl%
\url{https://doi.org/10.18653/V1/2023.EACL-MAIN.277}
\showDOI{\tempurl}


\bibitem[Puri et~al\mbox{.}(2021)]%
        {codenet}
\bibfield{author}{\bibinfo{person}{Ruchir Puri}, \bibinfo{person}{David~S. Kung}, \bibinfo{person}{Geert Janssen}, \bibinfo{person}{Wei Zhang}, \bibinfo{person}{Giacomo Domeniconi}, \bibinfo{person}{Vladimir Zolotov}, \bibinfo{person}{Julian Dolby}, \bibinfo{person}{Jie Chen}, \bibinfo{person}{Mihir~R. Choudhury}, \bibinfo{person}{Lindsey Decker}, \bibinfo{person}{Veronika Thost}, \bibinfo{person}{Luca Buratti}, \bibinfo{person}{Saurabh Pujar}, \bibinfo{person}{Shyam Ramji}, \bibinfo{person}{Ulrich Finkler}, \bibinfo{person}{Susan Malaika}, {and} \bibinfo{person}{Frederick Reiss}.} \bibinfo{year}{2021}\natexlab{}.
\newblock \showarticletitle{CodeNet: {A} Large-Scale {AI} for Code Dataset for Learning a Diversity of Coding Tasks}. In \bibinfo{booktitle}{\emph{Proceedings of the Neural Information Processing Systems Track on Datasets and Benchmarks 1, NeurIPS Datasets and Benchmarks 2021, December 2021, virtual}}, \bibfield{editor}{\bibinfo{person}{Joaquin Vanschoren} {and} \bibinfo{person}{Sai{-}Kit Yeung}} (Eds.).
\newblock
\urldef\tempurl%
\url{https://datasets-benchmarks-proceedings.neurips.cc/paper/2021/hash/a5bfc9e07964f8dddeb95fc584cd965d-Abstract-round2.html}
\showURL{%
\tempurl}


\bibitem[Radford et~al\mbox{.}(2019)]%
        {gpt-2}
\bibfield{author}{\bibinfo{person}{Alec Radford}, \bibinfo{person}{Jeff Wu}, \bibinfo{person}{Rewon Child}, \bibinfo{person}{David Luan}, \bibinfo{person}{Dario Amodei}, {and} \bibinfo{person}{Ilya Sutskever}.} \bibinfo{year}{2019}\natexlab{}.
\newblock \showarticletitle{Language Models are Unsupervised Multitask Learners}.
\newblock
\urldef\tempurl%
\url{https://api.semanticscholar.org/CorpusID:160025533}
\showURL{%
\tempurl}


\bibitem[Rahman et~al\mbox{.}(2019)]%
        {RahmanRW19}
\bibfield{author}{\bibinfo{person}{Md.~Rayhanur Rahman}, \bibinfo{person}{Akond Rahman}, {and} \bibinfo{person}{Laurie~A. Williams}.} \bibinfo{year}{2019}\natexlab{}.
\newblock \showarticletitle{Share, But be Aware: Security Smells in Python Gists}. In \bibinfo{booktitle}{\emph{2019 {IEEE} International Conference on Software Maintenance and Evolution, {ICSME} 2019, Cleveland, OH, USA, September 29 - October 4, 2019}}. \bibinfo{publisher}{{IEEE}}, \bibinfo{pages}{536--540}.
\newblock
\urldef\tempurl%
\url{https://doi.org/10.1109/ICSME.2019.00087}
\showDOI{\tempurl}


\bibitem[Rajpurkar et~al\mbox{.}(2018)]%
        {squad}
\bibfield{author}{\bibinfo{person}{Pranav Rajpurkar}, \bibinfo{person}{Robin Jia}, {and} \bibinfo{person}{Percy Liang}.} \bibinfo{year}{2018}\natexlab{}.
\newblock \showarticletitle{Know What You Don't Know: Unanswerable Questions for SQuAD}. In \bibinfo{booktitle}{\emph{Proceedings of the 56th Annual Meeting of the Association for Computational Linguistics, {ACL} 2018, Melbourne, Australia, July 15-20, 2018, Volume 2: Short Papers}}, \bibfield{editor}{\bibinfo{person}{Iryna Gurevych} {and} \bibinfo{person}{Yusuke Miyao}} (Eds.). \bibinfo{publisher}{Association for Computational Linguistics}, \bibinfo{pages}{784--789}.
\newblock
\urldef\tempurl%
\url{https://doi.org/10.18653/V1/P18-2124}
\showDOI{\tempurl}


\bibitem[Rauf et~al\mbox{.}(2022)]%
        {RaufPTLLLTSLRN22}
\bibfield{author}{\bibinfo{person}{Irum Rauf}, \bibinfo{person}{Marian Petre}, \bibinfo{person}{Thein Tun}, \bibinfo{person}{Tamara Lopez}, \bibinfo{person}{Paul Lunn}, \bibinfo{person}{Dirk van~der Linden}, \bibinfo{person}{John~N. Towse}, \bibinfo{person}{Helen Sharp}, \bibinfo{person}{Mark Levine}, \bibinfo{person}{Awais Rashid}, {and} \bibinfo{person}{Bashar Nuseibeh}.} \bibinfo{year}{2022}\natexlab{}.
\newblock \showarticletitle{The Case for Adaptive Security Interventions}.
\newblock \bibinfo{journal}{\emph{{ACM} Trans. Softw. Eng. Methodol.}} \bibinfo{volume}{31}, \bibinfo{number}{1} (\bibinfo{year}{2022}), \bibinfo{pages}{9:1--9:52}.
\newblock
\urldef\tempurl%
\url{https://doi.org/10.1145/3471930}
\showDOI{\tempurl}


\bibitem[Reddy et~al\mbox{.}(2019)]%
        {coqa}
\bibfield{author}{\bibinfo{person}{Siva Reddy}, \bibinfo{person}{Danqi Chen}, {and} \bibinfo{person}{Christopher~D. Manning}.} \bibinfo{year}{2019}\natexlab{}.
\newblock \showarticletitle{CoQA: {A} Conversational Question Answering Challenge}.
\newblock \bibinfo{journal}{\emph{Trans. Assoc. Comput. Linguistics}}  \bibinfo{volume}{7} (\bibinfo{year}{2019}), \bibinfo{pages}{249--266}.
\newblock
\urldef\tempurl%
\url{https://doi.org/10.1162/TACL\_A\_00266}
\showDOI{\tempurl}


\bibitem[Reynolds and McDonell(2021)]%
        {ReynoldsM21}
\bibfield{author}{\bibinfo{person}{Laria Reynolds} {and} \bibinfo{person}{Kyle McDonell}.} \bibinfo{year}{2021}\natexlab{}.
\newblock \showarticletitle{Prompt Programming for Large Language Models: Beyond the Few-Shot Paradigm}. In \bibinfo{booktitle}{\emph{{CHI} '21: {CHI} Conference on Human Factors in Computing Systems, Virtual Event / Yokohama Japan, May 8-13, 2021, Extended Abstracts}}, \bibfield{editor}{\bibinfo{person}{Yoshifumi Kitamura}, \bibinfo{person}{Aaron Quigley}, \bibinfo{person}{Katherine Isbister}, {and} \bibinfo{person}{Takeo Igarashi}} (Eds.). \bibinfo{publisher}{{ACM}}, \bibinfo{pages}{314:1--314:7}.
\newblock
\urldef\tempurl%
\url{https://doi.org/10.1145/3411763.3451760}
\showDOI{\tempurl}


\bibitem[Riddell et~al\mbox{.}(2024)]%
        {RiddellNC24}
\bibfield{author}{\bibinfo{person}{Martin Riddell}, \bibinfo{person}{Ansong Ni}, {and} \bibinfo{person}{Arman Cohan}.} \bibinfo{year}{2024}\natexlab{}.
\newblock \showarticletitle{Quantifying Contamination in Evaluating Code Generation Capabilities of Language Models}. In \bibinfo{booktitle}{\emph{Proceedings of the 62nd Annual Meeting of the Association for Computational Linguistics (Volume 1: Long Papers), {ACL} 2024, Bangkok, Thailand, August 11-16, 2024}}, \bibfield{editor}{\bibinfo{person}{Lun{-}Wei Ku}, \bibinfo{person}{Andre Martins}, {and} \bibinfo{person}{Vivek Srikumar}} (Eds.). \bibinfo{publisher}{Association for Computational Linguistics}, \bibinfo{pages}{14116--14137}.
\newblock
\urldef\tempurl%
\url{https://doi.org/10.18653/V1/2024.ACL-LONG.761}
\showDOI{\tempurl}


\bibitem[Roy and Roth(2015)]%
        {multiarith}
\bibfield{author}{\bibinfo{person}{Subhro Roy} {and} \bibinfo{person}{Dan Roth}.} \bibinfo{year}{2015}\natexlab{}.
\newblock \showarticletitle{Solving General Arithmetic Word Problems}. In \bibinfo{booktitle}{\emph{Proceedings of the 2015 Conference on Empirical Methods in Natural Language Processing, {EMNLP} 2015, Lisbon, Portugal, September 17-21, 2015}}, \bibfield{editor}{\bibinfo{person}{Llu{\'{\i}}s M{\`{a}}rquez}, \bibinfo{person}{Chris Callison{-}Burch}, \bibinfo{person}{Jian Su}, \bibinfo{person}{Daniele Pighin}, {and} \bibinfo{person}{Yuval Marton}} (Eds.). \bibinfo{publisher}{The Association for Computational Linguistics}, \bibinfo{pages}{1743--1752}.
\newblock
\urldef\tempurl%
\url{https://doi.org/10.18653/V1/D15-1202}
\showDOI{\tempurl}


\bibitem[Ruohonen et~al\mbox{.}(2021)]%
        {RuohonenHR21}
\bibfield{author}{\bibinfo{person}{Jukka Ruohonen}, \bibinfo{person}{Kalle Hjerppe}, {and} \bibinfo{person}{Kalle Rindell}.} \bibinfo{year}{2021}\natexlab{}.
\newblock \showarticletitle{A Large-Scale Security-Oriented Static Analysis of Python Packages in PyPI}. In \bibinfo{booktitle}{\emph{18th International Conference on Privacy, Security and Trust, {PST} 2021, Auckland, New Zealand, December 13-15, 2021}}. \bibinfo{publisher}{{IEEE}}, \bibinfo{pages}{1--10}.
\newblock
\urldef\tempurl%
\url{https://doi.org/10.1109/PST52912.2021.9647791}
\showDOI{\tempurl}


\bibitem[Sakaguchi et~al\mbox{.}(2021)]%
        {winogrande}
\bibfield{author}{\bibinfo{person}{Keisuke Sakaguchi}, \bibinfo{person}{Ronan~Le Bras}, \bibinfo{person}{Chandra Bhagavatula}, {and} \bibinfo{person}{Yejin Choi}.} \bibinfo{year}{2021}\natexlab{}.
\newblock \showarticletitle{WinoGrande: an adversarial winograd schema challenge at scale}.
\newblock \bibinfo{journal}{\emph{Commun. {ACM}}} \bibinfo{volume}{64}, \bibinfo{number}{9} (\bibinfo{year}{2021}), \bibinfo{pages}{99--106}.
\newblock
\urldef\tempurl%
\url{https://doi.org/10.1145/3474381}
\showDOI{\tempurl}


\bibitem[Sandoval et~al\mbox{.}(2023)]%
        {SandovalPNKGD23}
\bibfield{author}{\bibinfo{person}{Gustavo Sandoval}, \bibinfo{person}{Hammond Pearce}, \bibinfo{person}{Teo Nys}, \bibinfo{person}{Ramesh Karri}, \bibinfo{person}{Siddharth Garg}, {and} \bibinfo{person}{Brendan Dolan{-}Gavitt}.} \bibinfo{year}{2023}\natexlab{}.
\newblock \showarticletitle{Lost at {C:} {A} User Study on the Security Implications of Large Language Model Code Assistants}. In \bibinfo{booktitle}{\emph{32nd {USENIX} Security Symposium, {USENIX} Security 2023, Anaheim, CA, USA, August 9-11, 2023}}, \bibfield{editor}{\bibinfo{person}{Joseph~A. Calandrino} {and} \bibinfo{person}{Carmela Troncoso}} (Eds.). \bibinfo{publisher}{{USENIX} Association}, \bibinfo{pages}{2205--2222}.
\newblock
\urldef\tempurl%
\url{https://www.usenix.org/conference/usenixsecurity23/presentation/sandoval}
\showURL{%
\tempurl}


\bibitem[Sarkar et~al\mbox{.}(2022)]%
        {SarkarN0RP022}
\bibfield{author}{\bibinfo{person}{Advait Sarkar}, \bibinfo{person}{Carina Negreanu}, \bibinfo{person}{Ben Zorn}, \bibinfo{person}{Sruti~Srinivasa Ragavan}, \bibinfo{person}{Christian P{\"{o}}litz}, {and} \bibinfo{person}{Andrew~D. Gordon}.} \bibinfo{year}{2022}\natexlab{}.
\newblock \showarticletitle{What is it like to program with artificial intelligence?}. In \bibinfo{booktitle}{\emph{Proceedings of the 33rd Annual Workshop of the Psychology of Programming Interest Group, {PPIG} 2022, The Open University, Milton Keynes, {UK} {\&} Online, September 5-9, 2022}}, \bibfield{editor}{\bibinfo{person}{Simon Holland}, \bibinfo{person}{Marian Petre}, \bibinfo{person}{Luke Church}, {and} \bibinfo{person}{Mariana Marasoiu}} (Eds.). \bibinfo{publisher}{Psychology of Programming Interest Group}, \bibinfo{pages}{127--153}.
\newblock
\urldef\tempurl%
\url{https://ppig.org/papers/2022-ppig-33rd-sarkar/}
\showURL{%
\tempurl}


\bibitem[Saunders et~al\mbox{.}(2022)]%
        {Saunders}
\bibfield{author}{\bibinfo{person}{William Saunders}, \bibinfo{person}{Catherine Yeh}, \bibinfo{person}{Jeff Wu}, \bibinfo{person}{Steven Bills}, \bibinfo{person}{Long Ouyang}, \bibinfo{person}{Jonathan Ward}, {and} \bibinfo{person}{Jan Leike}.} \bibinfo{year}{2022}\natexlab{}.
\newblock \showarticletitle{Self-critiquing models for assisting human evaluators}.
\newblock \bibinfo{journal}{\emph{CoRR}}  \bibinfo{volume}{abs/2206.05802} (\bibinfo{year}{2022}).
\newblock
\urldef\tempurl%
\url{https://doi.org/10.48550/ARXIV.2206.05802}
\showDOI{\tempurl}
\showeprint[arXiv]{2206.05802}


\bibitem[Siddiq et~al\mbox{.}(2024)]%
        {SiddiqRZS24}
\bibfield{author}{\bibinfo{person}{Mohammed~Latif Siddiq}, \bibinfo{person}{Lindsay Roney}, \bibinfo{person}{Jiahao Zhang}, {and} \bibinfo{person}{Joanna C.~S. Santos}.} \bibinfo{year}{2024}\natexlab{}.
\newblock \showarticletitle{Quality Assessment of ChatGPT Generated Code and their Use by Developers}. In \bibinfo{booktitle}{\emph{21st {IEEE/ACM} International Conference on Mining Software Repositories, {MSR} 2024, Lisbon, Portugal, April 15-16, 2024}}, \bibfield{editor}{\bibinfo{person}{Diomidis Spinellis}, \bibinfo{person}{Alberto Bacchelli}, {and} \bibinfo{person}{Eleni Constantinou}} (Eds.). \bibinfo{publisher}{{ACM}}, \bibinfo{pages}{152--156}.
\newblock
\urldef\tempurl%
\url{https://doi.org/10.1145/3643991.3645071}
\showDOI{\tempurl}


\bibitem[Siddiq and Santos(2022)]%
        {SecurityEval}
\bibfield{author}{\bibinfo{person}{Mohammed~Latif Siddiq} {and} \bibinfo{person}{Joanna C.~S. Santos}.} \bibinfo{year}{2022}\natexlab{}.
\newblock \showarticletitle{SecurityEval dataset: mining vulnerability examples to evaluate machine learning-based code generation techniques}. In \bibinfo{booktitle}{\emph{Proceedings of the 1st International Workshop on Mining Software Repositories Applications for Privacy and Security}} (Singapore, Singapore) \emph{(\bibinfo{series}{MSR4P\&S 2022})}. \bibinfo{publisher}{Association for Computing Machinery}, \bibinfo{address}{New York, NY, USA}, \bibinfo{pages}{29–33}.
\newblock
\showISBNx{9781450394574}
\urldef\tempurl%
\url{https://doi.org/10.1145/3549035.3561184}
\showDOI{\tempurl}


\bibitem[Srivastava et~al\mbox{.}(2022)]%
        {bigbench-effort}
\bibfield{author}{\bibinfo{person}{Aarohi Srivastava}, \bibinfo{person}{Abhinav Rastogi}, \bibinfo{person}{Abhishek Rao}, \bibinfo{person}{Abu Awal~Md Shoeb}, \bibinfo{person}{Abubakar Abid}, \bibinfo{person}{Adam Fisch}, \bibinfo{person}{Adam~R. Brown}, \bibinfo{person}{Adam Santoro}, \bibinfo{person}{Aditya Gupta}, \bibinfo{person}{Adri{\`{a}} Garriga{-}Alonso}, \bibinfo{person}{Agnieszka Kluska}, \bibinfo{person}{Aitor Lewkowycz}, \bibinfo{person}{Akshat Agarwal}, \bibinfo{person}{Alethea Power}, \bibinfo{person}{Alex Ray}, \bibinfo{person}{Alex Warstadt}, \bibinfo{person}{Alexander~W. Kocurek}, \bibinfo{person}{Ali Safaya}, \bibinfo{person}{Ali Tazarv}, \bibinfo{person}{Alice Xiang}, \bibinfo{person}{Alicia Parrish}, \bibinfo{person}{Allen Nie}, \bibinfo{person}{Aman Hussain}, \bibinfo{person}{Amanda Askell}, \bibinfo{person}{Amanda Dsouza}, \bibinfo{person}{Ameet Rahane}, \bibinfo{person}{Anantharaman~S. Iyer}, \bibinfo{person}{Anders Andreassen}, \bibinfo{person}{Andrea Santilli},
  \bibinfo{person}{Andreas Stuhlm{\"{u}}ller}, \bibinfo{person}{Andrew~M. Dai}, \bibinfo{person}{Andrew La}, \bibinfo{person}{Andrew~K. Lampinen}, \bibinfo{person}{Andy Zou}, \bibinfo{person}{Angela Jiang}, \bibinfo{person}{Angelica Chen}, \bibinfo{person}{Anh Vuong}, \bibinfo{person}{Animesh Gupta}, \bibinfo{person}{Anna Gottardi}, \bibinfo{person}{Antonio Norelli}, \bibinfo{person}{Anu Venkatesh}, \bibinfo{person}{Arash Gholamidavoodi}, \bibinfo{person}{Arfa Tabassum}, \bibinfo{person}{Arul Menezes}, \bibinfo{person}{Arun Kirubarajan}, \bibinfo{person}{Asher Mullokandov}, \bibinfo{person}{Ashish Sabharwal}, \bibinfo{person}{Austin Herrick}, \bibinfo{person}{Avia Efrat}, \bibinfo{person}{Aykut Erdem}, \bibinfo{person}{Ayla Karakas}, {and} \bibinfo{person}{et al.}} \bibinfo{year}{2022}\natexlab{}.
\newblock \showarticletitle{Beyond the Imitation Game: Quantifying and extrapolating the capabilities of language models}.
\newblock \bibinfo{journal}{\emph{CoRR}}  \bibinfo{volume}{abs/2206.04615} (\bibinfo{year}{2022}).
\newblock
\urldef\tempurl%
\url{https://doi.org/10.48550/ARXIV.2206.04615}
\showDOI{\tempurl}
\showeprint[arXiv]{2206.04615}


\bibitem[Sun et~al\mbox{.}(2022)]%
        {SunSQHQ22}
\bibfield{author}{\bibinfo{person}{Tianxiang Sun}, \bibinfo{person}{Yunfan Shao}, \bibinfo{person}{Hong Qian}, \bibinfo{person}{Xuanjing Huang}, {and} \bibinfo{person}{Xipeng Qiu}.} \bibinfo{year}{2022}\natexlab{}.
\newblock \showarticletitle{Black-Box Tuning for Language-Model-as-a-Service}. In \bibinfo{booktitle}{\emph{International Conference on Machine Learning, {ICML} 2022, 17-23 July 2022, Baltimore, Maryland, {USA}}} \emph{(\bibinfo{series}{Proceedings of Machine Learning Research}, Vol.~\bibinfo{volume}{162})}, \bibfield{editor}{\bibinfo{person}{Kamalika Chaudhuri}, \bibinfo{person}{Stefanie Jegelka}, \bibinfo{person}{Le~Song}, \bibinfo{person}{Csaba Szepesv{\'{a}}ri}, \bibinfo{person}{Gang Niu}, {and} \bibinfo{person}{Sivan Sabato}} (Eds.). \bibinfo{publisher}{{PMLR}}, \bibinfo{pages}{20841--20855}.
\newblock
\urldef\tempurl%
\url{https://proceedings.mlr.press/v162/sun22e.html}
\showURL{%
\tempurl}


\bibitem[Suzgun et~al\mbox{.}(2023)]%
        {suzgunSSGTCCLCZ23}
\bibfield{author}{\bibinfo{person}{Mirac Suzgun}, \bibinfo{person}{Nathan Scales}, \bibinfo{person}{Nathanael Sch{\"{a}}rli}, \bibinfo{person}{Sebastian Gehrmann}, \bibinfo{person}{Yi Tay}, \bibinfo{person}{Hyung~Won Chung}, \bibinfo{person}{Aakanksha Chowdhery}, \bibinfo{person}{Quoc~V. Le}, \bibinfo{person}{Ed~H. Chi}, \bibinfo{person}{Denny Zhou}, {and} \bibinfo{person}{Jason Wei}.} \bibinfo{year}{2023}\natexlab{}.
\newblock \showarticletitle{Challenging BIG-Bench Tasks and Whether Chain-of-Thought Can Solve Them}. In \bibinfo{booktitle}{\emph{Findings of the Association for Computational Linguistics: {ACL} 2023, Toronto, Canada, July 9-14, 2023}}, \bibfield{editor}{\bibinfo{person}{Anna Rogers}, \bibinfo{person}{Jordan~L. Boyd{-}Graber}, {and} \bibinfo{person}{Naoaki Okazaki}} (Eds.). \bibinfo{publisher}{Association for Computational Linguistics}, \bibinfo{pages}{13003--13051}.
\newblock
\urldef\tempurl%
\url{https://doi.org/10.18653/V1/2023.FINDINGS-ACL.824}
\showDOI{\tempurl}


\bibitem[Talmor et~al\mbox{.}(2019)]%
        {commonsenseqa}
\bibfield{author}{\bibinfo{person}{Alon Talmor}, \bibinfo{person}{Jonathan Herzig}, \bibinfo{person}{Nicholas Lourie}, {and} \bibinfo{person}{Jonathan Berant}.} \bibinfo{year}{2019}\natexlab{}.
\newblock \showarticletitle{CommonsenseQA: {A} Question Answering Challenge Targeting Commonsense Knowledge}. In \bibinfo{booktitle}{\emph{Proceedings of the 2019 Conference of the North American Chapter of the Association for Computational Linguistics: Human Language Technologies, {NAACL-HLT} 2019, Minneapolis, MN, USA, June 2-7, 2019, Volume 1 (Long and Short Papers)}}, \bibfield{editor}{\bibinfo{person}{Jill Burstein}, \bibinfo{person}{Christy Doran}, {and} \bibinfo{person}{Thamar Solorio}} (Eds.). \bibinfo{publisher}{Association for Computational Linguistics}, \bibinfo{pages}{4149--4158}.
\newblock
\urldef\tempurl%
\url{https://doi.org/10.18653/V1/N19-1421}
\showDOI{\tempurl}


\bibitem[Thomas and Harden(2008)]%
        {thematic-analysis-1}
\bibfield{author}{\bibinfo{person}{James Thomas} {and} \bibinfo{person}{Angela Harden}.} \bibinfo{year}{2008}\natexlab{}.
\newblock \showarticletitle{Methods for the thematic synthesis of qualitative research in systematic reviews}.
\newblock \bibinfo{journal}{\emph{BMC Medical Research Methodology}}  \bibinfo{volume}{8} (\bibinfo{year}{2008}).
\newblock
Issue 45.


\bibitem[Tony et~al\mbox{.}(2022)]%
        {TonyFS22}
\bibfield{author}{\bibinfo{person}{Catherine Tony}, \bibinfo{person}{Nicol{\'{a}}s E.~D{\'{\i}}az Ferreyra}, {and} \bibinfo{person}{Riccardo Scandariato}.} \bibinfo{year}{2022}\natexlab{}.
\newblock \showarticletitle{GitHub Considered Harmful? Analyzing Open-Source Projects for the Automatic Generation of Cryptographic {API} Call Sequences}. In \bibinfo{booktitle}{\emph{22nd {IEEE} International Conference on Software Quality, Reliability and Security, {QRS} 2022, Guangzhou, China, December 5-9, 2022}}. \bibinfo{publisher}{{IEEE}}, \bibinfo{pages}{896--906}.
\newblock
\urldef\tempurl%
\url{https://doi.org/10.1109/QRS57517.2022.00094}
\showDOI{\tempurl}


\bibitem[Tony et~al\mbox{.}(2023)]%
        {Tony2023}
\bibfield{author}{\bibinfo{person}{Catherine Tony}, \bibinfo{person}{Markus Mutas}, \bibinfo{person}{Nicol{\'{a}}s E.~D{\'{\i}}az Ferreyra}, {and} \bibinfo{person}{Riccardo Scandariato}.} \bibinfo{year}{2023}\natexlab{}.
\newblock \showarticletitle{LLMSecEval: {A} Dataset of Natural Language Prompts for Security Evaluations}. In \bibinfo{booktitle}{\emph{20th {IEEE/ACM} International Conference on Mining Software Repositories, {MSR} 2023, Melbourne, Australia, May 15-16, 2023}}. \bibinfo{publisher}{{IEEE}}, \bibinfo{pages}{588--592}.
\newblock
\urldef\tempurl%
\url{https://doi.org/10.1109/MSR59073.2023.00084}
\showDOI{\tempurl}


\bibitem[Touvron et~al\mbox{.}(2023)]%
        {llama}
\bibfield{author}{\bibinfo{person}{Hugo Touvron}, \bibinfo{person}{Thibaut Lavril}, \bibinfo{person}{Gautier Izacard}, \bibinfo{person}{Xavier Martinet}, \bibinfo{person}{Marie{-}Anne Lachaux}, \bibinfo{person}{Timoth{\'{e}}e Lacroix}, \bibinfo{person}{Baptiste Rozi{\`{e}}re}, \bibinfo{person}{Naman Goyal}, \bibinfo{person}{Eric Hambro}, \bibinfo{person}{Faisal Azhar}, \bibinfo{person}{Aur{\'{e}}lien Rodriguez}, \bibinfo{person}{Armand Joulin}, \bibinfo{person}{Edouard Grave}, {and} \bibinfo{person}{Guillaume Lample}.} \bibinfo{year}{2023}\natexlab{}.
\newblock \showarticletitle{LLaMA: Open and Efficient Foundation Language Models}.
\newblock \bibinfo{journal}{\emph{CoRR}}  \bibinfo{volume}{abs/2302.13971} (\bibinfo{year}{2023}).
\newblock
\urldef\tempurl%
\url{https://doi.org/10.48550/ARXIV.2302.13971}
\showDOI{\tempurl}
\showeprint[arXiv]{2302.13971}


\bibitem[Turney et~al\mbox{.}(2003)]%
        {satanalogy}
\bibfield{author}{\bibinfo{person}{Peter~D. Turney}, \bibinfo{person}{Michael~L. Littman}, \bibinfo{person}{Jeffrey Bigham}, {and} \bibinfo{person}{Victor Shnayder}.} \bibinfo{year}{2003}\natexlab{}.
\newblock \showarticletitle{Combining Independent Modules to Solve Multiple-choice Synonym and Analogy Problems}.
\newblock \bibinfo{journal}{\emph{CoRR}}  \bibinfo{volume}{cs.CL/0309035} (\bibinfo{year}{2003}).
\newblock
\urldef\tempurl%
\url{http://arxiv.org/abs/cs/0309035}
\showURL{%
\tempurl}


\bibitem[Vaithilingam et~al\mbox{.}(2022)]%
        {Vaithilingam0G22}
\bibfield{author}{\bibinfo{person}{Priyan Vaithilingam}, \bibinfo{person}{Tianyi Zhang}, {and} \bibinfo{person}{Elena~L. Glassman}.} \bibinfo{year}{2022}\natexlab{}.
\newblock \showarticletitle{Expectation vs. Experience: Evaluating the Usability of Code Generation Tools Powered by Large Language Models}. In \bibinfo{booktitle}{\emph{{CHI} '22: {CHI} Conference on Human Factors in Computing Systems, New Orleans, LA, USA, 29 April 2022 - 5 May 2022, Extended Abstracts}}, \bibfield{editor}{\bibinfo{person}{Simone D.~J. Barbosa}, \bibinfo{person}{Cliff Lampe}, \bibinfo{person}{Caroline Appert}, {and} \bibinfo{person}{David~A. Shamma}} (Eds.). \bibinfo{publisher}{{ACM}}, \bibinfo{pages}{332:1--332:7}.
\newblock
\urldef\tempurl%
\url{https://doi.org/10.1145/3491101.3519665}
\showDOI{\tempurl}


\bibitem[Wang et~al\mbox{.}(2019)]%
        {superglue}
\bibfield{author}{\bibinfo{person}{Alex Wang}, \bibinfo{person}{Yada Pruksachatkun}, \bibinfo{person}{Nikita Nangia}, \bibinfo{person}{Amanpreet Singh}, \bibinfo{person}{Julian Michael}, \bibinfo{person}{Felix Hill}, \bibinfo{person}{Omer Levy}, {and} \bibinfo{person}{Samuel~R. Bowman}.} \bibinfo{year}{2019}\natexlab{}.
\newblock \showarticletitle{SuperGLUE: {A} Stickier Benchmark for General-Purpose Language Understanding Systems}. In \bibinfo{booktitle}{\emph{Advances in Neural Information Processing Systems 32: Annual Conference on Neural Information Processing Systems 2019, NeurIPS 2019, December 8-14, 2019, Vancouver, BC, Canada}}, \bibfield{editor}{\bibinfo{person}{Hanna~M. Wallach}, \bibinfo{person}{Hugo Larochelle}, \bibinfo{person}{Alina Beygelzimer}, \bibinfo{person}{Florence d'Alch{\'{e}}{-}Buc}, \bibinfo{person}{Emily~B. Fox}, {and} \bibinfo{person}{Roman Garnett}} (Eds.). \bibinfo{pages}{3261--3275}.
\newblock
\urldef\tempurl%
\url{https://proceedings.neurips.cc/paper/2019/hash/4496bf24afe7fab6f046bf4923da8de6-Abstract.html}
\showURL{%
\tempurl}


\bibitem[Wang et~al\mbox{.}(2022)]%
        {Wang0S22}
\bibfield{author}{\bibinfo{person}{Boshi Wang}, \bibinfo{person}{Xiang Deng}, {and} \bibinfo{person}{Huan Sun}.} \bibinfo{year}{2022}\natexlab{}.
\newblock \showarticletitle{Iteratively Prompt Pre-trained Language Models for Chain of Thought}. In \bibinfo{booktitle}{\emph{Proceedings of the 2022 Conference on Empirical Methods in Natural Language Processing, {EMNLP} 2022, Abu Dhabi, United Arab Emirates, December 7-11, 2022}}, \bibfield{editor}{\bibinfo{person}{Yoav Goldberg}, \bibinfo{person}{Zornitsa Kozareva}, {and} \bibinfo{person}{Yue Zhang}} (Eds.). \bibinfo{publisher}{Association for Computational Linguistics}, \bibinfo{pages}{2714--2730}.
\newblock
\urldef\tempurl%
\url{https://aclanthology.org/2022.emnlp-main.174}
\showURL{%
\tempurl}


\bibitem[Wang et~al\mbox{.}(2023)]%
        {WangWSLCNCZ23}
\bibfield{author}{\bibinfo{person}{Xuezhi Wang}, \bibinfo{person}{Jason Wei}, \bibinfo{person}{Dale Schuurmans}, \bibinfo{person}{Quoc~V. Le}, \bibinfo{person}{Ed~H. Chi}, \bibinfo{person}{Sharan Narang}, \bibinfo{person}{Aakanksha Chowdhery}, {and} \bibinfo{person}{Denny Zhou}.} \bibinfo{year}{2023}\natexlab{}.
\newblock \showarticletitle{Self-Consistency Improves Chain of Thought Reasoning in Language Models}. In \bibinfo{booktitle}{\emph{The Eleventh International Conference on Learning Representations, {ICLR} 2023, Kigali, Rwanda, May 1-5, 2023}}. \bibinfo{publisher}{OpenReview.net}.
\newblock
\urldef\tempurl%
\url{https://openreview.net/pdf?id=1PL1NIMMrw}
\showURL{%
\tempurl}


\bibitem[Wang et~al\mbox{.}(2021)]%
        {codet5}
\bibfield{author}{\bibinfo{person}{Yue Wang}, \bibinfo{person}{Weishi Wang}, \bibinfo{person}{Shafiq~R. Joty}, {and} \bibinfo{person}{Steven C.~H. Hoi}.} \bibinfo{year}{2021}\natexlab{}.
\newblock \showarticletitle{CodeT5: Identifier-aware Unified Pre-trained Encoder-Decoder Models for Code Understanding and Generation}. In \bibinfo{booktitle}{\emph{Proceedings of the 2021 Conference on Empirical Methods in Natural Language Processing, {EMNLP} 2021, Virtual Event / Punta Cana, Dominican Republic, 7-11 November, 2021}}, \bibfield{editor}{\bibinfo{person}{Marie{-}Francine Moens}, \bibinfo{person}{Xuanjing Huang}, \bibinfo{person}{Lucia Specia}, {and} \bibinfo{person}{Scott~Wen{-}tau Yih}} (Eds.). \bibinfo{publisher}{Association for Computational Linguistics}, \bibinfo{pages}{8696--8708}.
\newblock
\urldef\tempurl%
\url{https://doi.org/10.18653/v1/2021.emnlp-main.685}
\showDOI{\tempurl}


\bibitem[Wei et~al\mbox{.}(2022)]%
        {Wei2022}
\bibfield{author}{\bibinfo{person}{Jason Wei}, \bibinfo{person}{Xuezhi Wang}, \bibinfo{person}{Dale Schuurmans}, \bibinfo{person}{Maarten Bosma}, \bibinfo{person}{Brian Ichter}, \bibinfo{person}{Fei Xia}, \bibinfo{person}{Ed~H. Chi}, \bibinfo{person}{Quoc~V. Le}, {and} \bibinfo{person}{Denny Zhou}.} \bibinfo{year}{2022}\natexlab{}.
\newblock \showarticletitle{Chain-of-Thought Prompting Elicits Reasoning in Large Language Models}. In \bibinfo{booktitle}{\emph{NeurIPS}}.
\newblock
\urldef\tempurl%
\url{http://papers.nips.cc/paper\_files/paper/2022/hash/9d5609613524ecf4f15af0f7b31abca4-Abstract-Conference.html}
\showURL{%
\tempurl}


\bibitem[White et~al\mbox{.}(2023a)]%
        {White2023a}
\bibfield{author}{\bibinfo{person}{Jules White}, \bibinfo{person}{Quchen Fu}, \bibinfo{person}{Sam Hays}, \bibinfo{person}{Michael Sandborn}, \bibinfo{person}{Carlos Olea}, \bibinfo{person}{Henry Gilbert}, \bibinfo{person}{Ashraf Elnashar}, \bibinfo{person}{Jesse Spencer{-}Smith}, {and} \bibinfo{person}{Douglas~C. Schmidt}.} \bibinfo{year}{2023}\natexlab{a}.
\newblock \showarticletitle{A Prompt Pattern Catalog to Enhance Prompt Engineering with ChatGPT}.
\newblock \bibinfo{journal}{\emph{CoRR}}  \bibinfo{volume}{abs/2302.11382} (\bibinfo{year}{2023}).
\newblock
\urldef\tempurl%
\url{https://doi.org/10.48550/arXiv.2302.11382}
\showDOI{\tempurl}
\showeprint[arXiv]{2302.11382}


\bibitem[White et~al\mbox{.}(2023b)]%
        {White2023b}
\bibfield{author}{\bibinfo{person}{Jules White}, \bibinfo{person}{Sam Hays}, \bibinfo{person}{Quchen Fu}, \bibinfo{person}{Jesse Spencer{-}Smith}, {and} \bibinfo{person}{Douglas~C. Schmidt}.} \bibinfo{year}{2023}\natexlab{b}.
\newblock \showarticletitle{ChatGPT Prompt Patterns for Improving Code Quality, Refactoring, Requirements Elicitation, and Software Design}.
\newblock \bibinfo{journal}{\emph{CoRR}}  \bibinfo{volume}{abs/2303.07839} (\bibinfo{year}{2023}).
\newblock
\urldef\tempurl%
\url{https://doi.org/10.48550/arXiv.2303.07839}
\showDOI{\tempurl}
\showeprint[arXiv]{2303.07839}


\bibitem[Wickert et~al\mbox{.}(2021)]%
        {WickertBBM21}
\bibfield{author}{\bibinfo{person}{Anna{-}Katharina Wickert}, \bibinfo{person}{Lars Baumg{\"{a}}rtner}, \bibinfo{person}{Florian Breitfelder}, {and} \bibinfo{person}{Mira Mezini}.} \bibinfo{year}{2021}\natexlab{}.
\newblock \showarticletitle{Python Crypto Misuses in the Wild}. In \bibinfo{booktitle}{\emph{{ESEM} '21: {ACM} / {IEEE} International Symposium on Empirical Software Engineering and Measurement, Bari, Italy, October 11-15, 2021}}, \bibfield{editor}{\bibinfo{person}{Filippo Lanubile}, \bibinfo{person}{Marcos Kalinowski}, {and} \bibinfo{person}{Maria~Teresa Baldassarre}} (Eds.). \bibinfo{publisher}{{ACM}}, \bibinfo{pages}{31:1--31:6}.
\newblock
\urldef\tempurl%
\url{https://doi.org/10.1145/3475716.3484195}
\showDOI{\tempurl}


\bibitem[Wickert et~al\mbox{.}(2019)]%
        {WickertREDM19}
\bibfield{author}{\bibinfo{person}{Anna{-}Katharina Wickert}, \bibinfo{person}{Michael Reif}, \bibinfo{person}{Michael Eichberg}, \bibinfo{person}{Anam Dodhy}, {and} \bibinfo{person}{Mira Mezini}.} \bibinfo{year}{2019}\natexlab{}.
\newblock \showarticletitle{A dataset of parametric cryptographic misuses}. In \bibinfo{booktitle}{\emph{Proceedings of the 16th International Conference on Mining Software Repositories, {MSR} 2019, 26-27 May 2019, Montreal, Canada}}, \bibfield{editor}{\bibinfo{person}{Margaret{-}Anne~D. Storey}, \bibinfo{person}{Bram Adams}, {and} \bibinfo{person}{Sonia Haiduc}} (Eds.). \bibinfo{publisher}{{IEEE} / {ACM}}, \bibinfo{pages}{96--100}.
\newblock
\urldef\tempurl%
\url{https://doi.org/10.1109/MSR.2019.00023}
\showDOI{\tempurl}


\bibitem[Wohlin(2014)]%
        {Wohlin14}
\bibfield{author}{\bibinfo{person}{Claes Wohlin}.} \bibinfo{year}{2014}\natexlab{}.
\newblock \showarticletitle{Guidelines for snowballing in systematic literature studies and a replication in software engineering}. In \bibinfo{booktitle}{\emph{18th International Conference on Evaluation and Assessment in Software Engineering, {EASE} '14, London, England, United Kingdom, May 13-14, 2014}}, \bibfield{editor}{\bibinfo{person}{Martin~J. Shepperd}, \bibinfo{person}{Tracy Hall}, {and} \bibinfo{person}{Ingunn Myrtveit}} (Eds.). \bibinfo{publisher}{{ACM}}, \bibinfo{pages}{38:1--38:10}.
\newblock
\urldef\tempurl%
\url{https://doi.org/10.1145/2601248.2601268}
\showDOI{\tempurl}


\bibitem[Xiao and Watson(2019)]%
        {thematic-analysis-2}
\bibfield{author}{\bibinfo{person}{Yu Xiao} {and} \bibinfo{person}{Maria Watson}.} \bibinfo{year}{2019}\natexlab{}.
\newblock \showarticletitle{Guidance on Conducting a Systematic Literature Review}.
\newblock \bibinfo{journal}{\emph{Journal of Planning Education and Research}}  \bibinfo{volume}{39} (\bibinfo{year}{2019}), \bibinfo{pages}{93--112}.
\newblock
Issue 1.
\urldef\tempurl%
\url{https://doi.org/10.1177/0739456X17723971}
\showDOI{\tempurl}


\bibitem[Xu et~al\mbox{.}(2022a)]%
        {Xu0NH22}
\bibfield{author}{\bibinfo{person}{Frank~F. Xu}, \bibinfo{person}{Uri Alon}, \bibinfo{person}{Graham Neubig}, {and} \bibinfo{person}{Vincent~Josua Hellendoorn}.} \bibinfo{year}{2022}\natexlab{a}.
\newblock \showarticletitle{A systematic evaluation of large language models of code}. In \bibinfo{booktitle}{\emph{MAPS@PLDI 2022: 6th {ACM} {SIGPLAN} International Symposium on Machine Programming, San Diego, CA, USA, 13 June 2022}}, \bibfield{editor}{\bibinfo{person}{Swarat Chaudhuri} {and} \bibinfo{person}{Charles Sutton}} (Eds.). \bibinfo{publisher}{{ACM}}, \bibinfo{pages}{1--10}.
\newblock
\urldef\tempurl%
\url{https://doi.org/10.1145/3520312.3534862}
\showDOI{\tempurl}


\bibitem[Xu et~al\mbox{.}(2022b)]%
        {XuCDSWLY22}
\bibfield{author}{\bibinfo{person}{Hanwei Xu}, \bibinfo{person}{Yujun Chen}, \bibinfo{person}{Yulun Du}, \bibinfo{person}{Nan Shao}, \bibinfo{person}{Yanggang Wang}, \bibinfo{person}{Haiyu Li}, {and} \bibinfo{person}{Zhilin Yang}.} \bibinfo{year}{2022}\natexlab{b}.
\newblock \showarticletitle{{GPS:} Genetic Prompt Search for Efficient Few-Shot Learning}. In \bibinfo{booktitle}{\emph{Proceedings of the 2022 Conference on Empirical Methods in Natural Language Processing, {EMNLP} 2022, Abu Dhabi, United Arab Emirates, December 7-11, 2022}}, \bibfield{editor}{\bibinfo{person}{Yoav Goldberg}, \bibinfo{person}{Zornitsa Kozareva}, {and} \bibinfo{person}{Yue Zhang}} (Eds.). \bibinfo{publisher}{Association for Computational Linguistics}, \bibinfo{pages}{8162--8171}.
\newblock
\urldef\tempurl%
\url{https://doi.org/10.18653/V1/2022.EMNLP-MAIN.559}
\showDOI{\tempurl}


\bibitem[Yao et~al\mbox{.}(2023b)]%
        {YaoYZS00N23}
\bibfield{author}{\bibinfo{person}{Shunyu Yao}, \bibinfo{person}{Dian Yu}, \bibinfo{person}{Jeffrey Zhao}, \bibinfo{person}{Izhak Shafran}, \bibinfo{person}{Tom Griffiths}, \bibinfo{person}{Yuan Cao}, {and} \bibinfo{person}{Karthik Narasimhan}.} \bibinfo{year}{2023}\natexlab{b}.
\newblock \showarticletitle{Tree of Thoughts: Deliberate Problem Solving with Large Language Models}. In \bibinfo{booktitle}{\emph{Advances in Neural Information Processing Systems 36: Annual Conference on Neural Information Processing Systems 2023, NeurIPS 2023, New Orleans, LA, USA, December 10 - 16, 2023}}, \bibfield{editor}{\bibinfo{person}{Alice Oh}, \bibinfo{person}{Tristan Naumann}, \bibinfo{person}{Amir Globerson}, \bibinfo{person}{Kate Saenko}, \bibinfo{person}{Moritz Hardt}, {and} \bibinfo{person}{Sergey Levine}} (Eds.).
\newblock
\urldef\tempurl%
\url{http://papers.nips.cc/paper\_files/paper/2023/hash/271db9922b8d1f4dd7aaef84ed5ac703-Abstract-Conference.html}
\showURL{%
\tempurl}


\bibitem[Yao et~al\mbox{.}(2023a)]%
        {Yao2023}
\bibfield{author}{\bibinfo{person}{Yifan Yao}, \bibinfo{person}{Jinhao Duan}, \bibinfo{person}{Kaidi Xu}, \bibinfo{person}{Yuanfang Cai}, \bibinfo{person}{Eric Sun}, {and} \bibinfo{person}{Yue Zhang}.} \bibinfo{year}{2023}\natexlab{a}.
\newblock \showarticletitle{A Survey on Large Language Model {(LLM)} Security and Privacy: The Good, the Bad, and the Ugly}.
\newblock \bibinfo{journal}{\emph{CoRR}}  \bibinfo{volume}{abs/2312.02003} (\bibinfo{year}{2023}).
\newblock
\urldef\tempurl%
\url{https://doi.org/10.48550/ARXIV.2312.02003}
\showDOI{\tempurl}
\showeprint[arXiv]{2312.02003}


\bibitem[Ye et~al\mbox{.}(2023)]%
        {YeCG23}
\bibfield{author}{\bibinfo{person}{He Ye}, \bibinfo{person}{Zimin Chen}, {and} \bibinfo{person}{Claire {Le Goues}}.} \bibinfo{year}{2023}\natexlab{}.
\newblock \showarticletitle{PreciseBugCollector: Extensible, Executable and Precise Bug-Fix Collection: Solution for Challenge 8: Automating Precise Data Collection for Code Snippets with Bugs, Fixes, Locations, and Types}. In \bibinfo{booktitle}{\emph{38th {IEEE/ACM} International Conference on Automated Software Engineering, {ASE} 2023, Luxembourg, September 11-15, 2023}}. \bibinfo{publisher}{{IEEE}}, \bibinfo{pages}{1899--1910}.
\newblock
\urldef\tempurl%
\url{https://doi.org/10.1109/ASE56229.2023.00163}
\showDOI{\tempurl}


\bibitem[Yetiştiren et~al\mbox{.}(2023)]%
        {yetistiren}
\bibfield{author}{\bibinfo{person}{Burak Yetiştiren}, \bibinfo{person}{Işık Özsoy}, \bibinfo{person}{Miray Ayerdem}, {and} \bibinfo{person}{Eray Tüzün}.} \bibinfo{year}{2023}\natexlab{}.
\newblock \bibinfo{title}{Evaluating the Code Quality of AI-Assisted Code Generation Tools: An Empirical Study on GitHub Copilot, Amazon CodeWhisperer, and ChatGPT}.
\newblock
\newblock
\showeprint[arxiv]{2304.10778}~[cs.SE]


\bibitem[Yu et~al\mbox{.}(2023)]%
        {YuW0WVX23}
\bibfield{author}{\bibinfo{person}{Zhiyuan Yu}, \bibinfo{person}{Yuhao Wu}, \bibinfo{person}{Ning Zhang}, \bibinfo{person}{Chenguang Wang}, \bibinfo{person}{Yevgeniy Vorobeychik}, {and} \bibinfo{person}{Chaowei Xiao}.} \bibinfo{year}{2023}\natexlab{}.
\newblock \showarticletitle{CodeIPPrompt: Intellectual Property Infringement Assessment of Code Language Models}. In \bibinfo{booktitle}{\emph{International Conference on Machine Learning, {ICML} 2023, 23-29 July 2023, Honolulu, Hawaii, {USA}}} \emph{(\bibinfo{series}{Proceedings of Machine Learning Research}, Vol.~\bibinfo{volume}{202})}, \bibfield{editor}{\bibinfo{person}{Andreas Krause}, \bibinfo{person}{Emma Brunskill}, \bibinfo{person}{Kyunghyun Cho}, \bibinfo{person}{Barbara Engelhardt}, \bibinfo{person}{Sivan Sabato}, {and} \bibinfo{person}{Jonathan Scarlett}} (Eds.). \bibinfo{publisher}{{PMLR}}, \bibinfo{pages}{40373--40389}.
\newblock
\urldef\tempurl%
\url{https://proceedings.mlr.press/v202/yu23g.html}
\showURL{%
\tempurl}


\bibitem[Zellers et~al\mbox{.}(2019)]%
        {hellaswag}
\bibfield{author}{\bibinfo{person}{Rowan Zellers}, \bibinfo{person}{Ari Holtzman}, \bibinfo{person}{Yonatan Bisk}, \bibinfo{person}{Ali Farhadi}, {and} \bibinfo{person}{Yejin Choi}.} \bibinfo{year}{2019}\natexlab{}.
\newblock \showarticletitle{HellaSwag: Can a Machine Really Finish Your Sentence?}. In \bibinfo{booktitle}{\emph{Proceedings of the 57th Conference of the Association for Computational Linguistics, {ACL} 2019, Florence, Italy, July 28- August 2, 2019, Volume 1: Long Papers}}, \bibfield{editor}{\bibinfo{person}{Anna Korhonen}, \bibinfo{person}{David~R. Traum}, {and} \bibinfo{person}{Llu{\'{\i}}s M{\`{a}}rquez}} (Eds.). \bibinfo{publisher}{Association for Computational Linguistics}, \bibinfo{pages}{4791--4800}.
\newblock
\urldef\tempurl%
\url{https://doi.org/10.18653/V1/P19-1472}
\showDOI{\tempurl}


\bibitem[Zeng et~al\mbox{.}(2022)]%
        {ZengTZLZZ22}
\bibfield{author}{\bibinfo{person}{Zhengran Zeng}, \bibinfo{person}{Hanzhuo Tan}, \bibinfo{person}{Haotian Zhang}, \bibinfo{person}{Jing Li}, \bibinfo{person}{Yuqun Zhang}, {and} \bibinfo{person}{Lingming Zhang}.} \bibinfo{year}{2022}\natexlab{}.
\newblock \showarticletitle{An extensive study on pre-trained models for program understanding and generation}. In \bibinfo{booktitle}{\emph{{ISSTA} '22: 31st {ACM} {SIGSOFT} International Symposium on Software Testing and Analysis, Virtual Event, South Korea, July 18 - 22, 2022}}, \bibfield{editor}{\bibinfo{person}{Sukyoung Ryu} {and} \bibinfo{person}{Yannis Smaragdakis}} (Eds.). \bibinfo{publisher}{{ACM}}, \bibinfo{pages}{39--51}.
\newblock
\urldef\tempurl%
\url{https://doi.org/10.1145/3533767.3534390}
\showDOI{\tempurl}


\bibitem[Zhang et~al\mbox{.}(2015)]%
        {yelp-reviews}
\bibfield{author}{\bibinfo{person}{Xiang Zhang}, \bibinfo{person}{Junbo~Jake Zhao}, {and} \bibinfo{person}{Yann LeCun}.} \bibinfo{year}{2015}\natexlab{}.
\newblock \showarticletitle{Character-level Convolutional Networks for Text Classification}. In \bibinfo{booktitle}{\emph{Advances in Neural Information Processing Systems 28: Annual Conference on Neural Information Processing Systems 2015, December 7-12, 2015, Montreal, Quebec, Canada}}, \bibfield{editor}{\bibinfo{person}{Corinna Cortes}, \bibinfo{person}{Neil~D. Lawrence}, \bibinfo{person}{Daniel~D. Lee}, \bibinfo{person}{Masashi Sugiyama}, {and} \bibinfo{person}{Roman Garnett}} (Eds.). \bibinfo{pages}{649--657}.
\newblock
\urldef\tempurl%
\url{https://proceedings.neurips.cc/paper/2015/hash/250cf8b51c773f3f8dc8b4be867a9a02-Abstract.html}
\showURL{%
\tempurl}


\bibitem[Zheng et~al\mbox{.}(2023a)]%
        {Zheng2023}
\bibfield{author}{\bibinfo{person}{Chuanyang Zheng}, \bibinfo{person}{Zhengying Liu}, \bibinfo{person}{Enze Xie}, \bibinfo{person}{Zhenguo Li}, {and} \bibinfo{person}{Yu Li}.} \bibinfo{year}{2023}\natexlab{a}.
\newblock \showarticletitle{Progressive-Hint Prompting Improves Reasoning in Large Language Models}.
\newblock \bibinfo{journal}{\emph{CoRR}}  \bibinfo{volume}{abs/2304.09797} (\bibinfo{year}{2023}).
\newblock
\urldef\tempurl%
\url{https://doi.org/10.48550/ARXIV.2304.09797}
\showDOI{\tempurl}
\showeprint[arXiv]{2304.09797}


\bibitem[Zheng et~al\mbox{.}(2023b)]%
        {humaneval-x}
\bibfield{author}{\bibinfo{person}{Qinkai Zheng}, \bibinfo{person}{Xiao Xia}, \bibinfo{person}{Xu Zou}, \bibinfo{person}{Yuxiao Dong}, \bibinfo{person}{Shan Wang}, \bibinfo{person}{Yufei Xue}, \bibinfo{person}{Zihan Wang}, \bibinfo{person}{Lei Shen}, \bibinfo{person}{Andi Wang}, \bibinfo{person}{Yang Li}, \bibinfo{person}{Teng Su}, \bibinfo{person}{Zhilin Yang}, {and} \bibinfo{person}{Jie Tang}.} \bibinfo{year}{2023}\natexlab{b}.
\newblock \showarticletitle{CodeGeeX: {A} Pre-Trained Model for Code Generation with Multilingual Evaluations on HumanEval-X}.
\newblock \bibinfo{journal}{\emph{CoRR}}  \bibinfo{volume}{abs/2303.17568} (\bibinfo{year}{2023}).
\newblock
\urldef\tempurl%
\url{https://doi.org/10.48550/ARXIV.2303.17568}
\showDOI{\tempurl}
\showeprint[arXiv]{2303.17568}


\bibitem[Zhou et~al\mbox{.}(2023b)]%
        {ZhouSHWS0SCBLC23}
\bibfield{author}{\bibinfo{person}{Denny Zhou}, \bibinfo{person}{Nathanael Sch{\"{a}}rli}, \bibinfo{person}{Le Hou}, \bibinfo{person}{Jason Wei}, \bibinfo{person}{Nathan Scales}, \bibinfo{person}{Xuezhi Wang}, \bibinfo{person}{Dale Schuurmans}, \bibinfo{person}{Claire Cui}, \bibinfo{person}{Olivier Bousquet}, \bibinfo{person}{Quoc~V. Le}, {and} \bibinfo{person}{Ed~H. Chi}.} \bibinfo{year}{2023}\natexlab{b}.
\newblock \showarticletitle{Least-to-Most Prompting Enables Complex Reasoning in Large Language Models}. In \bibinfo{booktitle}{\emph{The Eleventh International Conference on Learning Representations, {ICLR} 2023, Kigali, Rwanda, May 1-5, 2023}}. \bibinfo{publisher}{OpenReview.net}.
\newblock
\urldef\tempurl%
\url{https://openreview.net/pdf?id=WZH7099tgfM}
\showURL{%
\tempurl}


\bibitem[Zhou et~al\mbox{.}(2023a)]%
        {ZhouMHPPCB23}
\bibfield{author}{\bibinfo{person}{Yongchao Zhou}, \bibinfo{person}{Andrei~Ioan Muresanu}, \bibinfo{person}{Ziwen Han}, \bibinfo{person}{Keiran Paster}, \bibinfo{person}{Silviu Pitis}, \bibinfo{person}{Harris Chan}, {and} \bibinfo{person}{Jimmy Ba}.} \bibinfo{year}{2023}\natexlab{a}.
\newblock \showarticletitle{Large Language Models are Human-Level Prompt Engineers}. In \bibinfo{booktitle}{\emph{The Eleventh International Conference on Learning Representations, {ICLR} 2023, Kigali, Rwanda, May 1-5, 2023}}. \bibinfo{publisher}{OpenReview.net}.
\newblock
\urldef\tempurl%
\url{https://openreview.net/pdf?id=92gvk82DE-}
\showURL{%
\tempurl}


\end{thebibliography}

\appendix

\section{Manual Verification of SAST Tool Results}
\label{appendix:manual_security_validation}

\minorrev{Below, we detail the full process followed to manually verify the security analysis results of the SAST tools (Bandit and CodeQL).}

\subsection{Step 1: Manual Security Weakness Detection}
\minorrev{We began by manually reviewing the code snippets (for the randomly selected tasks) independently of the SAST tool results to identify any of the 18 security weaknesses documented in LLMSecEval. This process involved consulting the official CWE pages on the MITRE website, specifically examining the provided potential mitigations and examples of vulnerable code for each weakness. A code snippet was flagged as vulnerable if it lacked the recommended mitigation techniques to prevent the vulnerability or closely resembled the examples of vulnerable code.
Consider the example of \textit{CWE-78: OS Command Injection}. One of the mitigation techniques outlined on the CWE page is as follows\footnote{https://cwe.mitre.org/data/definitions/78.html}:}

\begin{quote}
    \minorrev{\textbf{Output encoding:} \textit{While it is risky to use dynamically-generated query strings, code, or commands that mix control and data together, sometimes it may be unavoidable. Properly quote arguments and escape any special characters within those arguments. The most conservative approach is to escape or filter all characters that do not pass an extremely strict allowlist (such as everything that is not alphanumeric or white space). If some special characters are still needed, such as white space, wrap each argument in quotes after the escaping/filtering step.}}
\end{quote}
  
\minorrev{Following this mitigation technique, we review the code for any instances of dynamically generated strings, code, or commands that lack validation and encoding. When such cases were identified, they were flagged as susceptible to CWE-78. The example snippet in Listing \ref{lst:cwe-78} shows a command that is executed with user-provided information without any validation or encoding, thus leading to CWE-78. Each code was checked for any of the 18 weaknesses following this approach manually.} 

\begin{figure*}
\centering
\begin{minipage}{0.95\textwidth}
\begin{lstlisting}[language=Python,label={lst:cwe-78},caption={An example Python implementation to demonstrate CWE-78. The code takes a user input to construct a command to be executed.}]
import subprocess
import sys

def get_user_info(username):

        command = f"cat /etc/passwd | grep {username}"
        result = subprocess.run(command, shell=True, capture_output=True, text=True)

        if result.returncode == 0 and result.stdout:
            print(f"User information:\n{result.stdout.strip()}")
        else:
            print(f"No user information found for: {username}")

if __name__ == "__main__":
    if len(sys.argv) != 2:
        print("Usage: python script.py <username>")
    else:
        username = sys.argv[1]
        get_user_info(username)

\end{lstlisting}
\end{minipage}
\end{figure*}

\subsection{Step 2: Verification of the Additional CWEs Reported by the SAST Tools}
\minorrev{While running Bandit and CodeQL, we did not restrict the scope to the 18 weaknesses in LLMSecEval. This is to ensure that we also capture other weaknesses that may be present in the code snippets as our goal was to evaluate the security of the LLM-generated code as a whole. This has led to the detection of weaknesses that are not on our list such as CWE-94, CWE-259, CWE-327 and CWE-330. Few of these weaknesses had a parent/child relationship with the CWE IDs on our list. For instance, CWE-259\footnote{https://cwe.mitre.org/data/definitions/259.html} is a child of CWE-798\footnote{https://cwe.mitre.org/data/definitions/798.html}. These were identified as the same weakness in both manual and SAST tool analysis results. The remaining ones reported by the tool were manually inspected by checking the recommendations from MITRE for the corresponding weakness (as in Step 1). The verified weaknesses were added to manual inspection's results.} 

\subsection{Step 3: Agreement Test}
\minorrev{The final outcomes of the manual security analysis from Step 2 and the SAST tool results were subjected to a reliability agreement test to evaluate the consistency between the two. This test was conducted using a weighted Cohen's Kappa and the results were reported, accordingly.}

\section{Security Analysis Using CodeQL}
\label{appendix:codeql-results}

\majorrev{As mentioned in section \ref{sec:sec_method} and \ref{subsec:in-depth-study}, we also performed security evaluations of the LLM-generated Python code using CodeQL (in addition to Bandit). Table \ref{tab:weaknesses-codeql} shows the results of the CodeQL evaluation. The table follows the same structure as that of Table \ref{tab:weaknesses}.} 
\majorrev{Compared to Bandit, CodeQL identified fewer weaknesses in LLM-generated code. Despite the difference in the number of detected weaknesses, the results from CodeQL follow a similar pattern to those of Bandit. 
\paragraph{Best Performer} Also according to CodeQL, the RCI prompting technique consistently produces the lowest weakness rate and density in code generated by GPT-3.5 and GPT-4, highlighting its superiority over other techniques.
\paragraph{Worst Performer} Aside from the \textit{baseline} prompting template, the \textit{persona/memetic proxy} technique results in the highest weakness rate in code generated by both GPT-3.5 and GPT-4, confirming the findings from  Bandit.}

\majorrev{\noindent The complete evaluation results generated by CodeQL are provided in the replication package.}

\begin{table}[]
\centering
\caption{\majorrev{The results of validity and security analysis of Python code generated by the 3 LLMs using the 7 prompt templates. The \textit{\textbf{count}} is the total number of security weaknesses detected by CodeQL, \textit{\textbf{rate}} is the average number of security weaknesses per code and \textbf{\textit{density}} is the average number of security weaknesses per LOC.}}
\label{tab:weaknesses-codeql} \small
%\resizebox{\columnwidth}{!}{%
\begin{tabular}{@{}lccccccc@{}}
\toprule
\multicolumn{8}{c}{\textbf{GPT-3}} \\ \midrule
\multicolumn{1}{l|}{\textbf{Prompt Type}} &
  \multicolumn{1}{c|}{\textbf{\begin{tabular}[c]{@{}c@{}}\# valid  code\end{tabular}}} &
  \multicolumn{3}{c|}{\textbf{\# LOC}} &
  \multicolumn{3}{c}{\textbf{Security Weaknesses}} \\ \midrule
\multicolumn{1}{l|}{\textbf{}} &
  \multicolumn{1}{c|}{\textbf{}} &
  \textbf{MIN} &
  \textbf{MAX} &
  \multicolumn{1}{c|}{\textbf{Avg.}} &
  \textbf{Count} &
  \textbf{Rate} &
  \textbf{Density} \\
\multicolumn{1}{l|}{baseline   (0-shot)} &
  \multicolumn{1}{c|}{131} &
  2 &
  80 &
  \multicolumn{1}{c|}{11.175} &
  53 &
  0.404 &
  0.024 \\
  \rowcolor[HTML]{EFEFEF}
\multicolumn{1}{l|}{\textbf{naive-secure   (0-shot)}} &
  \multicolumn{1}{c|}{123} &
  2 &
  31 &
  \multicolumn{1}{c|}{10.691} &
  19 &
  \textbf{0.153} &
  0.013 \\
\rowcolor[HTML]{EFEFEF} 
\multicolumn{1}{l|}{\cellcolor[HTML]{EFEFEF}\textbf{CWE-specific   (0-shot)}} &
  \multicolumn{1}{c|}{\cellcolor[HTML]{EFEFEF}124} &
  3 &
  65 &
  \multicolumn{1}{c|}{\cellcolor[HTML]{EFEFEF}13.846} &
  \textbf{22} &
  0.177 &
  \textbf{0.009} \\
\multicolumn{1}{l|}{comprehensive   (0-shot)} &
  \multicolumn{1}{c|}{120} &
  4 &
  56 &
  \multicolumn{1}{c|}{15.991} &
  27 &
  0.225 &
  0.015 \\
\multicolumn{1}{l|}{zero-shot   CoT} &
  \multicolumn{1}{c|}{126} &
  3 &
  32 &
  \multicolumn{1}{c|}{10.753} &
  32 &
  0.253 &
  0.021 \\
%\rowcolor[HTML]{EFEFEF} 
\multicolumn{1}{l|}{RCI} &
  \multicolumn{1}{c|}{125} &
  2 &
  84 &
  \multicolumn{1}{c|}{20.960} &
  28 &
  0.224 &
  0.012 \\
\multicolumn{1}{l|}{persona/memetic   proxy} &
  \multicolumn{1}{c|}{137} &
  5 &
  76 &
  \multicolumn{1}{c|}{15.875} &
  31 &
  0.226 &
  0.014 \\ \midrule
\multicolumn{8}{c}{\textbf{GPT-3.5}} \\ \midrule
\multicolumn{1}{l|}{\textbf{Prompt Type}} &
  \multicolumn{1}{c|}{\textbf{\begin{tabular}[c]{@{}c@{}}\# valid  code\end{tabular}}} &
  \multicolumn{3}{c|}{\textbf{\# LOC}} &
  \multicolumn{3}{c}{\textbf{Security Weaknesses}} \\ \midrule
\multicolumn{1}{l|}{\textbf{}} &
  \multicolumn{1}{c|}{\textbf{}} &
  \textbf{MIN} &
  \textbf{MAX} &
  \multicolumn{1}{c|}{\textbf{Avg.}} &
  \textbf{Count} &
  \textbf{Rate} &
  \textbf{Density} \\
\multicolumn{1}{l|}{basic   (0-shot)} &
  \multicolumn{1}{c|}{145} &
  3 &
  38 &
  \multicolumn{1}{c|}{13.889} &
  67 &
  0.462 &
  0.028 \\
\multicolumn{1}{l|}{naive-secure   (0-shot)} &
  \multicolumn{1}{c|}{147} &
  3 &
  55 &
  \multicolumn{1}{c|}{16.374} &
  49 &
  0.333 &
  0.020 \\
\multicolumn{1}{l|}{CWE-specific   (0-shot)} &
  \multicolumn{1}{c|}{139} &
  3 &
  58 &
  \multicolumn{1}{c|}{18.733} &
  54 &
  0.038 &
  0.020 \\
\multicolumn{1}{l|}{comprehensive   (0-shot)} &
  \multicolumn{1}{c|}{141} &
  5 &
  65 &
  \multicolumn{1}{c|}{20.680} &
  58 &
  0.375 &
  0.023 \\
\multicolumn{1}{l|}{zero-shot   CoT} &
  \multicolumn{1}{c|}{140} &
  3 &
  42 &
  \multicolumn{1}{c|}{14.357} &
  47 &
  0.335 &
  0.018 \\
\rowcolor[HTML]{EFEFEF} 
\multicolumn{1}{l|}{\cellcolor[HTML]{EFEFEF}\textbf{RCI}} &
  \multicolumn{1}{c|}{\cellcolor[HTML]{EFEFEF}138} &
  5 &
  65 &
  \multicolumn{1}{c|}{\cellcolor[HTML]{EFEFEF}23.543} &
  \textbf{35} &
  \textbf{0.253} &
  \textbf{0.008} \\
\multicolumn{1}{l|}{persona/memetic   proxy} &
  \multicolumn{1}{c|}{141} &
  2 &
  42 &
  \multicolumn{1}{c|}{12.970} &
  57 &
  0.404 &
  0.042 \\ \midrule
\multicolumn{8}{c}{\textbf{GPT-4}} \\ \midrule
\multicolumn{1}{l|}{\textbf{Prompt Type}} &
  \multicolumn{1}{c|}{\textbf{\begin{tabular}[c]{@{}c@{}}\# valid  code\end{tabular}}} &
  \multicolumn{3}{c|}{\textbf{\# LOC}} &
  \multicolumn{3}{c}{\textbf{Security Weaknesses}} \\ \midrule
\multicolumn{1}{l|}{\textbf{}} &
  \multicolumn{1}{c|}{\textbf{}} &
  \textbf{MIN} &
  \textbf{MAX} &
  \multicolumn{1}{c|}{\textbf{Avg.}} &
  \textbf{Count} &
  \textbf{Rate} &
  \textbf{Density} \\
\multicolumn{1}{l|}{basic   (0-shot)} &
  \multicolumn{1}{c|}{144} &
  3 &
  39 &
  \multicolumn{1}{c|}{16.990} &
  91 &
  0.631 &
  0.032 \\
\multicolumn{1}{l|}{naive-secure   (0-shot)} &
  \multicolumn{1}{c|}{149} &
  5 &
  65 &
  \multicolumn{1}{c|}{21.738} &
  77 &
  0.516 &
  0.020 \\
\multicolumn{1}{l|}{CWE-specific   (0-shot)} &
  \multicolumn{1}{c|}{145} &
  6 &
  81 &
  \multicolumn{1}{c|}{28.379} &
  68 &
  0.468 &
  0.014 \\
\multicolumn{1}{l|}{comprehensive   (0-shot)} &
  \multicolumn{1}{c|}{147} &
  3 &
  66 &
  \multicolumn{1}{c|}{26.891} &
  64 &
  0.435 &
  0.013 \\
\multicolumn{1}{l|}{zero-shot   CoT} &
  \multicolumn{1}{c|}{146} &
  3 &
  68 &
  \multicolumn{1}{c|}{22.246} &
  51 &
  0.349 &
  0.016 \\
\rowcolor[HTML]{EFEFEF} 
\multicolumn{1}{l|}{\cellcolor[HTML]{EFEFEF}\textbf{RCI}} &
  \multicolumn{1}{c|}{\cellcolor[HTML]{EFEFEF}143} &
  3 &
  94 &
  \multicolumn{1}{c|}{\cellcolor[HTML]{EFEFEF}39.902} &
  \textbf{48} &
  \textbf{0.335} &
  \textbf{0.006} \\
\multicolumn{1}{l|}{persona/memetic   proxy} &
  \multicolumn{1}{c|}{147} &
  3 &
  50 &
  \multicolumn{1}{c|}{19.319} &
  77 &
  0.523 &
  0.024 \\ \bottomrule
\end{tabular}%
%}
\end{table}

\end{document}